\documentclass[12pt]{iopart}

\expandafter\let\csname equation*\endcsname=\relax 
\expandafter\let\csname endequation*\endcsname=\relax 
\usepackage{amsmath,amssymb,amsthm} % note amssymb will load amsfonts

\usepackage{graphicx}
\usepackage[font=small,labelfont=bf,justification=justified,format=plain]{caption} % 'format=plain' avoids hanging indentation
\DeclareUnicodeCharacter{2212}{-}

\makeatletter
\renewcommand*\env@matrix[1][\arraystretch]{%
	\edef\arraystretch{#1}%
	\hskip -\arraycolsep
	\let\@ifnextchar\new@ifnextchar
	\array{*\c@MaxMatrixCols c}}
\makeatother

\usepackage{multirow}

\usepackage{upgreek} % for \upmu

\usepackage{siunitx}

\usepackage{cite}

\begin{document}

\title[Nonlinear dynamics of Mott memristors]{Nonlinear dynamics and stability analysis of locally-active Mott memristors using a physics-based compact model}

\author{Wei Yi}

\address{HRL Laboratories, LLC., 3011 Malibu Canyon Rd, CA 90265, United States of America}
\ead{wyi@hrl.com}
\vspace{10pt}
\begin{indented}
\item[]September 2023 (Revised June 2024)
\end{indented}

\begin{abstract}
Locally-active memristors are a class of emerging nonlinear dynamic circuit elements that hold promise for scalable yet biomimetic neuromorphic circuits. Starting from a physics-based compact model, we performed small-signal linearization analyses and applied Chua's local activity theory to a one-dimensional locally-active vanadium dioxide Mott memristor based on an insulator-to-metal phase transition. This approach allows a connection between the dynamical behaviors of a Mott memristor and its physical device parameters as well as a complete mapping of the locally passive and edge of chaos domains in the frequency and current operating parameter space, which could guide materials and device development for neuromorphic circuit applications. We also examined the applicability of local analyses on a second-order relaxation oscillator circuit that consists of a voltage-biased vanadium dioxide memristor coupled to a parallel reactive capacitor element and a series resistor. We show that global nonlinear techniques, including nullclines and phase portraits, provide insights on instabilities and persistent oscillations near non-hyperbolic fixed points, such as a supercritical Hopf-like bifurcation from an unstable spiral to a stable limit cycle, with each of the three circuit parameters acting as a bifurcation parameter. The abruptive growth in the limit cycle resembles the Canard explosion phenomenon in systems exhibiting relaxation oscillations. Finally, we show that experimental limit cycle oscillations in a vanadium dioxide nano-device relaxation oscillator match well with SPICE simulations built upon the compact model.
\end{abstract}

\vspace{2pc}
\noindent{\it Keywords}: local activity, edge of chaos, memristor, vanadium dioxide, Mott transition, Hopf bifurcation, limit cycle
%\maketitle

\section[1. Introduction]{Introduction}
Recent years have witnessed a surge of interest in exploiting nonlinear dynamical phenomena in emerging devices for novel circuit applications such as neuromorphic computing. A subject that has been intensively studied is one-port (two-terminal) passive memristors that exhibit a pinched hysteresis which always passes through the origin in their current-voltage (I–V) loci, thus having a non-volatile memory effect~\cite{Strukov08,Dittmann19}. Passive memristors promise a scalable and energy-efficient venue to emulate biological synapses and realize computationally efficient neuromorphic learning rules~\cite{Kim15,Wang16,Covi18,Brivio21}. 

\begin{figure}[htb]
	\centering
	\includegraphics[width=0.9\linewidth]{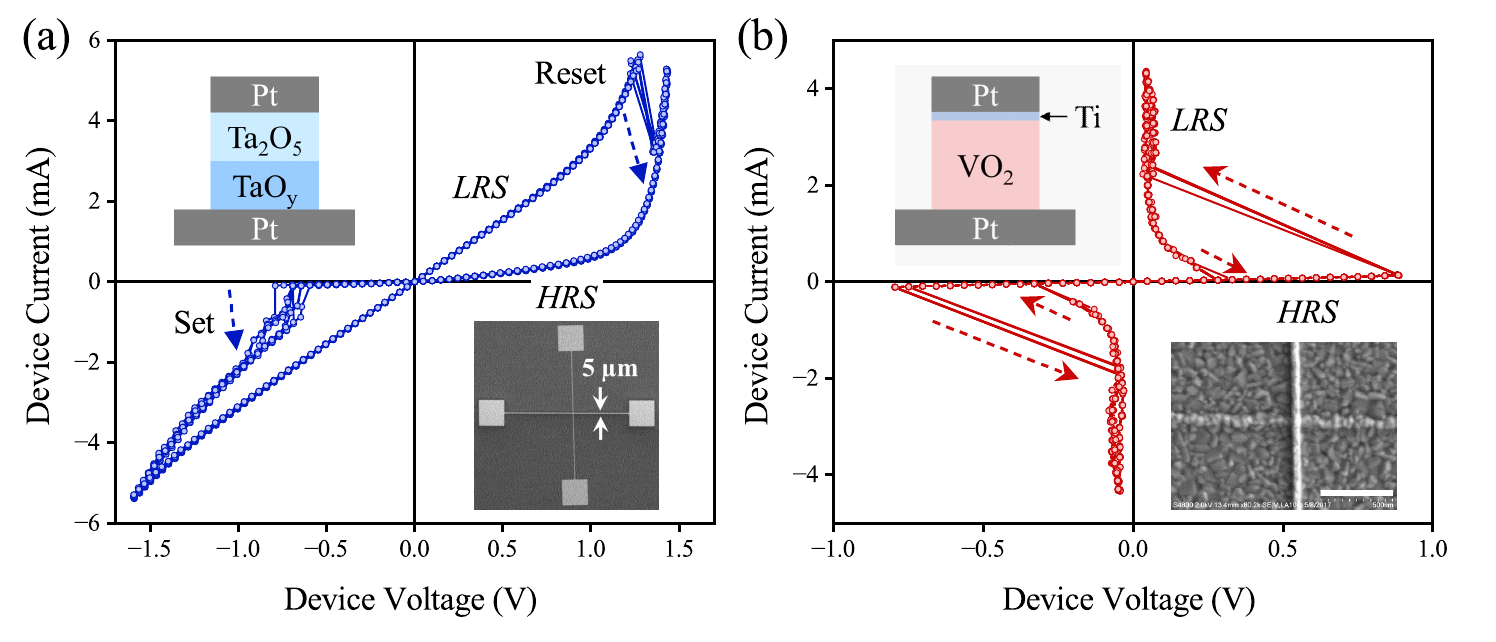}
	\caption{Experimental quasi-DC I–V curves for (a) a  TaO$_y$-Ta$_2$O$_5$ bilayer passive memristor, and (b) a VO$_2$ locally-active memristor, fabricated and characterized by HRL Laboratories, LLC. Resistance switching events are indicated by dashed arrows. Insets of (a): layer structure and optical image of the 5$\times$5~$\upmu$m$^2$ TaO$_y$-Ta$_2$O$_5$ ($y<2$) crossbar device. Insets of (b): layer structure and scanning electron micrograph of the 50$\times$50~nm$^2$ VO$_2$ nano-crossbar device (scale bar: 500~nm). Memristor crossbars are tested in a four-terminal Kelvin connection (see~\cite{Yi19} for details). The external voltage is swept at $\sim$1~V/s rate in the sequence of 0~$\rightarrow$~+$V_\text{p}$~$\rightarrow$~0~$\rightarrow$~–$V_\text{n}$~$\rightarrow$~0 (repeated 10 times). $V_\text{p}(V_\text{n})=2.5(2)$~V in (a) and $V_\text{p,n}=1.45$~V in (b). The metal electrodes contribute a series resistance of 600--800~$\Omega$. }
	\label{Fig1_ActPassMemRIV}
\end{figure}

Although a canonical memristor is a passive circuit element, any one-port device that exhibits a pinched hysteresis is an extended memristor~\cite{Chua14}, which includes a class of one-port devices that exhibit a non-monotonicity in their experimental quasi-direct current (quasi-DC) I–V curves --- a negative differential resistance (NDR). These devices typically exhibit a pronounced I–V hysteresis when driven by a voltage stimulus, however the hysteresis \textit{collapses} at a finite voltage, therefore they only have a transient (volatile) memory effect. Importantly, these passive one-port devices are \emph{locally active} within the NDR region and thus offers a gain of alternating-current (AC) signals --- a must-have capability for information processing and communication that has been dominated by transistors. Figure~\ref{Fig1_ActPassMemRIV} compares typical  quasi-DC I–V curves measured from a passive memristor and a locally-active memristor. Such a measurement varies V or I stimulus slowly and measures  time-averaged device responses, which could capture the resistance states before and after a resistance switching event (see arrows), but not the ultrafast switching dynamics that could occur within femtoseconds. Figure~\ref{Fig1_ActPassMemRIV}(a) shows a bipolar tantalum oxide (TaO$_y$-Ta$_2$O$_5$) passive memristor which switches from a low-resistance state (LRS) to a high-resistance state (HRS) if a sufficiently-large positive voltage is applied (Reset), and switches from the HRS back to the LRS if a sufficiently-large negative voltage is applied (Set). Both Reset and Set operations are nonvolatile, i.e.,  resistance changes are retained after powered off. In contrast, figure~\ref{Fig1_ActPassMemRIV}(b) shows a unipolar vanadium dioxide (VO$_2$) locally-active memristor, which abruptly switches from a HRS to a LRS if a sufficiently-large voltage is applied, regardless of its polarity. As voltage is reduced below a smaller threshold level, the device switches from the LRS back to the HRS --- resistance changes are volatile and get lost after powered off. Be mindful that the exemplary characteristics in figure~\ref{Fig1_ActPassMemRIV} are by no means exhaustive. Passive memristors can have either bipolar or unipolar non-volatile resistance switching behaviors, determined by intertwined ionic and electronic transport mechanisms within the nanoscale device volume~\cite{Waser09,Jeong12}. They may also exhibit a fading memory effect with its asymptotic behavior solely determined by the state dynamics irrespective of the initial condition~\cite{Ascoli16,Pershin19}.

We now turn our attention to those one-port devices that possess the peculiar I–V characteristics in figure~\ref{Fig1_ActPassMemRIV}(b). In fact, on-port devices with such switching characteristics have long been studied and put into engineering practice. They have been made out of many materials based on a variety of operating mechanisms. A familiar category is electro-thermal threshold switches such as ovonic threshold switches (OTS), which show rapid changes in resistance due to nonlinear interactions among local temperature, metastable structural change, and electrical conductivity~\cite{Ovshinsky68,Noe20}. Being a one-port device, locally-active memristors (LAMs) and passive memristors share the same level of 4F$^2$ (F: half pitch in lithography) scalability in a crossbar device geometry~\cite{Amsinck05}, resolving the trade-off between scalability and biological fidelity.

LAMs may be generally classified into two types, current-controlled (S-type) and voltage-controlled ($\lambda$-type), where the letter ``S'' and ``$\lambda$'' resembles the characteristic shape of NDR in the I–V curve plotted with voltage as the $x$ axis~\cite{Ridley63}. S-type LAMs are ``normally-off'' devices with a HRS when powered off. $\lambda$-type LAMs such as resonant tunneling diodes are ``normally-on'' devices with a LRS when powered off~\cite{Esaki58}. Therefore S-type LAMs are superior choices than $\lambda$-type LAMs with respect to the standby power consumption. Hereafter we focus our discussions on current-controlled LAMs.

A particularly interesting class of current-controlled LAMs is based on the insulator-to-metal phase transition (IMT) phenomena in strongly correlated materials that arise from a coupling between structural distortions (Peierls transition) and electronic instabilities (Mott transition)~\cite{Andrews19}. They possess several attractive features for circuit applications, such as ultralow (femtojoule) switching energy~\cite{Prinz20}, ultra-fast (tens of femtosecond) switching speed~\cite{Jager17}, and electroforming-free operations~\cite{Yi18}. We term all these IMT based LAMs ``Mott memristors'' without discerning the subtle differences in their phase transition mechanisms. Vanadium dioxide (VO$_2$) and niobium dioxide (NbO$_2$) are two intensively-studied Mott memristor materials among many others~\cite{Andrews19}.

For neuromorphic computing applications, circuits of self-excited oscillators and spiking neuron emulators have been built with one or more LAMs that are coupled with reactive elements (capacitors)~\cite{Farhat93,Moon15,Ignatov15,Stoliar17,Wang18}. An illustrative example is a scalable spiking neuron, which constitutes two oppositely energized (``polarized'' in neuroscience glossary) LAMs to mimic the voltage-gated potassium and sodium cell membrane ion channels. Coupled with parallel membrane capacitors and series load resistors, the composite circuit emulates a single-compartment nerve cell initiating all-or-nothing action potentials upon a suprathreshold stimulus~\cite{Yi18,Pickett13a}, or acting as a delayed buffer that allows bidirectional distortion-free propagation of action potentials if daisy chained~\cite{Pickett13b,Yi22}. Such a circuit topology bears some resemblance to a biologically-plausible Hodgkin-Huxley (HH) axon model~\cite{Chua12}, as well as the early 1960s proposals of ``neuristor'' axons utilizing nonscalable components such as inductors~\cite{Crane60,Crane62,Nagumo62}. Experimental spiking neurons built with VO$_2$ Mott memristors showed two dozens of biological neuronal temporal dynamics, including all three classes of neuronal excitability~\cite{Yi18}. Arguably, LAM-based neurons and passive memristor-based synapses form a self-sufficient basis to construct a transistorless neural network~\cite{Yi19}. 

Despite the wealth of experimental demonstrations, predictive modeling and analysis of LAM elements and circuits is nontrivial and remains a handicap for technology development. The difficulties are partially by virtue of fundamental mathematical challenges about nonlinear differential systems. One illustrious example is the second part of Hilbert's $16^{\text{th}}$ problem that questions whether there exists a finite upper bound for the number of limit cycles of planar polynomial differential systems. It remains unsolved today even for quadratic polynomials (degree $n=2$)~\cite{Ilyashenko02}. Qualitative local analysis, on the contrary, are facilitated by small-signal linearization techniques, which allow linear analysis to be applied to a nonlinear system near a \textit{hyperbolic} fixed point with all eigenvalues of the linearization having non-zero real parts~\cite{Perko91}. A key theoretical contribution made by Chua is the local activity (LA) theorem, which provides a rigorous mathematical definition of the LA as a necessary prerequisite for the emergence of complexity in nonlinear systems~\cite{Chua05}. Moreover, Chua provided a set of explicit and computable criteria in the parameter space, that allows identifying the \emph{edge of chaos} (EOC) region that is both locally active and locally stable, where most of the complexity phenomena emerge.

Mathematically rigorous but unphysical toy models of nonlinear dynamical elements were frequently used in the LA analysis procedure~\cite{Mannan16,Mannan17}. For engineering practice, such toy models could not offer a link between circuit or network level dynamics and measurable physical properties of constituent components. A recent review thoroughly elaborated the importance of applying appropriate device physics into mathematical memristor framework, and defining physically relevant model parameters to control the circuit dynamic behavior~\cite{Brown22a}.

The main objective of this manuscript is to apply relevant theoretical techniques to understand the dynamics and stability of nonlinear circuits that involve locally-active Mott memristors, and to map the conditions for the LA regime within the design parameter space~\cite{Messaris21,Ascoli21}. These theoretical techniques include essential local analysis methods such as the small-signal linearization and the LA theorem, and global techniques such as the nullclines and phase portraits. For engineering relevance, we base our analyses on an analytical one-dimensional (1D) Mott memristor compact model that is built on the laws of thermodynamics and only contains physically relevant device parameters. The model was developed by Pickett and Williams for NbO$_2$ Mott memristors~\cite{Pickett12}. Previously we have verified that it is also applicable to VO$_2$ Mott memristors after replacing the material properties~\cite{Oh10,Berglund69}, and our SPICE simulations based on this model faithfully reproduced most of the measured neuronal dynamics in neuron circuits built with VO$_2$ memristors~\cite{Yi18}. Here, we demonstrate that this physics-based compact model is mathematically tractable for applying the local and global analysis techniques, having closed-form expressions for all the important quantities involved in the analyses. It enables a connection between the system dynamics and component physical parameters to guide circuit designs and process development. The algorithmic analysis procedure we present using a VO$_2$ Mott memristor model is general in nature and suitable for analyzing other Mott memristor materials. Qualitatively, the predictions on the dynamics and stability by the present work are similar to those made by compact models based on different choices of state variable and kinetic function~\cite{Brown22a}. 

We focus on theoretical analyses and only included a cursory comparison between the model simulated and experimental characteristics of a VO$_2$ nano-device relaxation oscillator near the end. More detailed comparisons in the context of VO$_2$ Mott memristor neurons can be found in supplementary materials of ~\cite{Yi18}. It is understood that the compact model presented here has some simplifications and limitations. It is a nontrivial task to construct a computationally-efficient compact model for locally-active memristors with an appropriate balance between the dynamical fidelity and the computational complexity of solving the model equations. This is especially important for digital computer simulations of a scaled network that contains many instances of memristor elements, which could be costly in time and energy consumption.

The manuscript is organized as follows. After the introduction section, the first three sections (Section 2 -- 4) are dedicated to analyses of an isolated 1D Mott memristor. First (Section 2), we introduce the physics-based compact model and analyze the stability of an isolated 1D Mott memristor by examining its power-off plot and dynamic route map under constant input currents or voltages. This exercise confirms that the metallic state of a Mott memristor is unstable without power and is asymptotically stable with a finite input current. It also reveals that varying voltage as the bifurcation parameter leads to a supercritical saddle-node bifurcation. Then (Section 3) we solve its locus of steady states (fixed points) in the three-dimensional (3D) state space and their two-dimensional (2D) projections. Note that we use both fixed point and steady state for the same concept in an interchangeable manner, but avoid the term equilibrium unless the input is zero. See subsection 2.2 for an elaboration on this topic.

In Section 4, we apply local analysis techniques on an isolated Mott memristor, including linearization and small-signal analysis, pole-zero diagram, Chua's LA theorem, and frequency response. Its complex-domain ($s$-domain) equivalent circuit derived by Laplace transform contains three virtual elements --- a negative nonlinear capacitor in parallel with a negative nonlinear resistor, both in series with a positive nonlinear resistor. The negative $s$-domain capacitance gives rise to an apparent inductive response, similar to the memristive models of potassium and sodium ion channels~\cite{Chua12}. We found that an isolated Mott memristor about a fixed point dwells either in the locally passive (LP) or the EOC region. The EOC region coincides with the NDR region in its steady-state or DC I–V locus. Brown et al.~\cite{Brown22a} elaborated that for an extended electro-thermal memristor, the coincidence between NDR and EOC or LA regions is not guaranteed. Therefore NDR shall not be used as a sole signature for EOC. In our case, the crossover between the LP and EOC regions also manifests itself in the small-signal frequency response, which shows a sign reversal in the real part of the impedance (complexity) function Re$Z(s;Q)$ as predicted by the fourth LA criterion. In the frequency domain, an isolated Mott memristor is equivalent to a positive inductor in series with a resistor that is positive in the LP region and negative in the EOC domain. We derived the parametric Nyquist plot of the LP~$\leftrightarrow$~EOC crossover at a single current level, then extended it to a 2D color scale map of Re$Z(s;Q)$ to visualize the LP and EOC regions in the parameter space spanned by frequency and current, which is effectively a phase diagram for complexity. We also examined the scaling trend of the EOC region versus the device size, which shows that the VO$_2$ conduction channel radius is the relevant dimension for device miniaturization to enhance the EOC frequency regime.

Although an EOC region exists in an isolated 1D Mott memristor, the topological constraint limits the dynamics it can possess, making it impossible to exhibit persistent oscillations. In the last two sections (Section 5 -- 6), we lift the topological constraint for an isolated 1D Mott memristor by coupling it to one or more reactive elements, raising the system dimensionality and the dynamical complexity. For simplicity, we choose a DC voltage ($V_\text{dc}$) biased Pearson–Anson relaxation oscillator formed by a Mott memristor coupled with a parallel capacitor $C_\text{p}$ and a series resistor $R_\text{s}$ as an example 2D nonlinear system for our analysis~\cite{Pearson21}. The same analysis procedure can be applied to higher-dimensional systems, such as spiking neuron circuits consisting of two or more Mott memristors coupled with passive and reactive elements.

In Section 5 we first apply local linearization techniques on this example system, including the element combination approach, the Jacobian matrix method, and the trace-determinant plane classification, to study stability and qualitative behaviors about its hyperbolic fixed points. The element combination approach treats a Mott memristor in parallel with a capacitor as a composite second-order nonlinear element. The small-signal transfer function of the dimension-reduced system has a pair of complex conjugate poles. We derived the Nyquist plot and a 2D phase diagram of the system's poles visualizing the LP and EOC regions in the circuit parameter space. These results are corroborated by the the trace-determinant plane analysis of the Jacobian linearized 2D system, which reveals a stability-change bifurcation as the parametric (trace, determinant) locus crosses the zero-trace axis as one of the three circuit parameters is varied ($R_\text{s}$, $C_\text{p}$ and $V_\text{dc}$). However, analysis of stability behavior about a non-hyperbolic center requires additional theoretical tools, since the Hartman-Grobman theorem is not applicable with the loss of hyperbolicity~\cite{Hartman60,Grobman59}.

Finally, in Section 6 we apply several global methods, such as the nullclines and numerical phase portrait analyses, to understand qualitative behaviors about the non-hyperbolic centers in this example 2D nonlinear system. We found that each of the three circuit parameters ($R_\text{s}$, $C_\text{p}$ and $V_\text{dc}$) acts as a bifurcation parameter that switches the stability of a fixed point as the parametric (trace,~determinant) locus crosses a center. We verified that there exists a supercritical 2D Hopf-like bifurcation, i.e., the creation of a stable limit cycle encircling an unstable spiral as the fixed point switches its stability from stable to unstable. We also noticed that the limit cycle appears abruptively over an extremely thin bifurcation parameter interval, a phenomenon known as ``Canard explosion'' for relaxation oscillations in chemical and biological systems~\cite{Krupa01,Rotstein12}. This is a prominent distinction than the classical Hopf bifurcation, which predicts a gradual growth proportional to the square root of the bifurcation parameter. Each bifurcation parameter has a different bifurcation growth characteristics. We end the section with a comparison between experimental limit cycle characteristics of a VO$_2$ relaxation oscillator and SPICE simulations based on the Mott memristor model, showing excellent agreements between them.

We conclude the manuscript with brief remarks about the application implications of locally-active memristors and scalable neuromorphic dynamic neurons with high degree of complexity.

\section[2. Model]{One-dimensional locally-active Mott memristor}

\subsection[2.1]{The physics-based analytical compact model}
\begin{figure}[htb]
	\centering
	\includegraphics[width=0.5\linewidth]{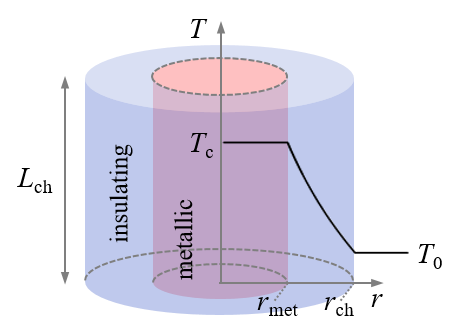}
	\caption{Schematics of the biphasic thermal model for a Mott memristor that undergoes an insulator-to-metal transition, illustrating a cylindrical conduction channel with a metallic-phase core surrounded by an insulating-phase shell. The model assumes that the metallic core is clamped to the transition temperature $T_\text{c}$, and the outer edge of the conduction channel is clamped to the ambient temperature $T_0$. The black solid line shows a calculated radial temperature profile. The top and bottom electrodes are not shown for clarity.}
	\label{Fig2_ModelDevice}
\end{figure}

The physics-based compact model for a 1D (one state variable) locally-active Mott memristor is biphasic in nature~\cite{Pickett12}. It assumes that once an IMT is triggered by Joule self-heating beyond a threshold level, metallic and insulating phases coexist in a constant-volume conduction channel defined by the top and bottom electrodes. For mathematical simplification, the conduction channel has an axial symmetry with a constant radius $r_\text{ch}$ along its length. An experimental crossbar device may have a square or rectangular cross section defined by its top and bottom electrodes. The insulating phase has much lower thermal and electrical conductivity than those of the metallic phase, therefore the core region turns metallic first and its radius $r_\text{met}$ grows as Joule heating increases. In analogy to the case of an ice-water mixture, the maximum temperature within the metallic core is capped to the transition temperature $T_\text{c}$ until the whole conduction channel turns metallic. The minimum temperature at the outer edge of the insulator shell is clamped to the ambient temperature $T_0$. The temperature rise required for IMT to occur is defined as $\Delta T=T_\text{c}-T_0$. With these assumptions, a radial temperature profile bounded between $T_0$ and $T_\text{c}$ is established across the insulating shell surrounding the metallic core. The schematics of this biphasic thermal model is shown in figure \ref{Fig2_ModelDevice}.

The state variable $x\triangleq r_\text{met}/r_\text{ch}$ is modeled as the dimensionless volumetric fraction of metallic phase in the conduction channel and is bounded between 0 and 1. The model derives that the temperature at a specific radius $T(r)$ is a nonlinear function of $x$ of the form $T(r)=T_0+\Delta T\ln(\frac{r}{r_\text{ch}})/\ln(x)$, $r_\text{met}\leq r\leq r_\text{ch}$.

Another assumption that the model makes for mathematical simplification is to ignore the axial heat exchanges with the electrodes and the associated temperature gradients near the top and bottom interfaces. Moreover, thermal and electrical conductivity of the insulating shell are approximated as constants regardless of the radial temperature gradient across it. This approximation holds true if neither of them varies significantly as temperature rises from $T_0$ to $T_\text{c}$. This is probably the case for VO$_2$ with a small $\Delta T$ ($T_\text{c}\approx340$~K, $\Delta T\approx43$~K)~\cite{Berglund69}, but becomes questionable for NbO$_2$ with a very large $\Delta T$ ($T_\text{c}\approx1080$~K, $\Delta T\approx784$~K)~\cite{Janninck66}.

The compact model consists of two coupled equations that satisfy the definition of a 1D extended memristor~\cite{Chua14}: a state-dependent instantaneous relationship between voltage and current in the form of Ohm's law (state-dependent Ohm's law), and a first-order ordinary differential equation (ODE) that determines the dynamics of the single state variable $x$ (state equation). The kinetic function that accounts for the state dynamics is a function of both the state variable $x$ and the input variable (voltage $v$ or current $i$). A Mott memristor therefore is a dynamical system --- a system whose state at a future time depends deterministically on its present sate and a physical law that governs its evolution over time.

Since Joule self-heating depends on the passage of current, a Mott memristor is a current-controlled memristor, and current $i$ instead of voltage $v$ is the appropriate input variable. The model equations take the following form
\begin{eqnarray}\label{eqarray1}
	\label{eqn1}
	v(x,i)=R_\text{ch}(x)i \\
	\label{eqn2}
	\frac{dx}{dt} \triangleq f_x(x,i)=\frac{i^2R_\text{ch}(x)-\Gamma_\text{th}(x)\Delta T}{H'(x)}
\end{eqnarray}
The single state variable $x\in(0,1)$ is a dimensionless quantity within the bounded open interval between 0 and 1. $f_x(x,i)$ is the kinetic function for the state variable $x$. The derivation of $f_x(x,i)$ is nontrivial and is the main task of building the compact model. For Mott memristors and generally electro-thermal memristors, $f_x(x,i)$ is derived from the first law of thermodynamics, which states that the change in the total enthalpy of a system $\Delta H$ equals the net heat flow $q_p$ into it at constant pressure: $\Delta H=q_p$. Therefore, their time derivatives are equal as well: $\frac{d\Delta H}{dt}=\frac{dq_p}{dt}$. This basic law forms the theoretical basis to interpret electro-thermal memristors wherein the local temperature change plays a key role. It is worth pointing out that since there is no explicit dependence on time $t$ in $f_x(x,i)$, this is an \emph{autonomous} system.

To simplify the expression for $f_x(x,i)$, three nonlinear auxiliary functions are defined: the state-dependent memristance function $R_\text{ch}(x)$, the state-dependent thermal conductance function $\Gamma_\text{th}(x)$, and $H'(x) \triangleq \frac{d\Delta H}{dx}$ is defined as the derivative of the total enthalpy change $\Delta H$ with regard to the state variable $x$
\begin{eqnarray}\label{eqarray2}
	\label{eqn3}
	R_\text{ch}(x)=\frac1{A(1+Bx^2)} \\
	\label{eqn4}
	\Gamma_\text{th}(x)\Delta T=-\frac C{\ln x} \\
	\label{eqn5}
	H'(x) \triangleq \frac{d\Delta H}{dx}=D\left(\frac{1-x^2+2x^2\ln x}{2x(\ln x)^2}+Ex\right)
\end{eqnarray}
where $A=\frac{\pi r_\text{ch}^2}{\rho_\text{ins}L_\text{ch}}$, $B=\frac{\rho_\text{ins}}{\rho_\text{met}}-1$, $C=2\pi L_\text{ch}\kappa \Delta T$, $D=\pi L_\text{ch}r_\text{ch}^2c_\text{p}\Delta T$, and $E=\frac{2\Delta h_\text{tr}}{c_\text{p}\Delta T}$ are constant coefficients whose values are determined by physical model parameters, including material properties and device geometry. Table~\ref{table1} lists values of these physical model parameters for the case of VO$_2$ material. The radius and length of the memristor conduction channel are device-dependent parameters and can be determined experimentally by the device geometry. Rest of the model parameters listed in Table 1 are electronic, thermal and phase transition properties of VO$_2$ material reported in literature\cite{Oh10,Berglund69}. Optimization of these material property-dependent parameters can be achieved by a calibration procedure with well-devised characterizations of VO$_2$ devices and least-square data fitting~\cite{Brown22b}, but is beyond the scope of this work.

\begin{table}
	\caption{\label{table1}Material properties and device parameters for a VO$_2$ Mott memristor model.} 
	%\begin{indented}
		\lineup
		%\item[]\begin{tabular}{@{}*{5}{c}}
		\begin{tabular}{@{}*{5}{c}}
			\br                              
			Model Property&Symbol&Value&Unit&Reference \\ 
			\mr
			Volumetric heat capacity&$c_\text{p}$&\num{3.30E6}&$\mathrm{Jm^{-3}K^{-1}}$&\cite{Oh10} \\
			Volumetric enthalpy change of IMT&$\Delta h_\text{tr}$&\num{2.35E8}&$\mathrm{Jm^{-3}}$&\cite{Berglund69} \\ 
			Thermal conductivity of insulating phase&$\kappa$ &3.5&$\mathrm{Wm^{-1}K^{-1}}$&\cite{Oh10} \\ 
			Electrical resistivity of metallic phase&$\rho_\text{met}$&\num{3.00E-6}&$\mathrm{\Omega m}$&\cite{Oh10,Berglund69} \\ 
			Electrical resistivity of insulating phase&$\rho_\text{ins}$&\num{1.00E-2}&$\mathrm{\Omega m}$&\cite{Berglund69} \\
			Temperature rise of IMT&$\Delta T$&43&K&\cite{Berglund69} \\
			Radius of the conduction channel&$r_\text{ch}$&\num{3.60E-8}&m&Experimental \\
			Length of the conduction channel&$L_\text{ch}$&\num{5.00E-8}&m&Experimental \\
			\br
		\end{tabular}
	%\end{indented}
\end{table}

Table~\ref{table2} lists values of model coefficients $A$, $B$, $C$, $D$, and $E$ for three arbitrarily-chosen VO$_2$ device sizes ---  the radius $r_\text{ch}$ and length $L_\text{ch}$ of the conduction channel. Coefficients $B$ and $E$ are dimensionless and device size-independent. Without loss of generality, these three device sizes are used throughout this manuscript to illustrate the scaling trend of a calculated quantity as the device size varies. If not mentioned explicitly, hereafter the modeled VO$_2$ device is the medium-sized one in table~\ref{table2} with $r_\text{ch}=36$~nm and $L_\text{ch}=50$~nm, and is referred to as the \textit{midsize} VO$_2$ Mott memristor or midsize VO$_2$ device.

\begin{table}
	\caption{\label{table2}Values of model coefficients for three VO$_2$ device sizes.} 
	%\begin{indented}
		\lineup
		\renewcommand{\arraystretch}{1.5}
		%\item[]\begin{tabular}{@{}*{6}{c}}
		\begin{tabular}{@{}*{6}{c}}
			\br                              
			Coefficient&Formula&Unit&$r_\text{ch}=10$~nm, &$r_\text{ch}=36$~nm, &$r_\text{ch}=56$~nm, \\ 
			& & &$L_\text{ch}=10$~nm&$L_\text{ch}=50$~nm&$L_\text{ch}=100$~nm \\
			\mr
			$A$&$\frac{\pi r_\text{ch}^2}{\rho_\text{ins}L_\text{ch}}$&Mho&\num{3.14159E-6}&\num{8.14301E-6}&\num{9.85203E-6} \\ 
			$B$&$\frac{\rho_\text{ins}}{\rho_\text{met}}-1$&Unitless&3332.3&3332.3&3332.3 \\ 
			$C$&$2\pi L_\text{ch}\kappa \Delta T$&Watt&\num{9.45619E-6}&\num{4.7281E-5}&\num{9.45619E-5} \\ 
			$D$&$\pi L_\text{ch}r_\text{ch}^2c_\text{p}\Delta T$&Joule&\num{4.45792E-16}&\num{2.88873E-14}&\num{1.398E-13} \\
			$E$&$\frac{2\Delta h_\text{tr}}{c_\text{p}\Delta T}$&Unitless&3.31219&3.31219&3.31219 \\
			\br
		\end{tabular}
		\renewcommand{\arraystretch}{1}
	%\end{indented}
\end{table}

A more general approach for physical modeling of an electro-thermal memristor treats the internal temperature as the sole state variable~\cite{Brown22a}. The kinetic function is derived from Newton’s law of cooling, which connects between the net heating power and a time-varying device internal temperature through a temperature-dependent thermal capacitance. There is clearly a benefit of adopting a universal state variable and a generalized formula of the kinetic function, albeit the temperature dependence of thermal capacitance is unknown and requires a model fitting with experimental characterizations, such as the temperature dependence of self-excited oscillation frequency in a memristor-based relaxation oscillator~\cite{Brown22b}. It is curious that both approaches can reach the same qualitative predictions about the system dynamics, despite the differences in the model assumptions, state variable, and kinetic function.

\subsection[2.2]{Stability analyses}
We start the stability analyses by looking at an isolated or uncoupled Mott memristor. The first step is to examine the stability of solutions for equation (\ref{eqn2}) by treating input current as a parameter with a zero or nonzero constant value, and plotting the kinetic function $f_x(x,i)$ as a function of the state variable $x$. If a solution $f_x(x,i)=0$ exists at a point $x_Q$, it is called a fixed point~\cite{Shashkin91}. This is because the state variable $x(t)$ with an initial condition $x(0)=x_Q$ remains unchanged at any future time, i.e., $x(t)=x_Q$ for $t>0$. Literature from different disciplines have adopted a variety of terminologies for the same concept, including stationary point, invariant point, equilibrium point, critical point, singular point and steady state point. These terms are generally exchangeable, but may be confusing if not carefully chosen. Especially the use of equilibrium point may cause misinterpretation by physical scientists for reasons we will elaborate below.

A system at equilibrium remains stable over time and does not require a net flow of energy or work to maintain that condition. A steady state also has stable internal states that remain unchanged over time. However, it requires a continuous energy input or work from the external environment to remain in a constant state. A memristor that has stable internal states while having a \emph{finite} current flowing through it is at a nonequilibrium steady state rather than at an equilibrium, since there is a net heat transfer $q_p$ into the memristor. In this manuscript, we mainly use the term of fixed point because of its prevalence in mathematics. Steady state will also be used as a descriptive term when it facilitates interpretation. For example, steady-state resistance is a preferred term than fixed-point resistance.

For a current-controlled memristor, current is the appropriate input variable for stability analysis. However, one can still treat voltage $v$ as an input and run the same type of analysis. Interestingly, doing so would result in a bifurcation -- a qualitative change in the solution of a nonlinear system incurred by a small change in a parameter, such as creation or annihilation of fixed points, or a change in their stability.

\subsubsection[2.2.1]{Power-off plot}
The question about whether or not a memristor is non-volatile can be answered by looking at the power-off plot (POP)~\cite{Chua15}. For a current-controlled memristor, its POP is the locus of the kinetic function $f_x(x,i)$ as a function of the state variable $x$ at zero input current, i.e., the locus of $f_x(x,0)$ vs.~$x$.

Setting input current to be zero in equation (\ref{eqn2}), one get
\begin{equation}\label{eqn6}
	f_x(x,0)=\frac{-\Gamma_\text{th}(x)\Delta T}{H'(x)}=\frac{2Cx\ln x}{D[1-x^2+2x^2\ln x+2E(x\ln x)^2]}
\end{equation}
If $f_x(x,0)$ has an intersection with the $x$-axis, then the intersection is an equilibrium point $x_Q$. The memristor state $x(t)$ with an initial state $x(0)=x_Q$ remains unchanged at any future time, i.e., $x(t)=x_Q$ for any $t>0$.

\begin{figure}[htb]
	\centering
		\includegraphics[width=0.7\linewidth]{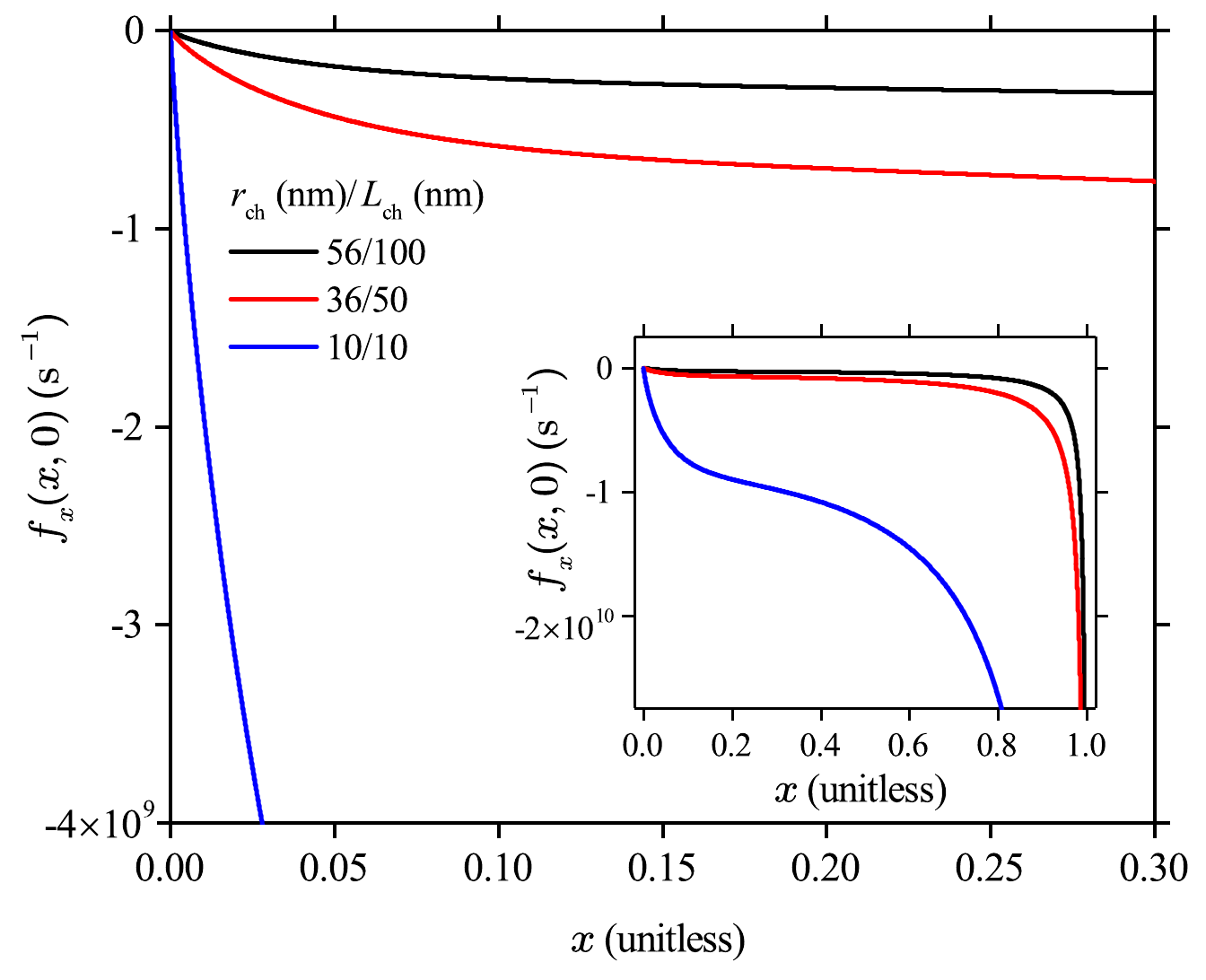}
	\caption{Analytically calculated power-off plots $f_x(x,0)$ vs.~$x$ for three different-sized VO$_2$ Mott memristors (as labeled) in the small $x$ region ($0<x<0.3$). Inset shows the same power-off plots for a much wider range of $0<x<1$.}
	\label{Fig3_POP}
\end{figure}

Figure \ref{Fig3_POP} shows that for a Mott memristor at zero input current, $f_x(x,0)$ remains negative for any state variable $x\in(0,1)$. It is plausible since if there were a finite fraction of the conduction channel in metallic phase at the beginning, it is unstable without the presence of Joule heating, and will always vanish over time. The memory effect in a Mott memristor is therefore transient or volatile in nature, and will be lost given long enough time after the removal of electrical power. Figure \ref{Fig3_POP} inset shows that the negative rate of change in $x$ increases dramatically as $x$ approaches 1.0 asymptotically. The calculations are performed using VO$_2$ model parameters, but this conclusion is generally applicable to other Mott memristor materials. 

\subsubsection[2.2.2]{Dynamic route map at constant input current}
\begin{figure}[htb]
	\centering
		\includegraphics[width=0.9\linewidth]{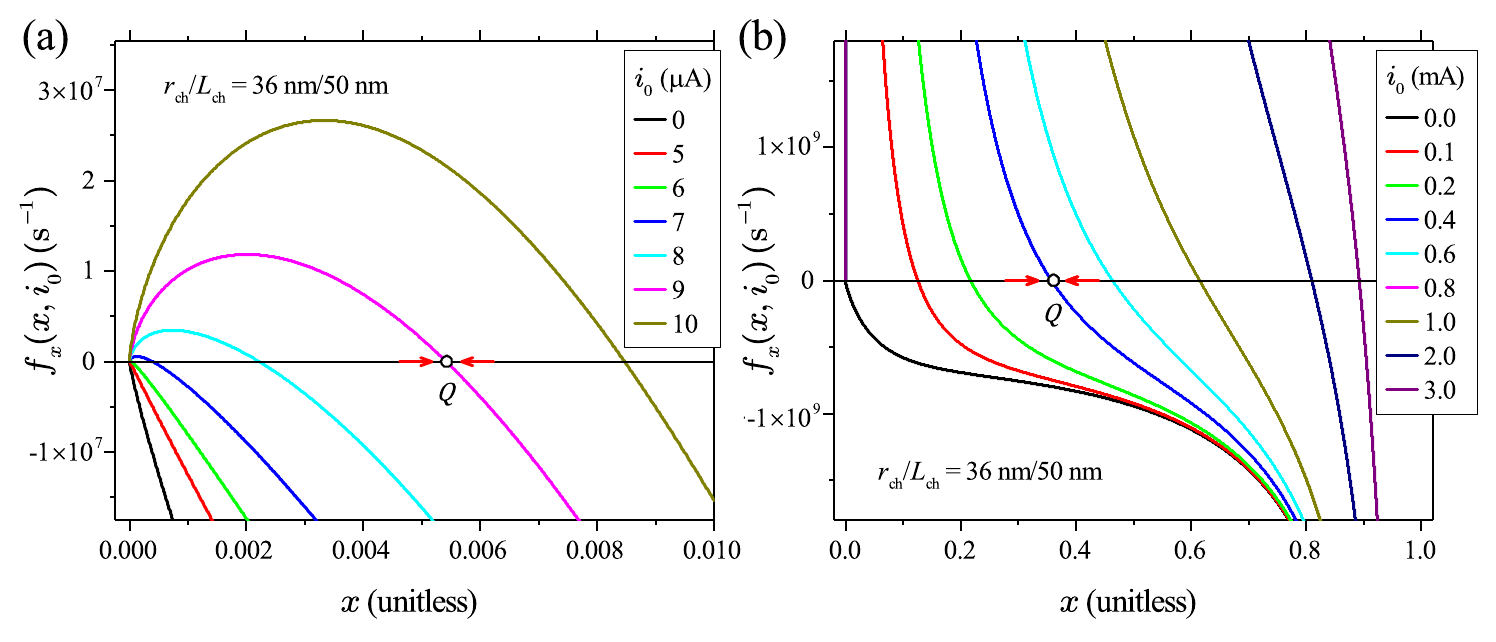}
	\caption{Analytically calculated dynamic route map of $f_x(x,i_0)$ at constant input current levels for the midsize VO$_2$ Mott memristor, plotted with (a) a narrow range of $i_0$ from 0 to 10~$\upmu$A, and (b) a wide range of $i_0$ from 0 to 3~mA. The open circle in (a) and (b) highlights a fixed point $Q$ where the $f_x(x,i_0)$ locus intersects the $x$-axis. Arrowheads show the direction of move for a solution $x(t)$ starting from an initial state located close to $Q$.}
	\label{Fig4_DR_IMode}
\end{figure}

If input current is fixed at a finite constant level $i_0\neq 0$, one can plot the dynamic route (DR) --- the locus of the kinetic function $f_x(x,i_0)$  as a function of the state variable $x$ at a constant input current $i_0$~\cite{Chua69}. A set of dynamic routes parameterized by input current (or voltage for a voltage-controlled memristor) is called a dynamic route map (DRM)~\cite{Chua18}. Rewriting $f_x(x,i)$ in equation (\ref{eqn2}) by replacing the auxiliary functions $R_\text{ch}(x)$, $\Gamma_\text{th}(x)$ and $H'(x)$ with their explicit expressions, one get
\begin{equation}\label{eqn7}
	f_x(x,i_0)=\frac{\frac{2x(\ln x)^2}{A(1+Bx^2)}i_0^2+2Cx\ln x}{D[1-x^2+2x^2\ln x+2E(x\ln x)^2]}
\end{equation}
As figure \ref{Fig4_DR_IMode}(a) shows, even a tiny input current of a few $\upmu$A creates a positive slope for the DR locus of the midsize VO$_2$ device, flipping the fourth-quadrant POP locus up into the first quadrant once a finite current is supplied. The slope of DR then levels off and becomes negative again as $x$ further increases. Consequently, a constant-current DR locus always intersects the $x$−axis at a single fixed point $x_Q$. This is confirmed by figure \ref{Fig4_DR_IMode}(b) which shows the DRM loci with a much wider range of current from 0 to 3~mA.

The theory of nonlinear dynamics tells us that the fixed point $x_Q$ is \emph{asymptotically stable} because the solution $x(t)$ starting from any initial state $x(0) \neq x_Q$ approaches the fixed point $x_Q$ as $t\to\infty$. For $x<x_Q$, $dx/dt>0$. For $x>x_Q$, $dx/dt<0$. The arrowhead pointing to the right indicates that the solution $x(t)$ starting from any initial state $x(0) \neq x_Q$ on the DR above the $x$-axis must move to the right of $x(0)$, because $dx/dt>0$ for $t>0$, as long as $x(t)$ lies above the $x$-axis. Conversely, the arrowhead pointing to the left indicates that the solution $x(t)$ starting from any initial state $x(0) \neq x_Q$ below the $x$-axis on the DR must move to the left of $x(0)$, because $dx/dt<0$ for $t>0$, as long as $x(t)$ lies below the $x$-axis.

\subsubsection[2.2.3]{Dynamic route map at constant input voltage: saddle-node bifurcation}
\begin{figure}[htb]
	\centering
		\includegraphics[width=0.9\linewidth]{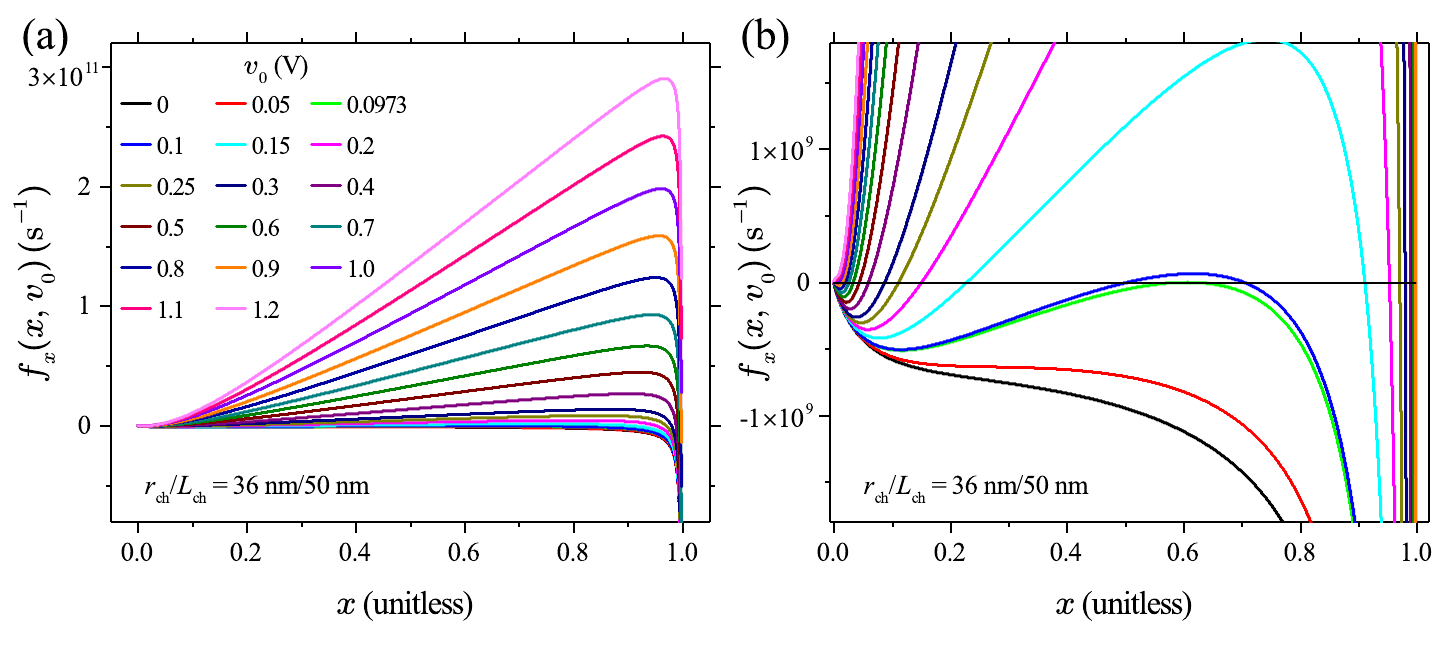}
	\caption{(a) Dynamic route map of $f_x(x,v_0)$ at constant input voltages in the range of 0 to 1.2~V, calculated for the midsize VO$_2$ Mott memristor. (b) is a zoomed portion of (a) to show that at $v_0>0.0973$~V, the DR locus intersects the $x$-axis at two distinctive locations. At $v_0=0.0973$~V, the DR locus becomes tangent to the $x$-axis with only one intersection point. At $v_0<0.0973$~V, the DR locus stays in the fourth quadrant and does not intersect the $x$-axis.}
	\label{Fig5_DR_Vmode}
\end{figure}

Although a Mott memristor is a current-controlled device, it is interesting to look at the state dynamics for the case that a constant finite input voltage is applied across it. Replacing current $i$ by voltage $v$ in equation (\ref{eqn2}), the kinetic function can be rewritten as a function of $x$ and $v$. At a constant input voltage $v_0$, it takes the form
\begin{equation}\label{eqn8}
	f_x(x,v_0)=\frac{1}{H'(x)}\left(\frac{v_0^2}{R_\text{ch}(x)}-\Gamma_\text{th}(x)\Delta T\right)=\frac{2Ax(\ln x)^2(1+Bx^2)v_0^2+2Cx\ln x}{D[1-x^2+2x^2\ln x+2E(x\ln x)^2]}
\end{equation}
Figure \ref{Fig5_DR_Vmode}(a) shows the DRM loci of $f_x(x,v_0)$ vs.~$x$ at constant $v_0$ levels from 0 to 1.2~V at 0.1~V interval for the midsize VO$_2$ device. Figure \ref{Fig5_DR_Vmode}(b) is a zoomed view which reveals three behaviorally distinctive regions determined by the amplitude of $v_0$. At a very small $v_0<0.0973$~V, the DR locus stays in the fourth quadrant and does not intersect with the $x$-axis. In other words, $f_x(x,v_0)<0$ is satisfied at any $x\in(0,1)$. It indicates that at such small input voltages, even if the initial condition is a metallic phase, a Mott memristor always return to the insulating state after a finite time. Physically speaking, the Joule heating level at such small voltages is too small to sustain a metallic filament at the IMT critical temperature against the heat loss. At $v_0=0.0973$~V, the DR locus becomes tangent to the $x$-axis with only one intersection point close to $x_0=0.606$. At a $v_0>0.0973$~V, the DR locus ``swings'' from the fourth quadrant to the first quadrant, then it swings back to the fourth quadrant, intersecting the $x$-axis at two distinctive points to the left and right of $x_0$.

For a 1D nonlinear ODE system, a saddle-node (tangent) bifurcation is the generic bifurcation in which the number of fixed points changes as some parameter is varied. If additional conditions are met, a transcritical or pitchfork bifurcation may occur. A simple example of a saddle-node bifurcation is $dx/dt=\mu\pm x^2$, where $\mu$ is the bifurcation parameter and the sign determines if it is supercritical ($\mu-x^2$) or subcritical ($\mu+x^2$). For the supercritical case, as $\mu$ increases through $\mu_0=0$ (the bifurcation value), the number of fixed points changes from zero to one then two. If $\mu<\mu_0$, $dx/dt$ is always negative and no fixed point exists. At $\mu=\mu_0$, there is one non-hyperbolic, semi-stable fixed point ($x=0$). At $\mu>\mu_0$, a pair of stable ($x=\sqrt{\mu}$) and unstable ($x=-\sqrt{\mu}$) hyperbolic fixed points are created.

\begin{figure}[htb]
	\centering
		\includegraphics[width=0.9\linewidth]{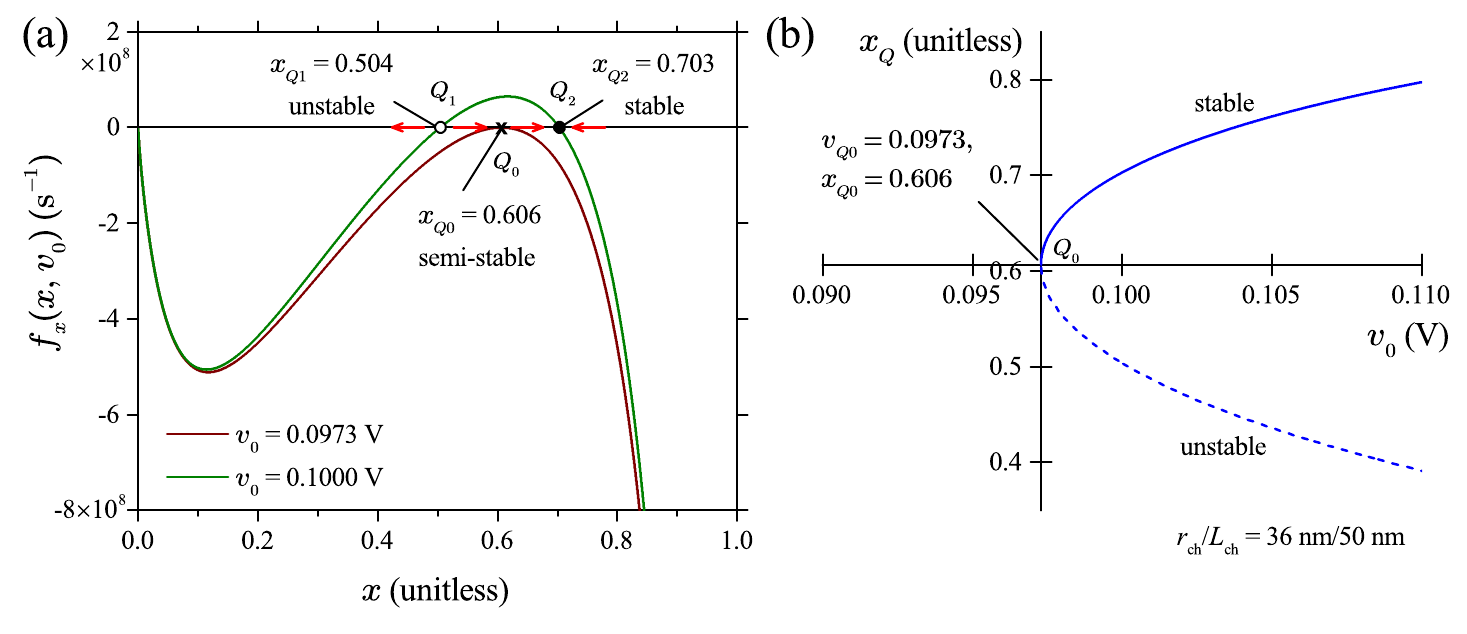}
	\caption{(a) Dynamic routes of $f_x(x,v_0)$ at constant input voltages of 0.0973~V and 0.1~V, calculated for the midsize VO$_2$ Mott memristor. At $v_0=0.0973$~V, the single intersection point $Q_0$ ($\times$) with the $x$-axis is a semi-stable fixed point. At $v_0=0.1$~V, the left intersection point $Q_1$ ($\circ$) with the $x$-axis is an unstable fixed point, and the right intersection point $Q_2$ ($\bullet$) with the $x$-axis is a stable fixed point. Arrowheads show the direction of move for a solution $x(t)$ starting from a initial state located close to a fixed point. (b) Bifurcation diagram of the same device, showing a 1D supercritical saddle-node bifurcation with $v_0$ as the bifurcation parameter. Solid (dashed) line shows the stable (unstable) solutions of fixed points $x_Q$.}
	\label{Fig6_SaddleNodeBifur}
\end{figure}

Figure \ref{Fig6_SaddleNodeBifur} illustrates that if a VO$_2$ Mott memristor is biased by a constant voltage $v_0$, a small change in $v_0$ as the bifurcation parameter results in a supercritical saddle-node bifurcation. For the midsize VO$_2$ device, the bifurcation value for $v_0$ is approximately 0.0973~V. Figure \ref{Fig6_SaddleNodeBifur}(a) re-plots two of the DRM loci in figure \ref{Fig5_DR_Vmode} at $v_0$ levels of 0.0973~V and 0.1~V. At $v_0=0.0973$~V, there is a single semi-stable fixed point $Q_0$ ($\times$). Raising the input voltage by a tiny amount to $v_0=0.1$~V results in a qualitative change in the solution structure and creates a pair of fixed points --- the left one $Q_1$ ($\circ$) is unstable, and the right one $Q_2$ ($\bullet$) is stable. The stability of a fixed point is told by the arrowheads indicating the direction of move for a solution $x(t)$ starting from a initial state located close to it. Figure \ref{Fig6_SaddleNodeBifur}(b) plots the bifurcation diagram of the 1D saddle-node bifurcation with the input voltage as the bifurcation parameter. Solid line and dashed line show the stable and unstable solutions of fixed points $x_Q$, respectively.

\section[3. DC IV]{Loci of steady states}
In the present approach, the internal temperature is embedded in the biphasic model and not treated as a state variable. The set of all fixed points $(x_Q,i_Q,v_Q)$ in the 3D $(x,i,v)$ state space that satisfy the instantaneous relationship $v_Q=R_\text{ch}(x_Q)i_Q$ and $(dx/dt)|_Q=0$ is defined as the steady-state or DC locus of a Mott memristor. Solving the steady-state locus of an isolated Mott memristor is among the first steps for the local linearization analysis. Henceforth, both the $(x_Q,i_Q,v_Q)$ locus and its 2D projections are called the steady-state loci without discerning the dimensional difference.

To obtain the steady-state $(x_Q,i_Q,v_Q)$ locus, one can first define a sequence of $i_Q \in \mathbb{R}$, then find the solutions of the state variable $x=x_Q(i_Q)$ numerically. This is achieved by setting the numerator in equation (\ref{eqn7}) to be zero, which gives an equation $CA(1+Bx_Q^2)=-i_Q^2\ln x_Q$ that can be solved numerically. After solving $x_Q(i_Q)$, voltage $v_Q$ can be calculated by the Ohm's law relationship $v_Q(i_Q)=R_\text{ch}(x_Q(i_Q))i_Q$.

However, there is a much easier way to obtain the steady-state $(x_Q,i_Q,v_Q)$ locus. Instead of numerically solving the value of $x_Q$ from a given $i_Q$, one can first define a sequence of $x_Q\in(0,1)$, then calculate $i_Q(x_Q)$ analytically using the following formula
\begin{equation}\label{eqn9}
	i_Q(x_Q)=\sqrt{\frac{-CA(1+Bx_Q^2)}{\ln x_Q}}
\end{equation}
Voltage $v_Q$ is then calculated by the Ohm's law $v_Q(x_Q)=R_\text{ch}(x_Q)i_Q(x_Q)$. The sequence of $x_Q$ can be chosen to be evenly spaced on a linear scale or a logarithmic scale, depending on how fast these functions vary with $x_Q$. We verified that steady states calculated by both methods are consistent with each other. The analytical method is used for discussions hereafter.

\begin{figure}[htb]
	\centering
		\includegraphics[width=0.9\linewidth]{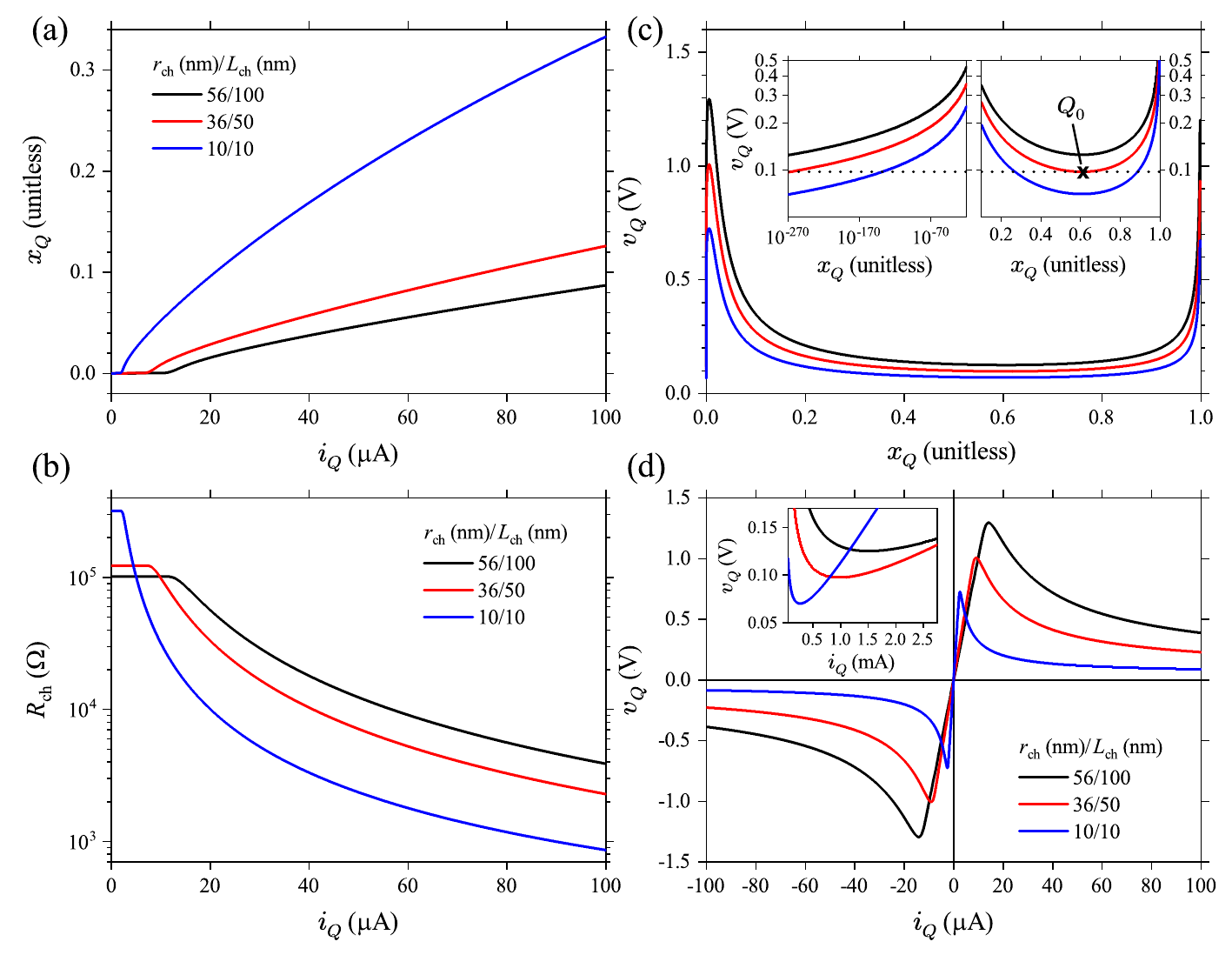}
	\caption{(a) Loci of the steady-state current $i_Q(x_Q)$ calculated by equation (\ref{eqn9}) and transposed to $x_Q(i_Q)$ with current as the independent variable. (b) Loci of the memristance function $R_\text{ch}(x_Q)$ vs.~$i_Q(x_Q)$. (c) Loci of the steady-state voltage $v_Q(x_Q)$. Insets are the very-small $x_Q$ (left) and halfway regions (right). Dashed line is $v_0=0.0973$~V. $Q_0$ ($\times$) is the semi-stable fixed point shown in figure~\ref{Fig6_SaddleNodeBifur}. (d) Loci of the steady-state $(i_Q,v_Q)$ showing the zero-crossing property of memristors and a PDR-to-NDR crossover at $i_Q\gtrsim2.522~\upmu$A, 9.077~$\upmu$A and 14.122~$\upmu$A respectively for the three VO$_2$ device sizes as labeled. Inset reveals another NDR-to-PDR crossover at $i_Q\gtrsim269.77~\upmu$A, 971.18~$\upmu$A and 1510.73~$\upmu$A respectively on the same three loci.}
	\label{Fig7_StatIV}
\end{figure}

Figure \ref{Fig7_StatIV}(a) shows the steady-state loci of $(x_Q,i_Q)$ calculated by equation (\ref{eqn9}) for three different VO$_2$ device sizes, plotted as $x_Q(i_Q)$ since Mott memristors are current-controlled devices. At small currents, the fraction of metallic phase $x_Q$ remains negligibly small. $x_Q$ starts to grow with current in a sublinear fashion once $i_Q$ exceeds a size-dependent threshold level. The current threshold grows with the device size and is at $\upmu$A level for the shown device sizes.

Figure \ref{Fig7_StatIV}(b) shows the loci of the memristance function $R_\text{ch}(x_Q)$ vs.~$i_Q$, which reveal that $R_\text{ch}(x_Q)$ has a similar crossover characteristics at the same $i_Q$ thresholds. At small currents, $R_\text{ch}(x_Q)$ remains elevated with negligible current dependence. Once $i_Q$ exceeds a size-dependent threshold, $R_\text{ch}(x_Q)$ drops rapidly with current in a nonlinear fashion. For the midsize VO$_2$ device ($r_\text{ch}=36$~nm, $L_\text{ch}=50$~nm), $R_\text{ch}(x_Q)$ drops by more than 3 decades from 122.8~k$\Omega$ to 97~$\Omega$ as $i_Q$ increases from 0 to 1~mA.

Figure~\ref{Fig7_StatIV}(c) shows the steady-state loci of $(x_Q,v_Q)$ plotted as $v_Q(x_Q)$, which resemble the shape of a left handled cup. The open left handle is nearly vertical, i.e., at very small $x_Q$ levels a tiny change in $x_Q$ will cause a large change in $v_Q$. Figure~\ref{Fig7_StatIV}(c) left inset plots the extremely-small $x_Q$ region of the $(x_Q,v_Q)$ loci in log-log scale, which reveals that at a given device size, there is a corresponding asymptotic lower bound of steady-state $v_Q$ as $x_Q$ approaches zero. For the midsize VO$_2$ device ($r_\text{ch}=36$~nm, $L_\text{ch}=50$~nm), the $v_Q$ lower bound turns out to be 0.0973~V (dashed line). Figure~\ref{Fig7_StatIV}(c) right inset plots the halfway $x_Q$ region in linear scale, which shows that the $v_Q=0.0973$~V horizontal line is tangent with the $(x_Q,v_Q)$ locus at its trough of $Q_0=(0.60628, 0.0973~\text{V})$ (marked as $\times$), the same semi-stable fixed point $Q_0$ found by the DR analysis. A slight increase in $v_Q$ would bifurcate $Q_0$ into a pair of fixed points on its left and right. The left inset also tells that in this case, another fixed point would emerge at an extremely-small $x_Q$ level (at $v_Q=0.2$~V, $x_Q$ is only $10^{-63}$), i.e., an insulating steady state exists at a finite voltage. These observations corroborate our previous DR analysis in figure~\ref{Fig6_SaddleNodeBifur}. All three $(x_Q,v_Q)$ loci have a sharp peak at $x_Q=0.00567$ and a rounded trough at $x_Q=0.60628$, resembling the shape of a cup. Notably the $x_Q$ coordinates of these two extrema are size-independent.

Figure \ref{Fig7_StatIV}(d) shows the steady-state loci of $(i_Q,v_Q)$ plotted as $v_Q(i_Q)$. As current-controlled memristors, the $(i_Q,v_Q)$ loci are letter ``N'' shaped when plotted with current as the $x$ axis. They are symmetric with respect to the origin in the first and third quadrants. Therefore one only needs to analyze the first-quadrant halves. Each $(i_Q,v_Q)$ locus has three distinctive regions: A lower positive differential resistance (PDR) region from 0 to a critical current $i_{c1}$. An NDR region between $i_{c1}$ and a second critical current $i_{c2}$  (see inset). An upper PDR region for even higher currents. Therefore $i_{c1}$ and $i_{c2}$ produce a local maximum and minimum in the $(i_Q,v_Q)$ loci. For the shown device sizes, values of $i_{c1}$ ($i_{c2}$) are 2.522~$\upmu$A (269.77~$\upmu$A), 9.077~$\upmu$A (971.18~$\upmu$A) and 14.122~$\upmu$A (1510.73~$\upmu$A), respectively. Figure~\ref{Fig7_StatIV}(d) also shows that the steady-state or DC loci of $(i_Q,v_Q)$ always pass through the origin (0, 0), satisfying the zero-crossing property of memristors.

It should be noted that the volumetric enthalpy change of IMT $\Delta h_\text{tr}$ only shows up in the denominator of the kinetic function $f_x(x,i)$ via coefficient $E$. Therefore it has no effect in determining the steady-state $(i_Q,v_Q)$ loci. The main effect of IMT on the shape of steady-state $(i_Q,v_Q)$ is applied via coefficient $B$ --- the coefficient of the quadratic nonlinearity in memristance function. Coefficient $B$ is approximately the electrical resistivity ratio $\frac{\rho_\text{ins}}{\rho_\text{met}}$ between the insulating and metallic phases.

\begin{figure}[htb]
	\centering
		\includegraphics[width=0.7\linewidth]{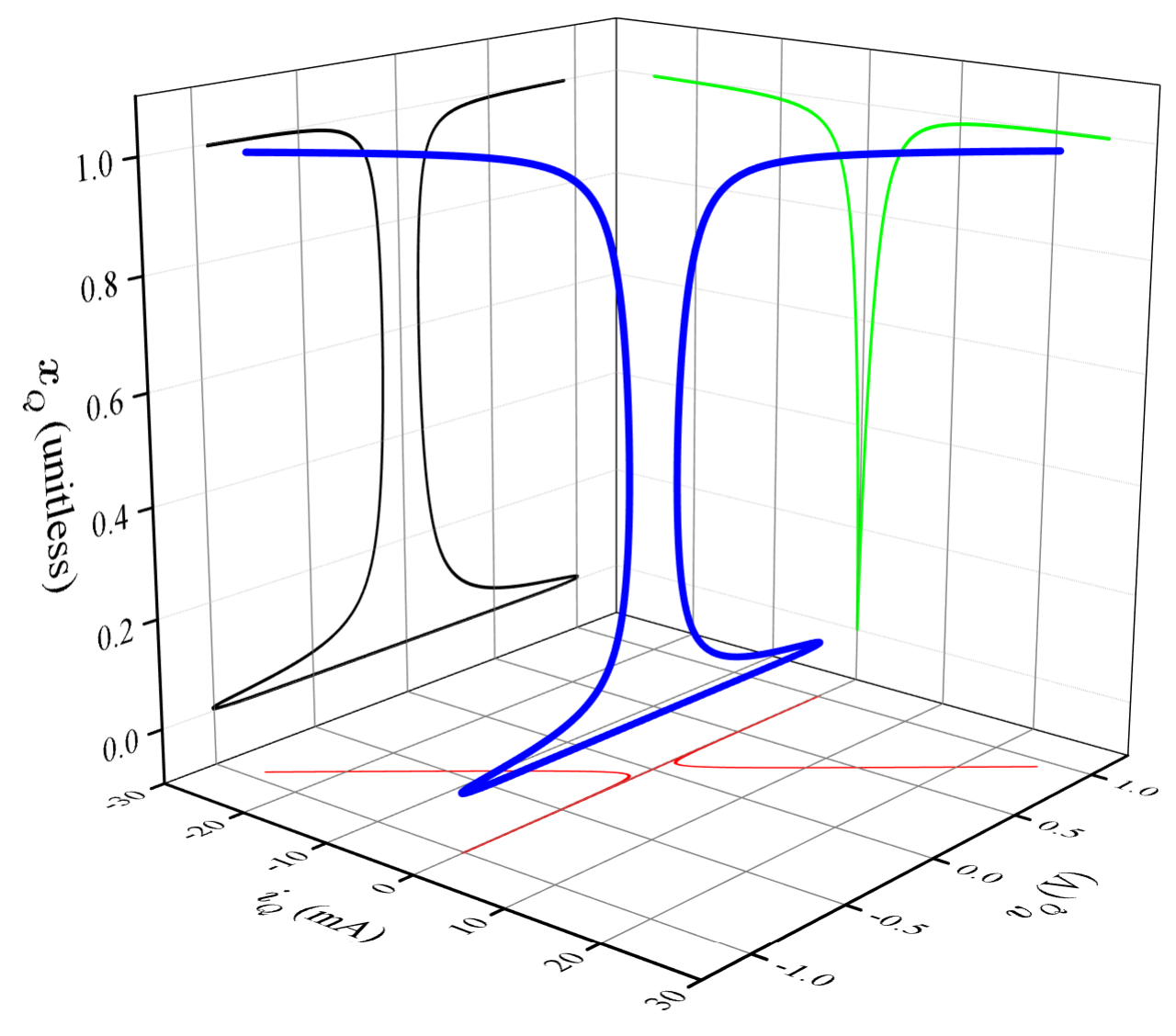}
	\caption{Locus of fixed points $(x_Q,i_Q,v_Q)$ in the 3D state space of $(x,i,v)$ calculated for the midsize VO$_2$ Mott memristor (blue line) and its 2D projections (green, black, and red lines).}
	\label{Fig8_3DStatXIV}
\end{figure}

\begin{figure}[htb]
	\centering
		\includegraphics[width=0.7\linewidth]{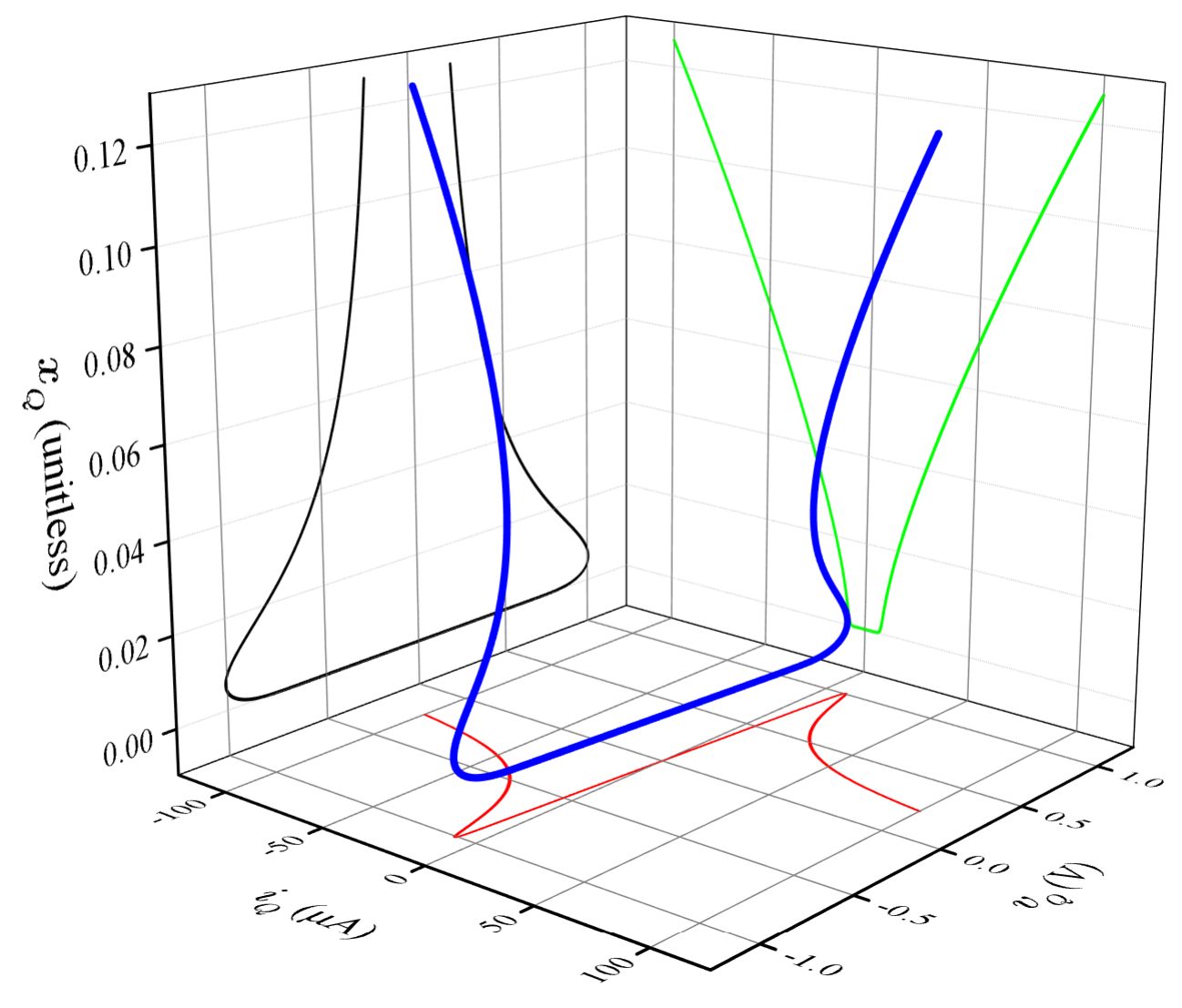}
	\caption{A zoomed view of figure \ref{Fig8_3DStatXIV} to visualize the low-current region of the fixed-point locus $(x_Q,i_Q,v_Q)$ calculated for the midsize VO$_2$ Mott memristor. Its 2D projections (green, black, and red lines) can now be compared with the loci of $i_Q(x_Q)$, $v_Q(x_Q)$, and $(i_Q,v_Q)$ shown in figure \ref{Fig7_StatIV}.}
	\label{Fig9_3DStatXIVzoom}
\end{figure}

The sets of loci plotted in figures \ref{Fig7_StatIV}(a), \ref{Fig7_StatIV}(c) and \ref{Fig7_StatIV}(d) are 2D projections of the steady-state loci $(x_Q,i_Q,v_Q)$ in the 3D state space. Figure \ref{Fig8_3DStatXIV} shows the locus of $(x_Q,i_Q,v_Q)$ calculated for the midsize VO$_2$ device ($r_\text{ch}=36$~nm, $L_\text{ch}=50$~nm). It looks somewhat like a twisted handle of a binder clip. The two open legs of the clip are rotated out of the plane defined by the looped clip head. Figure \ref{Fig9_3DStatXIVzoom} is a zoomed view of figure \ref{Fig8_3DStatXIV} to visualize the low-current region of the same $(x_Q,i_Q,v_Q)$ locus, so that its 2D projections onto the $(i,x)$, $(v,x)$ and $(i,v)$ planes can now be directly compared with the loci shown in figures \ref{Fig7_StatIV}(a), \ref{Fig7_StatIV}(c) and \ref{Fig7_StatIV}(d).

\section[4. LA Isolated]{Local analysis of an isolated Mott memristor}

\subsection[4.1]{Linearization and small-signal analysis}
Chua's LA theory outlines an algorithmic analysis procedure on nonlinear dynamical electronic circuits using equivalent linearized circuits~\cite{Chua05}. The linearized LA analysis examine the locus of fixed points of the composite circuit, the fluctuations around these fixed points, and their Laplace transforms. To explore the complex phenomena of nonlinear dynamical circuits, one can simply apply the LA criteria to access the active parameter domain rather than applying a time-consuming trial-and-error search in the parameter space. A good illustration of this procedure is the memristive HH axon circuit model~\cite{Chua12}. Here we apply the local linearization analysis and the LA theory to an isolated VO$_2$ Mott memristor to gain insights on its behavior near fixed points.

\subsubsection[4.1.1]{Linearization around a fixed point}
Considering a fixed point $Q$ with a coordinate $(x_Q,i_Q)$ on the steady-state locus of an isolated Mott memristor, one can expand voltage $v$ about the fixed point $(x_Q,i_Q)$ in a Taylor series
\begin{equation}\label{eqn10}
	v(x_Q+\delta x,i_Q+\delta i) = v_Q + i_QR'_\text{ch}(x_Q)\delta x + R_\text{ch}(x_Q)\delta i + \text{h.o.t.}
\end{equation}
where $R'_\text{ch}(x_Q) \triangleq \frac{dR_\text{ch}}{dx}|_Q$ and $\text{h.o.t.}$ denotes higher-order terms in $\delta x$ and $\delta i$. Neglecting $\text{h.o.t.}$, we get a linear equation
\begin{equation}\label{eqn11}
	\delta v = i_QR'_\text{ch}(x_Q)\delta x + R_\text{ch}(x_Q)\delta i = a_{11}(Q)\delta x + a_{12}(Q)\delta i
\end{equation}
where coefficients $a_{11}(Q) \triangleq i_QR'_\text{ch}(x_Q)$ and $a_{12}(Q) \triangleq R_\text{ch}(x_Q)$. Similarly, the kinetic function $f_x(x,i)$ can be expanded about the fixed point $(x_Q,i_Q)$ in a Taylor series
\begin{equation}\label{eqn12}
	f_x(x_Q+\delta x,i_Q+\delta i) = f_x(x_Q,i_Q) + \frac{\partial f_x(x,i)}{\partial x}|_Q\delta x + \frac{\partial f_x(x,i)}{\partial i}|_Q\delta i + \text{h.o.t.}
\end{equation}
Note that $f_x(x_Q,i_Q)=0$ since it is a fixed point $(x_Q,i_Q)$ on the steady-state locus. Neglecting $\text{h.o.t.}$, we can recast the nonlinear state equation $\frac{dx}{dt}=f_x(x,i)$ into the following linear differential equation
\begin{equation}\label{eqn13}
	\frac{d}{dt}(\delta x) = \frac{\partial f_x(x,i)}{\partial x}|_Q\delta x + \frac{\partial f_x(x,i)}{\partial i}|_Q\delta i = b_{11}(Q)\delta x + b_{12}(Q)\delta i
\end{equation}
where coefficients $b_{11}(Q) \triangleq \frac{\partial f_x(x,i)}{\partial x}|_Q$ and $b_{12}(Q) \triangleq \frac{\partial f_x(x,i)}{\partial i}|_Q$.
Applying Equations (\ref{eqn2}) and (\ref{eqn3}), one can easily obtain the expressions for the following three linear-term coefficients
\begin{eqnarray}\label{eqarray3}
	\label{eqn14}
	a_{11}(Q) = -\frac{2Bi_Qx_Q}{A\left(1+Bx_Q^2\right)^2} \\
	\label{eqn15}
	a_{12}(Q) = R_\text{ch}(x_Q) = \frac{1}{A\left(1+Bx_Q^2\right)} \\
	\label{eqn16}
	b_{12}(Q) = \frac{4x_Q(\ln x_Q)^2i_Q}{DA\left(1+Bx_Q^2\right)\left[1-x_Q^2+2x_Q^2\ln x_Q+2E(x_Q\ln x_Q)^2\right]}
\end{eqnarray}
To obtain the expression for $b_{11}(Q)$, we rewrite $f_x(x,i)$ as $f_x(x,i)=\frac{i^2X(x)+Y(x)}{Z(x)}$, where the three auxiliary functions are defined as $X(x)=\frac{2x(\ln x)^2}{A(1+Bx^2)}$, $Y(x)=2Cx\ln x$, and $Z(x)=D[1-x^2+2x^2\ln x+2E(x\ln x)^2]$.
Applying the quotient rule $\frac{d}{dx}\frac{X(x)}{Z(x)}=\frac{X'(x)Z(x)-X(x)Z'(x)}{Z(x)^2}$, we get
\begin{equation}\label{eqn17}
	b_{11}(Q) = i_Q^2\left. \frac{X'(x)Z(x)-X(x)Z'(x)}{Z(x)^2}\right|_Q+\left.\frac{Y'(x)Z(x)-Y(x)Z'(x)}{Z(x)^2}\right|_Q
\end{equation}
% Using either \left or \right on a period means the automatic delimiter sizing takes place, but only one delimiter is shown
%
The formulas for $X'(x) \triangleq \frac{dX(x)}{dx}$, $Y'(x) \triangleq \frac{dY(x)}{dx}$ and $Z'(x) \triangleq \frac{dZ(x)}{dx}$ are $X'(x)=\frac{2\ln x\left(2Bx^2-Bx^2\ln x+\ln x+2\right)}{A\left(1+Bx^2\right)^2}$, $Y'(x)=2C(\ln x+1)$ and $Z'(x)=4Dx\ln x[1+E(\ln x+1)]$.

\begin{figure}[htb]
	\centering
		\includegraphics[width=0.9\linewidth]{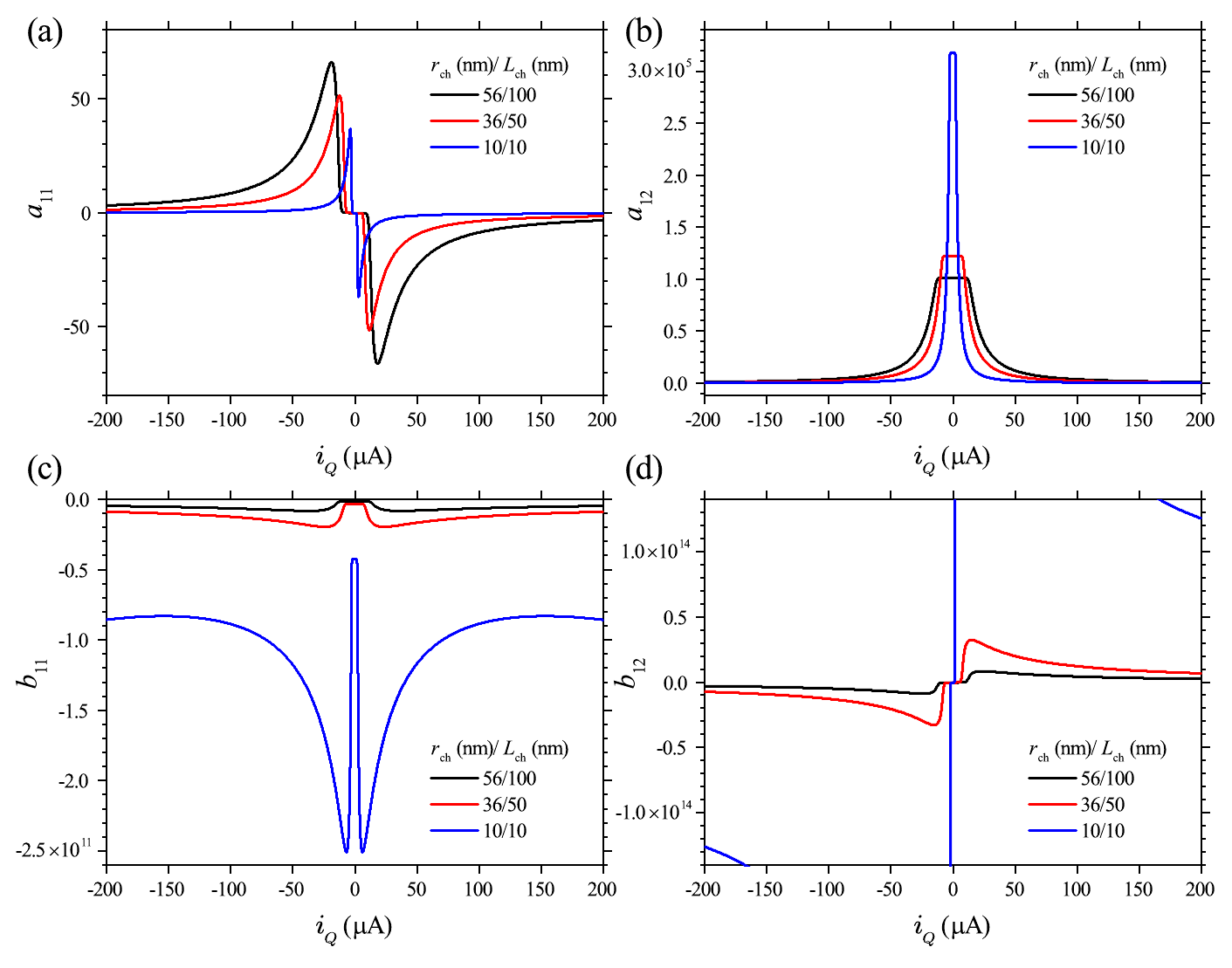}
	\caption{Current dependences of the linear-term coefficients (a) $a_{11}$, (b) $a_{12}$, (c) $b_{11}$, and (d) $b_{12}$ in the linearized expressions of voltage $v$ and the kinetic function $f_x(x,i)$ about a fixed point on the steady-state locus of three different-sized VO$_2$ Mott memristors as labeled.}
	\label{Fig10_a11a12b11b12}
\end{figure}

Figures \ref{Fig10_a11a12b11b12}(a)--\ref{Fig10_a11a12b11b12}(d) plot the current dependence of the linear-term coefficients $a_{11}$, $a_{12}$, $b_{11}$ and $b_{12}$ calculated by equations (\ref{eqn14})--(\ref{eqn17}) for three different VO$_2$ device sizes. They show that coefficients $a_{11}$  and $b_{12}$ are odd functions of the driving current, while coefficients $b_{11}$ and $a_{12}$ are even functions of the driving current. $a_{12}$ is the same as the memristance $R_\text{ch}$ and is always positive. In contrast, $b_{11}$ is always negative.

\subsubsection[4.1.2]{Complex-domain equivalent circuit}

Many insights can be gained about an isolated Mott memristor in the realm of complex analysis. As the second step of the local analysis, we can obtain its complex-domain equivalent circuit by the linear Laplace transform $\hat{f}(s) \triangleq \int_{0^-}^\infty \! f(t)e^{-st} \, dt$
% Use \! to bring the function closer to the integral sign and use \, to push the differential farther away
that maps a function $f(t)$ in the time domain to a function $\hat{f}(s)$ in the complex domain $\mathbb{C}$, whose elements are complex frequencies $s=\sigma+i\omega$. The complex domain is also known as the $s$ domain. A direct benefit of the Laplace transform is that it turns a differential equation into an algebraic equation.

Taking the Laplace transforms of equations (\ref{eqn11}) and (\ref{eqn13}), we obtain
\begin{eqnarray}\label{eqarray4}
	\label{eqn18}
	\hat{v}(s)=a_{11}(Q)\hat{x}(s)+a_{12}(Q)\hat{i}(s) \\
	\label{eqn19}
	s\hat{x}(s)=b_{11}(Q)\hat{x}(s)+b_{12}(Q)\hat{i}(s)
\end{eqnarray}
where $\hat{x}(s)$, $\hat{v}(s)$ and $\hat{i}(s)$ denote the Laplace transforms of $\delta x(t)$, $\delta v(t)$ and $\delta i(t)$, respectively. Solving equation (\ref{eqn19}) for $\hat{x}(s)$, we obtain
\begin{equation}\label{eqn20}
	\hat{x}(s)=\frac{b_{12}(Q)\hat{i}(s)}{s-b_{11}(Q)}
\end{equation}
Substituting equation (\ref{eqn20}) for $\hat{x}(s)$ in equation (\ref{eqn18}) and solving for the impedance function $Z(s;Q) \triangleq \hat{v}(s)/\hat{i}(s)$, we obtain the $s$−domain impedance function
\begin{equation}\label{eqn21}
	Z(s;Q)=\frac{a_{11}(Q)b_{12}(Q)}{s-b_{11}(Q)}+a_{12}(Q)
\end{equation}

For a current-controlled memristor, the impedance function $Z(s;Q)$ in equation (\ref{eqn21}) is the proper choice for its transfer function $H(s;Q)$. For a voltage-controlled memristor, admittance function $Y(s;Q)$ should be used. Chua pointed out that for a 1D system with just one port state variable, its transfer function is also the scalar \textit{complexity function} that forms the basis for the LA analysis~\cite{Chua05}. In Chua's original LA formulations for reaction-diffusion systems, a port state variable of a ``reaction'' cell (equivalent to a lumped circuit element) interacts with the neighboring cells via an energy or matter flow such as diffusion. Whereas a nonport state variable describes isolated internal dynamics and does not interact with other cells. The concept of LA is defined concerning only port state variables. Clearly the state variable $x$ in the Mott memristor model is a port state variable, since it interacts with a coupled circuit element through a current (energy) flow.

Since the $s$-domain representation of a capacitor looks like a ``resistance'' $1/sC$, one can recast the small-signal impedance function $Z(s;Q)$ of a Mott memristor about a fixed point $Q$ as an equivalent circuit that consists of three virtual elements: a capacitor $C_1$ in parallel with a resistor $R_1$, both of them in series with a second resistor $R_2$
\begin{equation}\label{eqn22}
	Z(s;Q)=\frac{(\frac{1}{sC_1})R_1}{(\frac{1}{sC_1})+R_1}+R_2
\end{equation}
where
\begin{eqnarray}\label{eqarray5}
	\label{eqn23}
	R_1 \triangleq -\frac{a_{11}(Q)b_{12}(Q)}{b_{11}(Q)} \\
	\label{eqn24}
	R_2 \triangleq a_{12}(Q) = R_\text{ch}(x_Q) \\
	\label{eqn25}
	C_1 \triangleq \frac{1}{a_{11}(Q)b_{12}(Q)}	
\end{eqnarray}

\begin{figure}[htb]
	\centering
	\includegraphics[width=0.9\linewidth]{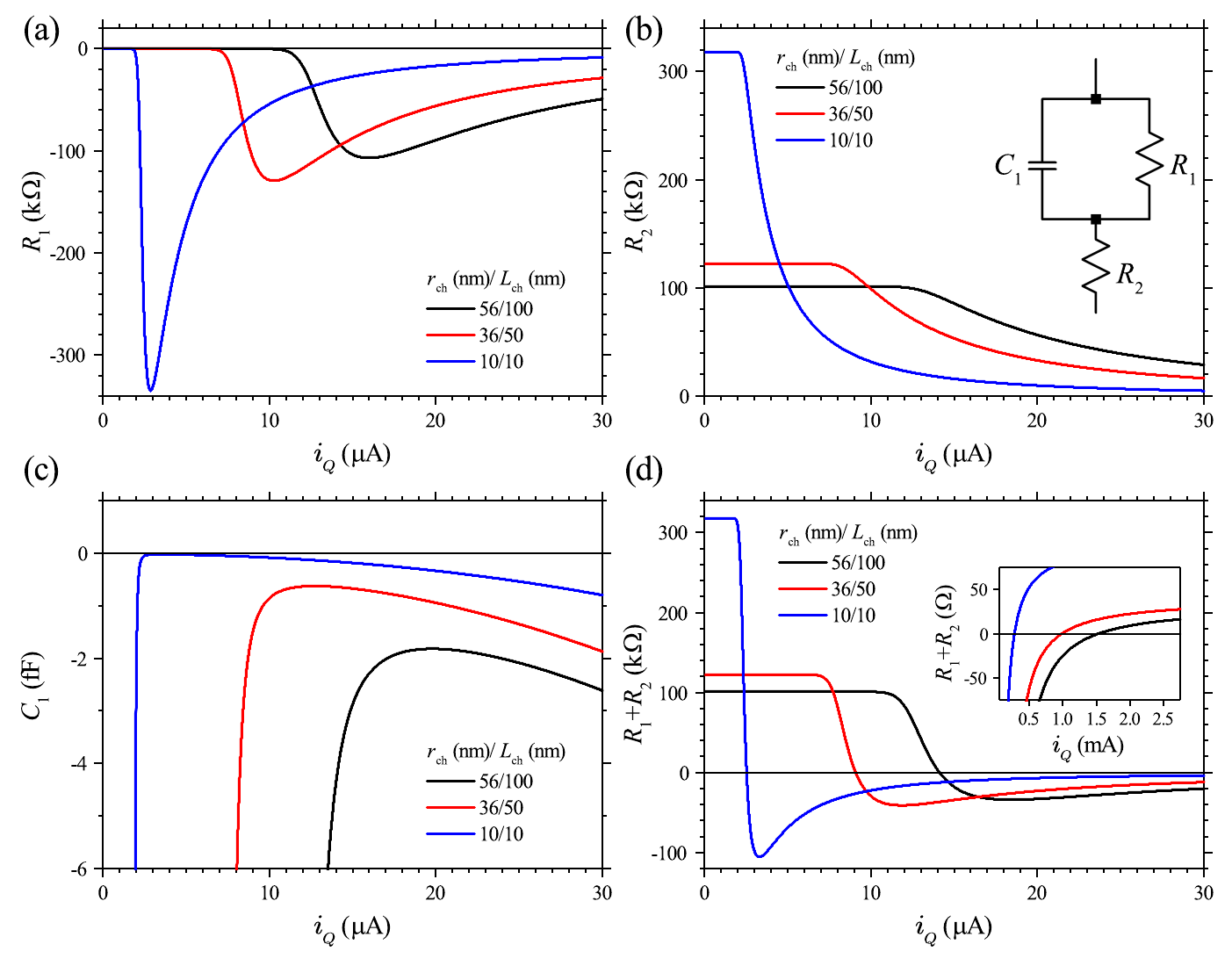}
	\caption{Current dependences of the three virtual circuit elements (a) $R_1$, (b) $R_2$ and (c) $C_1$ comprising the $s$-domain impedance function $Z(s;Q)$ about a fixed point on the steady-state $(i_Q,v_Q)$ loci of three different-sized VO$_2$ Mott memristors as labeled. $R_1$ and $C_1$ remain negative at any current for all three device sizes. Inset of (b) shows the small-signal equivalent circuit in the $s$-domain. (d) Current dependence for the sum of the two resistances $(R_1+R_2)$. $(R_1+R_2)$ turns negative as current exceeds a size-dependent limit. Inset of (d) shows that $(R_1+R_2)$ becomes positive again as current exceeds a much larger size-dependent limit. These two current limits are identical with the critical currents $i_{c1}$ at the lower PDR to NDR and $i_{c2}$ at the NDR to upper PDR transitions on the steady-state $(i_Q,v_Q)$ loci, respectively (see figure \ref{Fig7_StatIV}(d) and inset).}
	\label{Fig11_C1R1R2}
\end{figure}

Figures \ref{Fig11_C1R1R2}(a)--\ref{Fig11_C1R1R2}(c) plot the current dependence of the three virtual circuit elements $R_1$, $R_2$ and $C_1$ calculated by equations (\ref{eqn23})--(\ref{eqn25}) for three different VO$_2$ device sizes. They are all even functions of the driving current, so we only plot the positive $x$-axis halves. First thing to notice is that $R_1$ and $C_1$ stay negative at any current for all three device sizes calculated. In contrast, $R_2$ remains positive at any current. Note that $R_2$ is the same as $a_{12}$ and $R_\text{ch}(x_Q)$. Therefore, in the $s$-domain a Mott memristor can be modeled as a nonlinear positive resistor in series with a composite reactive element consisting of a nonlinear negative capacitor and a nonlinear negative resistor placed in parallel. This small-signal equivalent circuit in the $s$-domain is shown in figure \ref{Fig11_C1R1R2}(b) inset.

Since a negative capacitance translates to a positive frequency-dependent inductive reactance, it means that a Mott memristor (or generally a current-controlled LAM) has an apparent inductive reactance without involving a magnetic field. In physiology, an anomalous inductive reactance was observed as early as 1930s in voltage clamp measurements of the squid giant axon~\cite{Cole41}, but this perplexing phenomenon was not understood until Chua's memristive formulation for the potassium and sodium ion channels ~\cite{Chua12}.

Figure \ref{Fig11_C1R1R2}(d) plots the current dependence for the sum of the two resistances $(R_1+R_2)$. At small currents, $(R_1+R_2)$ is positive and remains nearly constant. As current increases, $(R_1+R_2)$ drops abruptly and becomes negative as current exceeds a limit that is identical with the critical current $i_{c1}$ for the lower PDR to NDR transition on the steady-state $(i_Q,v_Q)$ loci (see figure \ref{Fig7_StatIV}(d)), at 2.522~$\upmu$A, 9.077~$\upmu$A and 14.122~$\upmu$A respectively for the three device sizes. The negative $(R_1+R_2)$ then starts to rise with current. Inset of figure \ref{Fig11_C1R1R2}(d) shows that $(R_1+R_2)$ becomes positive again as current exceeds a much larger limit that is identical with the critical current $i_{c2}$ for the NDR to upper PDR transition on the steady-state $(i_Q,v_Q)$ loci (see figure \ref{Fig7_StatIV}(d) inset), at 269.77~$\upmu$A, 971.18~$\upmu$A and 1510.73~$\upmu$A respectively for the same three devices. The one-to-one correspondence between the sign of $(R_1+R_2)$ and the sign of the slope on the steady-state $(i_Q,v_Q)$ loci indicates that the three-element equivalent circuit shown in figure \ref{Fig11_C1R1R2}(b) inset is the proper small-signal representation of a Mott memristor in the $s$-domain.

\subsection[4.2]{Pole-zero diagram and Chua’s local activity theorem}

\subsubsection[4.2.1]{Poles and zeros of the transfer function}

For a dynamical system, the poles and zeros of its transfer function $H(s;Q)$ in the $s$-domain can tell about important characteristics of the system response without solving the complete differential equations. The first step of pole-zero analysis is to rewrite the $s$−domain small−signal transfer function $H(s;Q)$ as a rational function of $s$, i.e., a ratio of two polynomials. For the case of a 1D current-controlled Mott memristor, both the denominator and numerator $s$ polynomials have a degree of $n=1$, therefore its impedance function $Z(s;Q)$ is written as
\begin{equation}\label{eqn26}
	Z(s;Q)=\frac{b_{1}s+b_0}{a_{1}s+a_0}
\end{equation}
where all of the coefficients of $s$ polynomials in the denominator and numerator are real numbers. Using equation (\ref{eqn22}),  the expressions for these four coefficients are derived as
\begin{eqnarray}\label{eqarray6}
	\label{eqn27}
	a_0=1 \\
	\label{eqn28}
	a_1=R_{1}C_1 \\
	\label{eqn29}
	b_0=R_1+R_2 \\
	\label{eqn30}
	b_1=R_{1}R_{2}C_1
\end{eqnarray}

\begin{figure}[htb]
	\centering
	\includegraphics[width=0.9\linewidth]{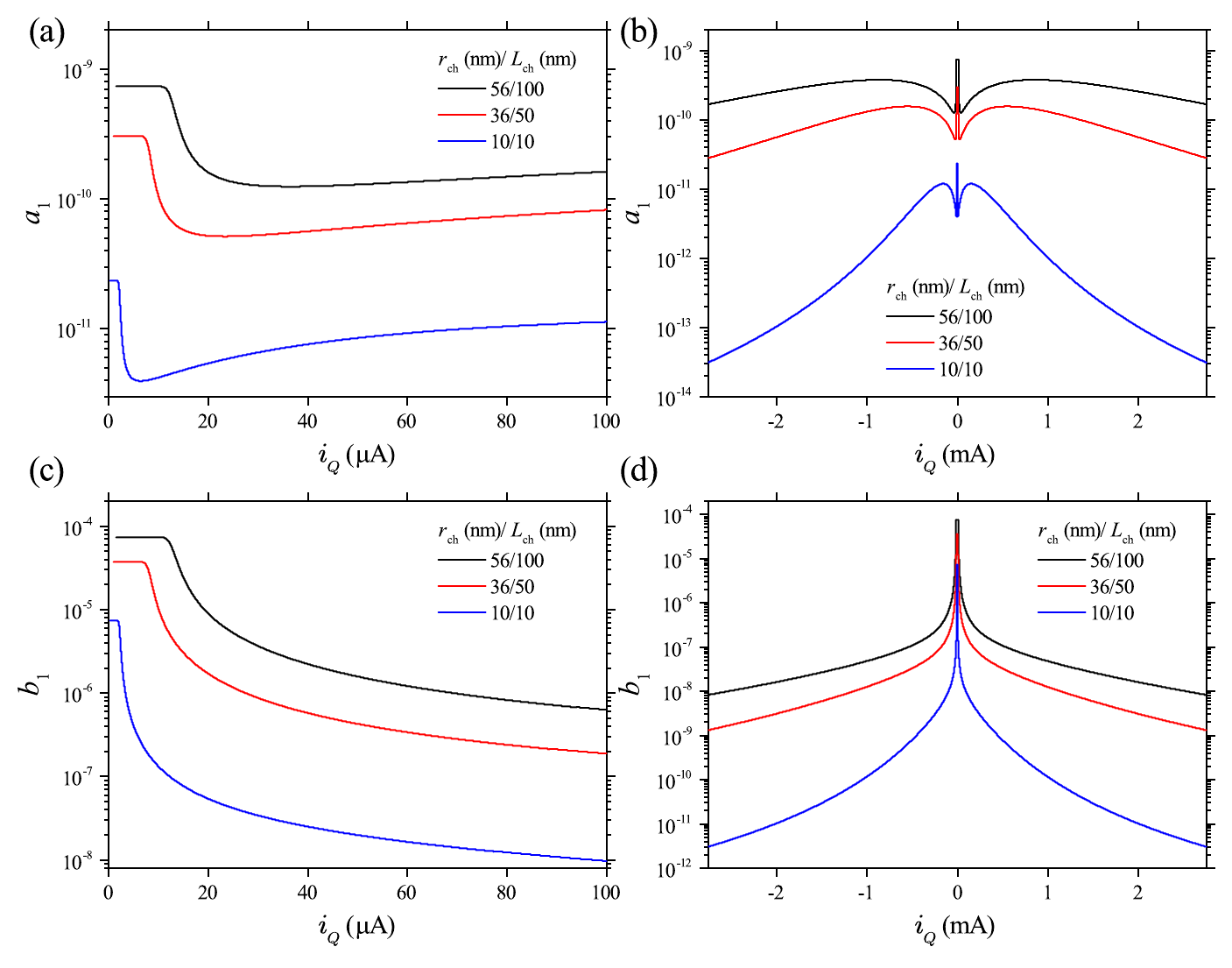}
	\caption{Current dependences for two of the four $s$-polynomial coefficients (a) $a_1$ (low-current part), (b) $a_1$ (wide range), (c) $b_1$ (low-current part) and (d) $b_1$ (wide range) in the rational-function representation of the impedance function $Z(s;Q)$ about a fixed point on the steady-state $(i_Q,v_Q)$ locus of three different-sized VO$_2$ Mott memristors as labeled. The other two coefficients are $a_0=1$ and $b_0=R_1+R_2$ (see figure \ref{Fig11_C1R1R2}(d)).}
	\label{Fig12_a1b1}
\end{figure}

Since $a_0$ is a constant and $b_0=R_1+R_2$ is already discussed (see figure \ref{Fig11_C1R1R2}(d)), we only need to look at $a_1=R_{1}C_1$ and $b_1=R_{1}R_{2}C_1$. Both of them are even functions of the input current. Their current dependence are plotted in figure \ref{Fig12_a1b1} for three different VO$_2$ device sizes.

A rational transfer function can be further rewritten in a factored or pole-zero form by expressing the $s$ polynomials in the denominator and numerator as products of linear factors. The roots of the denominator polynomial are the poles, and the roots of the numerator polynomials are the zeros. For any polynomial with real coefficients, its roots are either real or complex conjugate pairs. 

For an isolated 1D Mott memristor, there is just one pole and one zero. To obtain the expressions for the zero and the pole of $Z(s;Q)$, recast equation \ref{eqn26} as
\begin{equation}\label{eqn31}
	Z(s;Q)=\frac{k(s-z)}{(s-p)}
\end{equation}
where $k=b_1/a_1=R_2$ is a positive real coefficient, $z$ and $p$ denote respectively the zero and the pole of $Z(s;Q)$. The expressions for $z$ and $p$ are
\begin{eqnarray}\label{eqarray7}
	\label{eqn32}
	z=-\frac{b_0}{b_1}=-\frac{R_1+R_2}{R_{1}R_{2}C_1} \\
	\label{eqn33}
	p=-\frac{a_0}{a_1}=-\frac{1}{R_{1}C_1}=b_{11}
\end{eqnarray}

\begin{figure}[htb]
	\centering
	\includegraphics[width=0.9\linewidth]{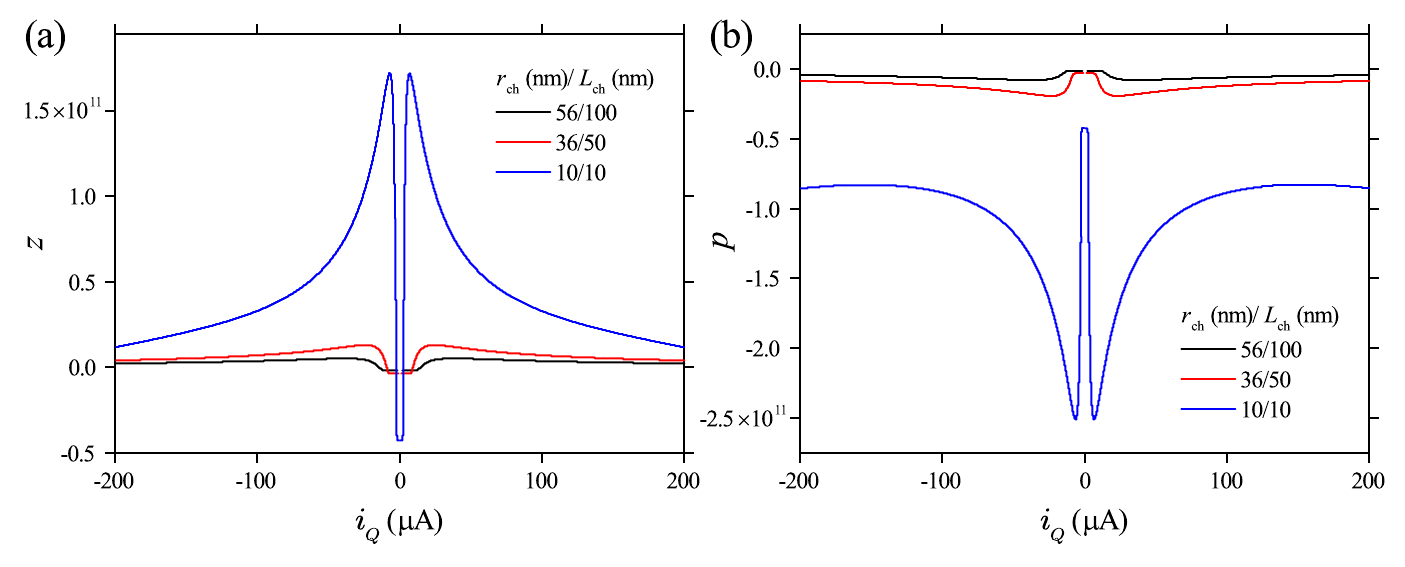}
	\caption{Current dependence of (a) the zero $z$ and (b) the pole $p$ of the $s$-domain impedance function $Z(s;Q)$ about a fixed point on the steady-state $(i_Q,v_Q)$ locus of a VO$_2$ Mott memristor, calculated for three different device sizes as labeled.}
	\label{Fig13_ZeroPole-I}
\end{figure}

Figure \ref{Fig13_ZeroPole-I} show the loci of the zero $z$ and the pole $p$ versus input current for three different VO$_2$ device sizes. It is conspicuous that both $z$ and $p$ are located on the real axis in the complex plane, and both are even functions of current. It may be noted that $p$ is already plotted in figure \ref{Fig10_a11a12b11b12}(c) in the form of $b_{11}$. It is replotted as figure \ref{Fig13_ZeroPole-I}(b) for a side-by-side comparison with $z$.

\begin{figure}[htb]
	\centering
	\includegraphics[width=0.9\linewidth]{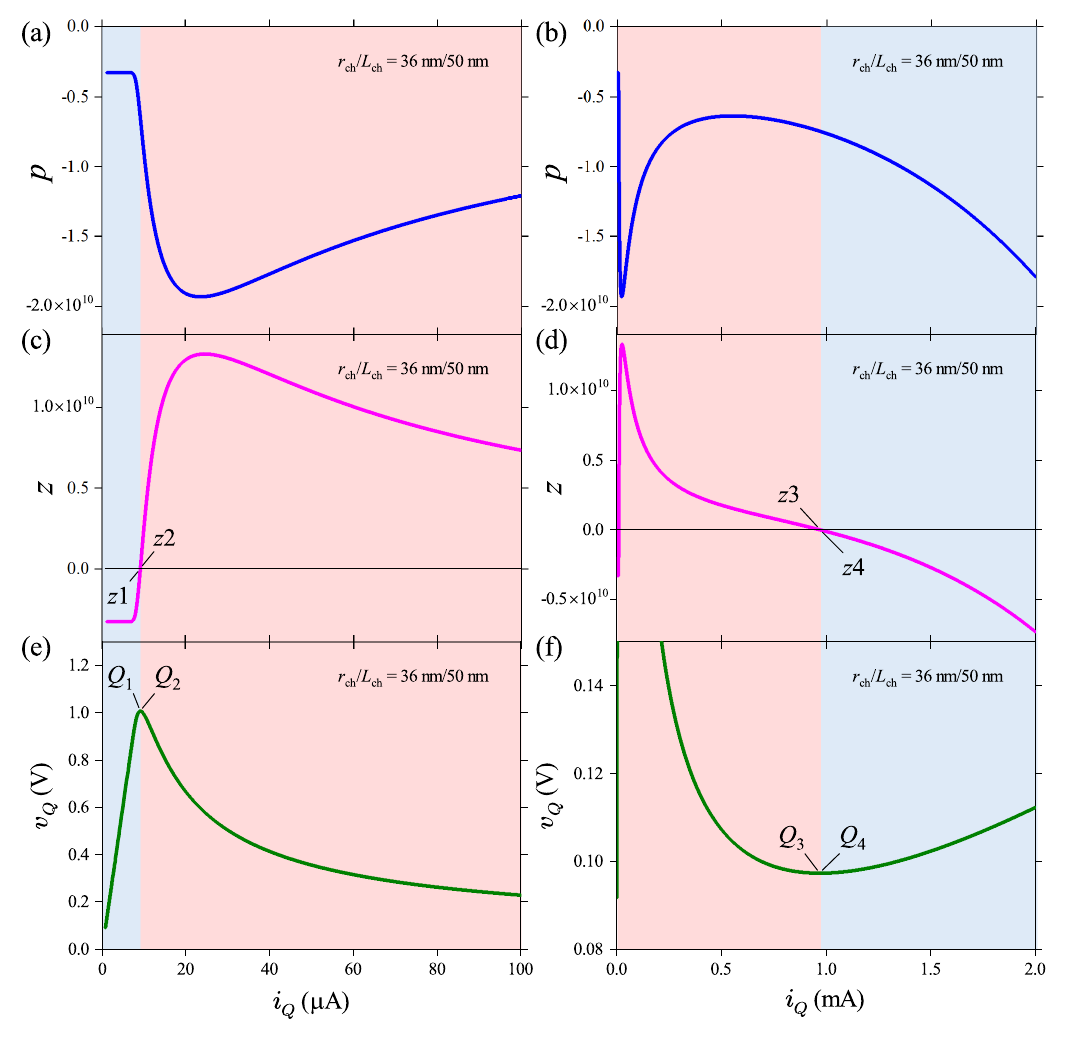}
	\caption{Current dependence of the pole $p$ of $Z(s;Q)$ in (a) and (b), and the zero $z$ in (c) and (d) for the midsize VO$_2$ Mott memristor. $p$ remains negative at any current. For $z$, (c) shows a $(-)\rightarrow(+)$ sign reversal at 9.077~$\upmu$A, as indicated by a pair of nearby zeros $\{z1,z2\}$ with opposite sign; (d) shows a $(+)\rightarrow(-)$ sign reversal at 971.18~$\upmu$A, as indicated by a pair of nearby zeros $\{z3,z4\}$ with opposite sign. (e) and (f) show the corresponding parts of the steady-state $(i_Q,v_Q)$ locus of the same device. The crossovers between the NDR region (red) and the lower and upper PDR regions (blue) at the local voltage extrema coincide with the sign reversals in $z$, as indicated by the locations of $\{Q_1,Q_2\}$ and $\{Q_3,Q_4\}$ pairs of fixed points on the $(i_Q,v_Q)$ locus.}
	\label{Fig14_SignOfZero}
\end{figure}

Since both the zero and the pole of $Z(s;Q)$ are located on the real axis, one can simply look at their signs to tell about the local dynamical behaviors about a fixed point, as it will be discussed in the next subsection. Generally speaking, for a 1D uncoupled Mott memristor, its pole $p$ (or $b_{11}$) remains negative at any current level. In contrast, its zero $z$ has two sign reversals at two distinctive input current levels. These characteristics are illustrated in figure \ref{Fig14_SignOfZero} which shows the current dependence of $p$ and $z$ of $Z(s;Q)$ calculated for the midsize VO$_2$ device ($r_\text{ch}=36$~nm, $L_\text{ch}=50$~nm).

Figures~\ref{Fig14_SignOfZero}(a) and \ref{Fig14_SignOfZero}(b) show the loci of $p(i_Q)$ for the low-current part (up to 100~$\upmu$A) and a wide range (up to 2~mA), respectively. $p$ has a non-monotonic current dependence, but $p<0$ always holds true. Figure \ref{Fig14_SignOfZero}(c) shows that $z$ is initially negative at small currents, then it turns positive if the current is higher than $\approx$9.077~$\upmu$A, as indicated by a pair of nearby fixed points $\{Q_1,Q_2\}$ across zero. Their coordinates $[x_Q,i_Q,v_Q, \text{Re}(z)]$ are $[0.00566,9.075~\upmu\text{A},1.007~\text{V},\num{-5.882E6}]$ for $Q_1$ and $[0.00567,9.078~\upmu\text{A},1.007~\text{V},\num{3.388E6}]$ for $Q_2$.  Figure~\ref{Fig14_SignOfZero}(d) shows that $z$ becomes negative again if current exceeds $\approx$971.18~$\upmu$A, as indicated by a pair of nearby fixed points $\{Q_3,Q_4\}$ across zero. Their coordinates are $[0.60628,971.171~\upmu\text{A},0.097~\text{V},\num{3.854E4}]$ for $Q_3$ and $[0.60629,971.203~\upmu\text{A},0.097~\text{V},\num{-8.477E4}]$ for $Q_4$. The two critical currents for sign reversals in $z$ match exactly with the ones that separate the NDR region from the lower and upper PDR regions on the steady-state $(i_Q,v_Q)$ locus of the same device, as shown in figure \ref{Fig14_SignOfZero}(e) and \ref{Fig14_SignOfZero}(f). The coincidences are confirmed by examining the locations of the same two pairs of nearby fixed points $\{Q_1,Q_2\}$ and $\{Q_3,Q_4\}$ on the $(i_Q,v_Q)$ locus.

\subsubsection[4.2.2]{Chua’s local activity theorem}

Chua’s local analysis method established a practical set of criteria to classify the dynamics of an isolated or uncoupled nonlinear circuit element about its fixed points. The approach to determine if a linearized element is LP or locally active about a fixed point $Q=(x_Q,i_Q,v_Q)$, is to check if its output fluctuations in response to small input fluctuations dissipate over time, or get amplified otherwise. For the discussion, we choose the example of a 1D current-controlled memristor with current as the input and voltage as the output. Their roles are exchanged for a voltage-controlled memristor. Mathematically, with a homogeneous initial condition $(\delta x(0), \delta i(0), \delta v(0))=\textbf{0}$ (no fluctuation at $t=0$), a linearized element is LP if and only if (iff) the fluctuation energy integrated over time remains positive 
\begin{equation}\label{eqn34}
	\text{LP} \Leftrightarrow \int_{0}^{t'} \delta i(t)\cdot\delta v(t) dt \geq 0
\end{equation}
for any finite time interval $t'>0$. The uncoupled element is locally active at a fixed point $Q$, iff there exists an input fluctuation $\delta i(t)$ and a finite time $0<T<\infty$, such that the integrated fluctuation energy becomes \textit{negative}. For a multidimensional element, the fluctuation power to be integrated is a scalar ``dot'' product between the two vectors $\delta \textbf{i}(t)$ and $\delta \textbf{v}(t)$.

However, it is not practical to inspect the time-domain integral for all possible input fluctuations. By applying the Laplace transform, Chua derived a mathematically equivalent but much more practical formula for the local passivity theorem in the complex domain. For the 1D scalar case, the necessary and sufficient condition for an uncoupled 1D circuit element to be LP is that its complexity function or transfer function $H(s;Q)$ is a positive real (PR) function, which satisfies both (1) $\text{Im}[H(s;Q)]=0$ if $\text{Im}[s]=0$, and (2) $\text{Re}[H(s;Q)]\geq0$ if $\text{Re}[s]\geq0$. Condition (1) is always satisfied since $H(s;Q)$ is a rational function. Condition (2) means that the closed right half plane (RHP) of $s$ maps into the closed RHP of $H(s;Q)$. A simple example for a PR function is $H(s;Q) = a + bs + cs^{-1}$, where $a$, $b$, and $c\geq0$.

Chua proved the following local passivity theorem as a practical test for the PR condition: An uncoupled 1D circuit element is LP at a fixed point \textit{iff all of the following four criteria are satisfied}
\renewcommand{\labelenumi}{\roman{enumi})}
\begin{enumerate}
	\item $H(s;Q)$ has no poles in the open RHP ($\text{Re}(s)>0$).
	\item $H(s;Q)$ has no higher-order poles (degree $n\geq2$) on the imaginary axis (Im axis).
	\item If $H(s;Q)$ has a simple pole $s=i\omega_\text{p}$ on the Im axis, then the residue of $H(s;Q)$ at $i\omega_\text{p}$ must be a PR number.
	\item The Im axis (excluding poles) maps into the closed RHP of $H(s;Q)$, i.e., $\text{Re}[H(i\omega;Q)]\geq0$ for all $\omega\in(-\infty,\infty)$ where $s=i\omega$ is not a pole.
\end{enumerate}
The LA theorem is derived by negating any one of the above conditions. That is to say, an uncoupled 1D circuit element is locally active at a fixed point \textit{iff any one of the following four criteria is satisfied}
\renewcommand{\labelenumi}{\roman{enumi})}
\begin{enumerate}
	\item $H(s;Q)$ has a pole in the open RHP $\text{Re}(s)>0$.
	\item $H(s;Q)$ has a higher-order pole (degree $n\geq2$) on the Im axis.
	\item $H(s;Q)$ has a simple pole on the Im axis with negative-real or complex residue.
	\item At least some points on the Im axis map into the open left half plane (LHP) of $H(s;Q)$, i.e., $\text{Re}[H(i\omega;Q)]<0$ for some $\omega\in(-\infty,\infty)$.
\end{enumerate}
For a system of higher dimensions, Chua proved a similar set of four test criteria for LA, wherein the complexity function $H(s;Q)$ for an 1D element is replaced by the complexity matrix for a multidimensional element.

As elaborated in Ref.~\cite{Brown22a}, the local stability of a fixed point is a property that is independent of the local activity of dynamics about it. Near a fixed point, an isolated memristor may have four possible combinations of local stability and local activity properties that can lead to persistent or decaying dynamics. Since the condition of both LP and locally unstable is unphysical to realize, in general one only needs to consider three possible scenarios: LP and stable, locally active and stable which is termed as edge of chaos (EOC) by Chua, and locally active but unstable (LA$\backslash$EOC).

For a 1D uncoupled memristor, if its transfer function has a positive coefficient ($k>0$ in equation~(\ref{eqn31})), the dynamical class about a fixed point is told by where the pole and zero of its transfer function are located in the complex plane, as specified below
\renewcommand{\labelenumi}{\roman{enumi})}
\begin{enumerate}
	\item Locally Passive $\Leftrightarrow$ pole in open LHP ($\text{Re}(p)<0$) and zero in closed LHP ($\text{Re}(z)\leq 0$)
	\item Edge of Chaos $\Leftrightarrow$ pole in open LHP ($\text{Re}(p)<0$) and zero in open RHP ($\text{Re}(z)>0$)
	\item Locally Active but Unstable $\Leftrightarrow$ pole in closed RHP ($\text{Re}(p)\geq 0$)
\end{enumerate}
Plots of $\text{Im}(p)$ versus $\text{Re}(p)$ and $\text{Im}(z)$ versus $\text{Re}(z)$, known as pole-zero diagram, thus offer a graphical determination of local steady-state dynamics without resorting to time-domain integration.

As discussed previously, for a current-controlled Mott memristor both the pole and the zero are located on the real axis. The pole $p$ of its impedance function $Z(s;Q)$ is always in the open LHP ($\text{Re}(p)<0$), therefore it does not possess the LA$\backslash$EOC dynamics in (iii). On the other hand, the zero $z$ of $Z(s;Q)$ can reside in either the closed LHP or the open RHP, depending on the input current amplitude. In figure \ref{Fig14_SignOfZero}, we have shown that $\text{Re}(z)$ flips its sign twice depending on the input current, and the two sign reversals in $\text{Re}(z)$ coincide with the crossovers between the NDR region and the lower and upper PDR regions on the steady-state $(i_Q,v_Q)$ locus.

\subsubsection[4.2.3]{Pole-zero diagram}

\begin{figure}[htb]
	\centering
	\includegraphics[width=0.9\linewidth]{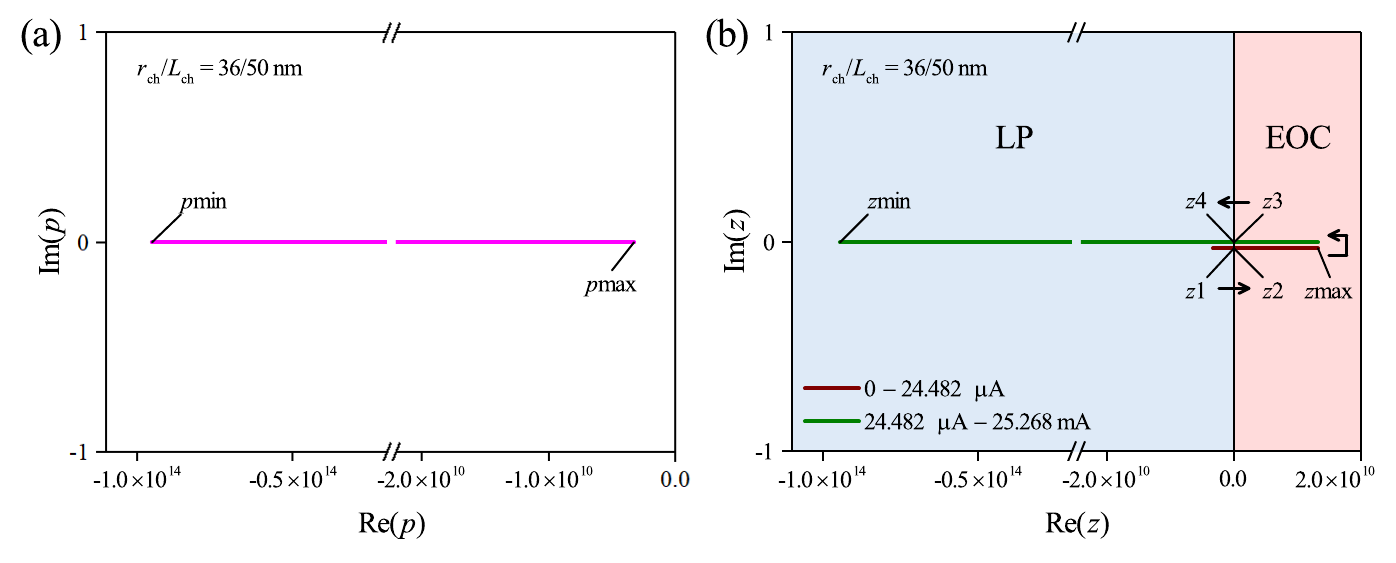}
	\caption{Locations of (a) the pole $p$ and (b) the zero $z$ in the complex plane as functions of input current, calculated for the midsize VO$_2$ Mott memristor. Both $p$ and $z$ are located on the real axis. $p$ remains in LHP with its minimal and maximal calculated values indicated by $p_\text{min}$ and $p_\text{max}$. $z$ is located in LHP at $i_Q=0$ and shifts to the right with current, crossing the imaginary axis into RHP at 9.077~$\upmu$A as indicated by $\{z1,z2\}$. It continues shift to the right until reaching $z_\text{max}$ at 24.482~$\upmu$A, then shifts to the left with current and reenters LHP at 971.18~$\upmu$A as indicated by $\{z3,z4\}$. The minimum of $z$ is approximately at $z_\text{min}$. The part of $z$ locus between 0 and 24.482~$\upmu$A (brown) is shifted vertically for clarity. The LP and EOC regions are highlighted by blue and red colors, respectively.}
	\label{Fig15_ZeroPole_Im-Re}
\end{figure}

In figure \ref{Fig15_ZeroPole_Im-Re}, we visualize the evolution of $p$ and $z$ locations in the complex plane as functions of the input current for the current-controlled midsize VO$_2$ Mott memristor. Figure \ref{Fig15_ZeroPole_Im-Re}(a) shows that $\text{Re}(p)<0$ is always satisfied. The coordinates $[x_Q,i_Q,v_Q, \text{Re}(p)]$ for the minimal and maximal calculated values of $p$, labeled as $p_\text{min}$ and $p_\text{max}$, are $[0.998,25.268~\text{mA},0.935~\text{V},\num{-9.543E13}]$ and $[\num{1E-145},1.074~\upmu\text{A},0.132~\text{V},\num{-3.273E9}]$, respectively. As may be noted, $p_\text{min}$ and $p_\text{max}$ in our calculations are not the actual bounds of $p$, since $x_Q$ can approach very closely to its asymptotes 0 and 1 but will never touch them.

Figure \ref{Fig15_ZeroPole_Im-Re}(b) shows that the zero $z$ is located in the LHP at zero current, and it shifts to the right as current increases. $z$ crosses the Im axis into the RHP at a critical current of 9.077~$\upmu$A, as indicated by a pair of nearby fixed points $\{z1,z2\}$ on the opposite side of the Im axis (the same ones as shown in figure \ref{Fig14_SignOfZero}). $z$ continues shifting to the right with current until it reaches a maximum value at $z_\text{max}$ with a coordinate of $[0.03549,24.482~\upmu\text{A},0.578~\text{V},\num{1.327E10}]$. Then it reverses course and shifts to the left with current. $z$ crosses the Im axis again and returns to the LHP at a second critical current of 971.18~$\upmu$A, as indicated by a pair of nearby fixed points $\{z3,z4\}$ on the opposite side of the Im axis (the same ones as shown in figure \ref{Fig14_SignOfZero}). Continuously increasing the current will drive $x_Q$ asymptotically toward 1 and further decrease $z$. We stop the calculation at $z_\text{min}$ with a coordinate of $[0.998,25.268~\text{mA},0.935~\text{V},\num{-9.467E13}]$. We use the same blue and red colors as shown in figure \ref{Fig14_SignOfZero} to highlight the LP and EOC regions, respectively.

Applying the pole-zero diagram LA criteria specified above, we conclude that an uncoupled 1D Mott memristor about a fixed point either belongs to the LP class or the EOC class, but can never belong to the LA$\backslash$EOC class. For the midsize VO$_2$ device, the LP $\rightarrow$ EOC transition occurs at $(x_Q,i_Q,v_Q)\approx(0.00567,9.076~\upmu\text{A},1.007~\text{V})$. The EOC $\rightarrow$ LP transition occurs at $(x_Q,i_Q,v_Q)\approx(0.60629,971.2~\upmu\text{A},0.0973~\text{V})$.

\subsection[4.3]{Frequency response}
An important question now arises: for an uncoupled 1D Mott memristor that is current-biased in the EOC region (coinciding with its NDR region), will it remain to be locally active so that it can amplify a small sinusoidal input fluctuation at arbitrarily-high frequencies? Otherwise, is there a finite upper limit for the input fluctuation frequency, beyond which the element cannot provide an AC signal gain anymore? In this section, we move the small-signal analysis to the frequency domain, which allows us to apply the fourth criterion in Chua's LA theorem to get the answers.

For dynamical systems, it is useful to study the system’s frequency response. In small-signal analysis, this is performed by applying a single-frequency sinusoidal fluctuation of current input $i(t)=I\text{sin}\omega t$ with an angular frequency $\omega=2\pi f$, where $f$ is the frequency of the sinewave. The amplitude $I\ll1$ is very small to satisfy the small-signal condition. For a 1D Mott memristor about a fixed point $Q$, substituting $s=i\omega$ for the complex frequency $s$ in the small-signal impedance $Z(s;Q)$ in equation (\ref{eqn26}) and rearranging into its real and imaginary parts, we get 
\begin{equation}\label{eqn35}
	Z(i\omega;Q)=\left[\frac{a_{0}b_0+a_{1}b_1\omega^2}{a_0^2+a_1^2\omega^2}\right]+i\left[\frac{(a_{0}b_1-a_{1}b_0)\omega}{a_0^2+a_1^2\omega^2}\right]
\end{equation}
The functions $\text{Re}Z(i\omega;Q)$ and $\text{Im}Z(i\omega;Q)$ are the real and imaginary parts of the frequency response expressed in terms of the small-signal impedance $Z(i\omega;Q)$, both expressed as rational functions of $\omega$:
\begin{eqnarray}\label{eqarray8}
	\label{eqn36}
	\text{Re}Z(i\omega;Q)=\frac{a_{0}b_0+a_{1}b_1\omega^2}{a_0^2+a_1^2\omega^2} \\
	\label{eqn37}
	\text{Im}Z(i\omega;Q)=\frac{(a_{0}b_1-a_{1}b_0)\omega}{a_0^2+a_1^2\omega^2}
\end{eqnarray}
where the coefficients $a_0$, $a_1$, $b_0$ and $b_1$ are given in equations (\ref{eqn27})--(\ref{eqn30}).

Figure \ref{Fig16_ReZImZ-freq}(a) and \ref{Fig16_ReZImZ-freq}(b) plot the frequency dependence of $\text{Re}Z(i\omega;Q)$ and $\text{Im}Z(i\omega;Q)$ (also referred to as Re$Z$ and Im$Z$ hereafter) at different steady-state current levels between 2~$\upmu$A and 10~$\upmu$A for the midsize VO$_2$ Mott memristor. We replaced angular frequency $\omega$ with frequency $f$ as the $x$-axis for engineering convenience. Notice that positive and negative frequencies refer to the opposite directions of rotation for the complex exponential $e^{i\omega t}$ vector in the complex plane. Re$Z$ is an even function of frequency, while Im$Z$ is an odd function of frequency. At small currents, Re$Z$ is in the order of $10^5~\Omega$ and shows very weak frequency dependence. Increasing current will ``pull'' it toward negative direction and develop a dip centered at zero frequency. The higher the current is, the stronger the frequency dependence becomes.

Frequency response of Re$Z$ shows a dramatic change as current increases from 9~$\upmu$A to 10~$\upmu$A. From the pole-zero diagram analysis, we already know that for the midsize VO$_2$ Mott memristor, the critical current at the LP (lower PDR) to EOC (NDR) crossover is $i_{c1}\approx9.077$~$\upmu$A. At $i_Q=9$~$\upmu$A, Re$Z$ still remains positive at any frequency, but its minimum at zero frequency is very close to the origin. At $i_Q=10$~$\upmu$A, Re$Z$ turns negative at frequencies lower than a limit $\left|f_\text{max}\right|\approx0.88$~GHz, indicating that the element is locally active within certain frequency upper bound. This $(+)\rightarrow(-)$ sign reversal in Re$Z$ is yet another hallmark of the LP $\rightarrow$ EOC transition, and brings us new information on \textit{the boundary of the EOC region in the frequency domain}.

The value of $f_\text{max}$ can be derived from Chua's fourth LA criterion. For an uncoupled 1D current-driven memristor in the frequency domain,  $\text{Re}Z(i\omega;Q)<0$ for some finite angular frequencies $\omega\in(-\infty,\infty)$ is a sufficient condition for it to be LA. From equation \ref{eqn35}, this means $a_{0}b_0+a_{1}b_1\omega^2<0$, or $\omega^2<\frac{-a_{0}b_0}{a_{1}b_1}$. Therefore, a 1D uncoupled Mott memristor is locally active if the angular frequency is lower than an upper bound specified as
\begin{equation}\label{eqn38}
	\left|\omega\right|<\omega_\text{max}=\sqrt{\frac{-a_{0}b_0}{a_{1}b_1}}
\end{equation}
which also requires that $\frac{a_{0}b_0}{a_{1}b_1}<0$ so that $\omega_\text{max}$ is a real number.

At small currents, Im$Z$ is both very small and shows very weak frequency dependence. Increasing current will make its amplitude and frequency dependence more substantial. The amplitude of Im$Z$ first rises quickly with frequency before reaching a peak at a characteristic frequency $f_\text{p}$, then it falls with frequency and approaches the $x$-axis asymptotically. $\left|f_\text{p}\right|$ increases with current and reaches 1.51~GHz at $i_Q=10$~$\upmu$A.  Inset of figure \ref{Fig16_ReZImZ-freq}(b) is the same frequency dependence of Im$Z$ plotted in log-log scale, which shows that Im$Z$ is proportional to frequency for $\left|f\right|<\left|f_\text{p}\right|$ and inversely proportional to frequency for $\left|f\right|>\left|f_\text{p}\right|$.

\subsubsection[4.3.1]{Nyquist plot}

It is instructive to plot the locus of $\text{Im}Z(i\omega;Q)$ vs.~$\text{Re}Z(i\omega;Q)$ in Cartesian coordinates with $\omega$ indicated as a parameter. Such a parametric plot is called a Nyquist plot, which is a graphic technique used to provide intuition on the stability of a dynamical system.

\begin{figure}[htb]
	\centering
	\includegraphics[width=0.9\linewidth]{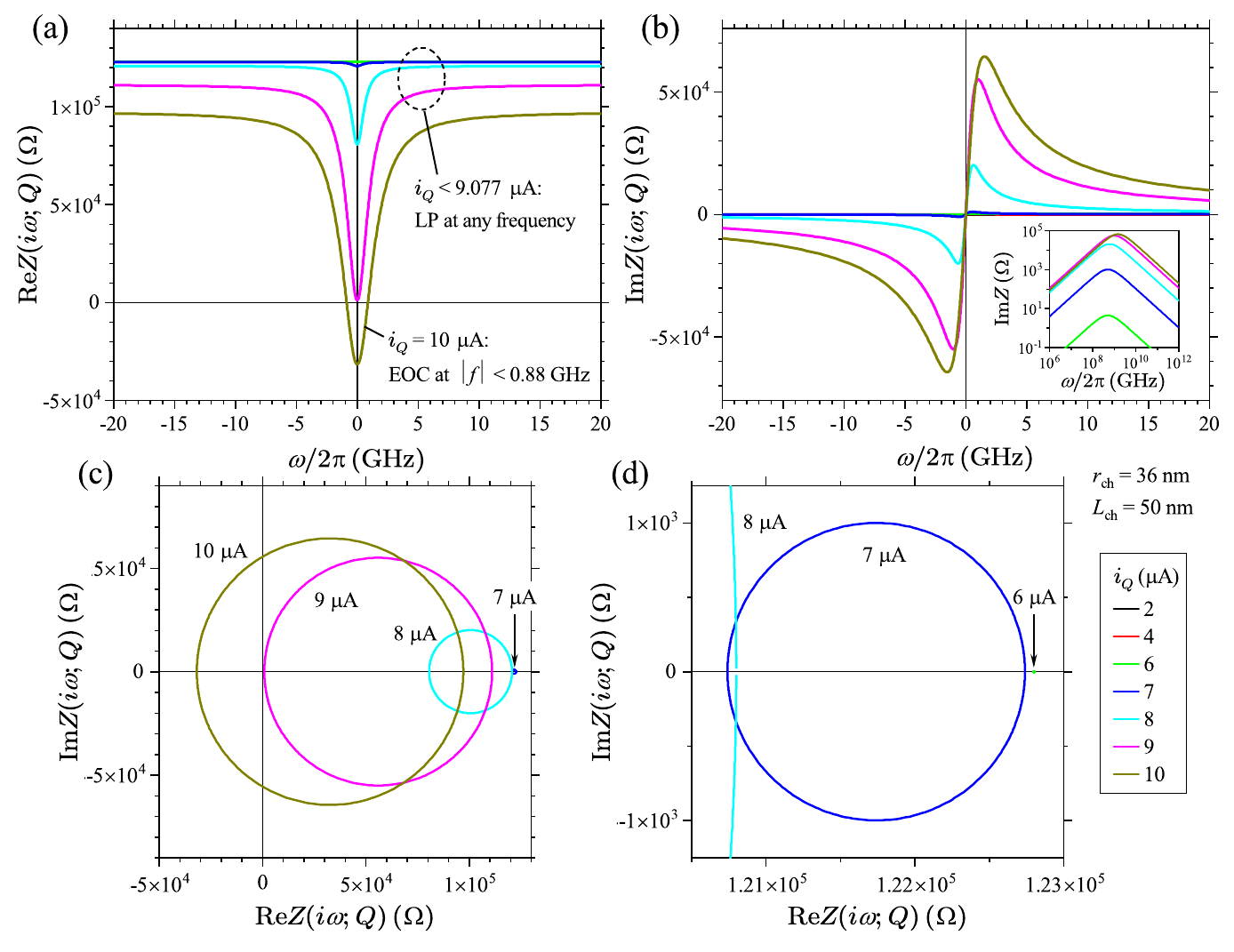}
	\caption{Small-signal impedance $Z(i\omega;Q)$ frequency response of the midsize VO$_2$ Mott memristor biased at constant steady-state currents in the range of 2~$\upmu$A to 10~$\upmu$A. (a) $\text{Re}Z(i\omega;Q)$ vs.~$\omega/2\pi$ (in GHz). At $i_Q<9.077$~$\upmu$A, Re$Z>0$ at any frequency and the memristor remains LP. At $i_Q=10$~$\upmu$A, an EOC region exists with Re$Z<0$ at $\left|f\right|<0.88$~GHz. (b) $\text{Im}Z(i\omega;Q)$ vs.~$\omega/2\pi$. Inset is a part of the same figure plotted in log-log scale. (c) Nyquist plot $\text{Im}Z(i\omega;Q)$ vs.~$\text{Re}Z(i\omega;Q)$. (d) Zoomed portion of (c) to reveal the loci at smaller currents.}
	\label{Fig16_ReZImZ-freq}
\end{figure}

Figure \ref{Fig16_ReZImZ-freq}(c) shows the loci of Nyquist plot for the same VO$_2$ device as shown in figure \ref{Fig16_ReZImZ-freq}(a) and \ref{Fig16_ReZImZ-freq}(b). Figure \ref{Fig16_ReZImZ-freq}(d) is a zoomed portion of it to reveal those much smaller loci at $i_Q\leq7$~$\upmu$A. Ostensibly,  the locus of small-signal $\text{Im}Z(i\omega;Q)$ vs.~$\text{Re}Z(i\omega;Q)$ at a finite steady-state current appears to be a circle centered on the $x$-axis. Increasing current will inflate the radius of the circle and move its center toward the negative direction. Points in the upper half-plane correspond to positive frequencies, and those in the lower half-plane correspond to negative frequencies. Increasing the frequency modulus will move a point in the right direction along the upper or lower arm of the locus. A closer look finds that the left half of the locus intersects the $x$-axis at zero frequency. For the right half, the distance between the locus and the $x$-axis approaches 0 as $\left|f\right|\rightarrow\infty$, but there is no intersection at any finite frequency. In other words, the $x$-axis is a horizontal asymptote for the right half of the locus. Therefore the locus of $\text{Im}Z(i\omega;Q)$ vs.~$\text{Re}Z(i\omega;Q)$is actually an open set of points rather than a closed loop. At $i_Q\approx9.077$~$\upmu$A, the locus crosses the $y$-axis into the LHP, as illustrated by the two loci at 9~$\upmu$A and 10~$\upmu$A. Therefore, Nyquist plot provides another visualization of the LP $\rightarrow$ EOC transition as the steady-state current increases.

\begin{figure}[htb]
	\centering
	\includegraphics[width=0.7\linewidth]{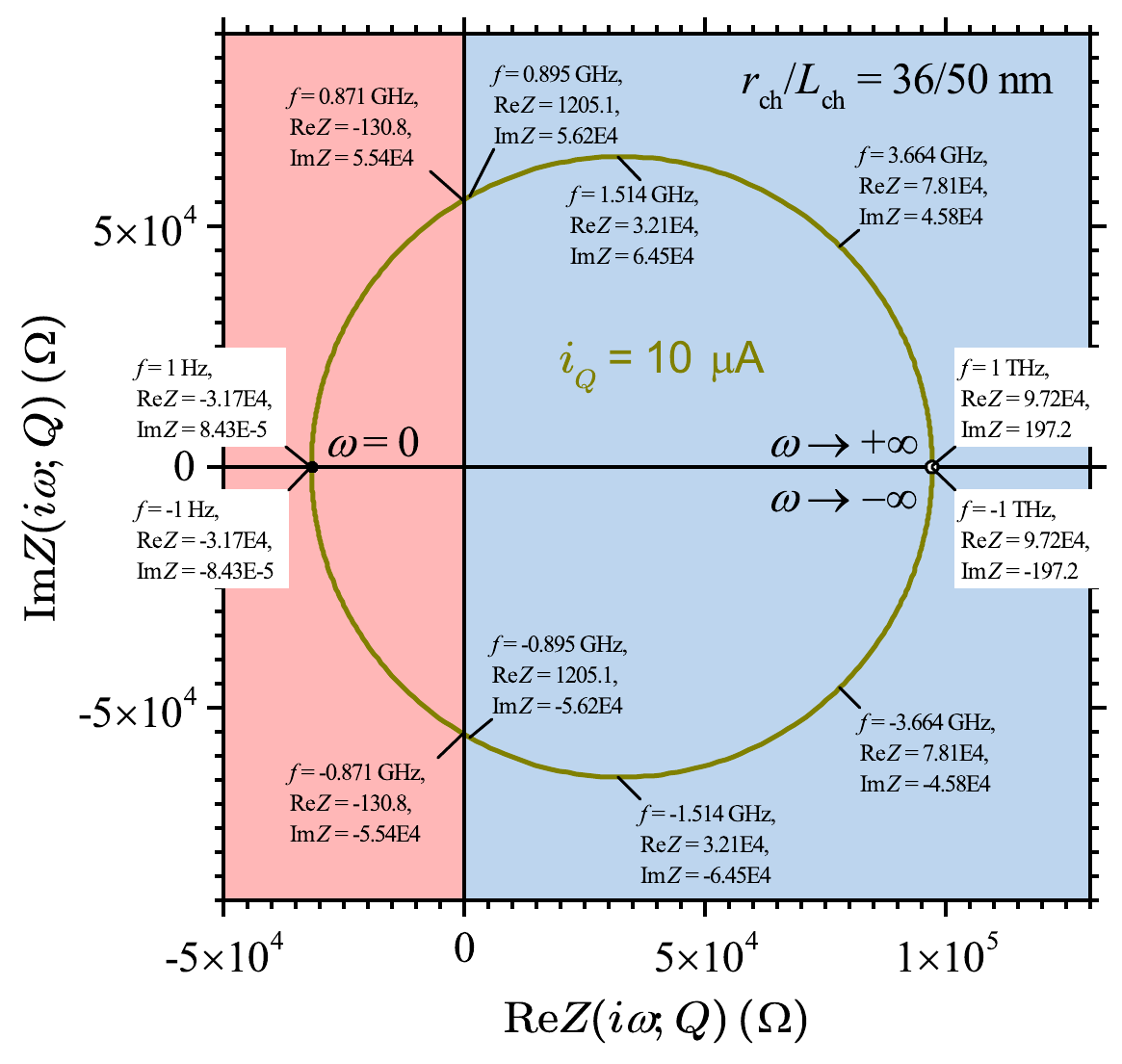}
	\caption{Nyquist plot $\text{Im}Z(i\omega;Q)$ vs.~$\text{Re}Z(i\omega;Q)$ of the midsize VO$_2$ Mott memristor at a constant steady-state current $i_Q=10$~$\upmu$A. The LP and EOC regions are highlighted by blue and red colors, respectively.}
	\label{Fig17_NyquistPlot-EOC}
\end{figure}

Figure \ref{Fig17_NyquistPlot-EOC} is an annotated Nyquist plot for the Im$Z$ vs.~Re$Z$ locus of the same VO$_2$ device at $i_Q=10$~$\upmu$A, highlighting several key points on the locus. We use the same blue and red colors as shown in figure \ref{Fig14_SignOfZero} and figure \ref{Fig15_ZeroPole_Im-Re} to represent the LP and EOC regions. Clearly, the lower half of the locus is a reflection of the upper half of it over the $x$-axis, by negating the values of Im$Z$ and frequency at the same Re$Z$ value. The solid dot ($\bullet$) at Re$Z$ $\approx\num{-3.17E4}$ represents the $x$-intercept of the locus at zero frequency, as indicated by a pair of nearby points at $f=1$~Hz and $f=-1$~Hz. The open circle ($\circ$) at Re$Z$ $\approx\num{9.72E4}$ represents the $x$-asymptote of the locus as $\left|f\right|\rightarrow\infty$, as indicated by a pair of nearby points at $f=1$~THz and $f=-1$~THz. The two pairs of points at $f=\pm0.871$~GHz and $f=\pm0.895$~GHz indicate the crossover from the EOC (red) region to the LP (blue) region as frequency exceeds 0.88~GHz. 

\subsubsection[4.3.2]{Frequency-domain equivalent circuit}

The frequency-domain equivalent circuit of an isolated Mott memristor can be readily obtained by substituting $a_0$, $a_1$, $b_0$ and $b_1$ in Formula (\ref{eqn35}) of $Z(i\omega;Q)$ with $R_1$, $R_2$ and $C_1$ using equations (\ref{eqn27})--(\ref{eqn30}). The real part of $Z(i\omega;Q)$ in Formula (\ref{eqn36}), now defined as the frequency-domain resistance function, takes the form of
\begin{equation}\label{eqn39}
	R_\omega(\omega,Q)\triangleq\text{Re}Z(i\omega;Q)=\frac{(R_1+R_2)+(R_{1}C_1)^2R_2\omega^2}{1+(R_{1}C_1)^2\omega^2}
\end{equation}
which can be further rewritten by replacing $C_1$ with $-1/b_{11}R_1$ using equation (\ref{eqn33})
\begin{equation}\label{eqn40}
	R_\omega(\omega,Q)=\frac{b_{11}^2(R_1+R_2)+R_2\omega^2}{b_{11}^2+\omega^2}
\end{equation}
The sign of $R_\omega(\omega,Q)$ can be either positive or negative, depending on the $(\omega,Q)$ coordinate. $R_\omega(\omega,Q)\geq0$ maps to the LP region, and $R_\omega(\omega,Q)<0$ maps to the EOC region. The angular frequency formula (\ref{eqn38}) to satisfy Chua's fourth LA criterion now becomes
\begin{equation}\label{eqn41}
	\left|\omega\right|<\omega_\text{max}=-b_{11}\sqrt{\frac{-(R_1+R_2)}{R_1}}
\end{equation}
Since the memristance $R_2$ is always positive, this indicates that $(R_1+R_2)$ must be negative for $\omega_\text{max}$ to be a real number. From the previous discussion of figure \ref{Fig11_C1R1R2}(d), $(R_1+R_2)<0$ maps into the NDR (EOC) region on the steady-state $(i_Q,v_Q)$ locus.

We now look at the imaginary part of $Z(i\omega;Q)$. By substituting $a_0$, $a_1$, $b_0$ and $b_1$ in Formula (\ref{eqn37}) of $\text{Im}Z(i\omega;Q)$ with $R_1$, $R_2$ and $C_1$, we rewrite $\text{Im}Z(i\omega;Q)$ as
\begin{equation}\label{eqn42}
	\text{Im}Z(i\omega;Q)\triangleq L_\omega(\omega,Q)\omega = \left[\frac{-R_1^2C_1}{1+(R_{1}C_1)^2\omega^2}\right]\omega
\end{equation}
where $L_\omega(\omega,Q)$ is defined as the frequency-domain inductance function. Evidently, the sign of $L_\omega(\omega,Q)$ is determined by the sign of $C_1$. Since $C_1$ remains negative at any fixed point $Q$ (see discussion on figure \ref{Fig11_C1R1R2}(c)), $L_\omega(\omega,Q)$ is always positive, regardless of the location of $Q$ in the LP or EOC region. Therefore, \textit{the frequency-domain reactance of an isolated Mott memristor is always inductive}, causing its voltage output to lead a sinusoidal current input in phase. $\text{Im}Z(i\omega;Q)$ can be further rewritten by replacing $C_1$ with $-1/b_{11}R_1$
\begin{equation}\label{eqn43}
	L_\omega(\omega,Q)\omega = \left(\frac{b_{11}R_1}{b_{11}^2+\omega^2}\right)\omega
\end{equation}
Finally, the frequency-domain small-signal impedance function is expressed as
\begin{equation}\label{eqn44}
	Z(i\omega;Q) = R_\omega(\omega,Q) + iL_\omega(\omega,Q)\omega
\end{equation}
Therefore, in frequency domain one can treat an uncoupled Mott memristor as a positive inductor in series with a resistor that is negative up to certain maximum frequency (the EOC region) and positive beyond it (the LP region)~\cite{Liang22}.

\subsubsection[4.3.3]{Phase diagram for complexity}

The fourth criterion in Chua's LA theorem tells that a negative real part of the complexity function of an uncoupled 1D circuit element at some finite frequencies is a sufficient condition for it to be locally active. For a current−driven memristor, its complexity function is the impedance function $Z(i\omega;Q)$. Since $Z(i\omega;Q)$ depends on both the angular frequency $\omega$ and the steady-state current $i_Q$, plotting Re$Z$ as a color scale with current and frequency as the $(x,y)$ coordinate provides a visualization of the LP and EOC regions in the operating parameter space. The $\text{Re}Z=0$ contour outlines the border between them. One could call such a 2D graphical representation of Re$Z$ a \emph{phase diagram for complexity}.

\begin{figure}[htb]
	\centering
	\includegraphics[width=0.9\linewidth]{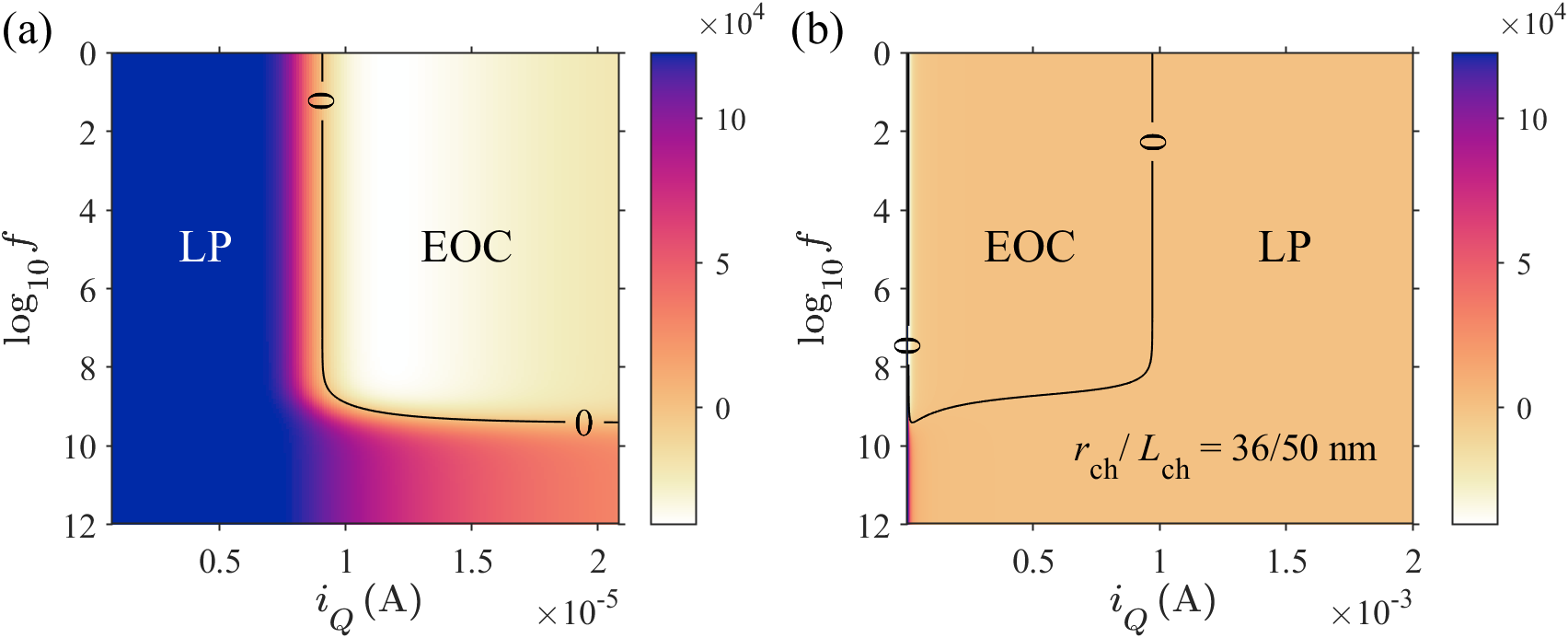}
	\caption{2D color scale map of $\text{Re}Z(i_Q,f)$ for the midsize VO$_2$ Mott memristor as a visualization of the LP and EOC regions in the frequency and current parameter space, showing (a) the low-current region of it up to $i_Q=20.8$~$\upmu$A, and (b) a wide-range map plotted up to $i_Q=2$~mA. Frequencies are plotted in logarithmic scale.}
	\label{Fig18_ReZ-Colormap}
\end{figure}

In figure \ref{Fig18_ReZ-Colormap}, we plot the 2D color scale map of $\text{Re}Z(i_Q,f)$ for the midsize VO$_2$ Mott memristor. Figure \ref{Fig18_ReZ-Colormap}(a) is the low−current region of it plotted up to 20.8~$\upmu$A. It shows that at lower frequencies,  the LP $\rightarrow$ EOC transition occurs at a nearly frequency-independent critical current $i_{c1}\approx9.077$~$\upmu$A, as indicated by an almost vertical $\text{Re}Z=0$ contour. At frequencies higher than $\sim$0.88~GHz, the critical current increases drastically, consequently the direction of the $\text{Re}Z=0$ contour turns almost parallel to the current axis. Figure \ref{Fig18_ReZ-Colormap}(b) is the same color scale Re$Z$ map with a much wider current range up to 2~mA, revealing an EOC $\rightarrow$ LP transition that occurs at a nearly constant critical current $i_{c2}\approx971.18$~$\upmu$A at low frequencies. The direction of the $\text{Re}Z=0$ contour shows a similar crossover from nearly vertical at frequencies lower than $\sim$0.88~GHz to almost horizontal at higher frequencies.

\begin{figure}[htb]
	\centering
	\includegraphics[width=0.9\linewidth]{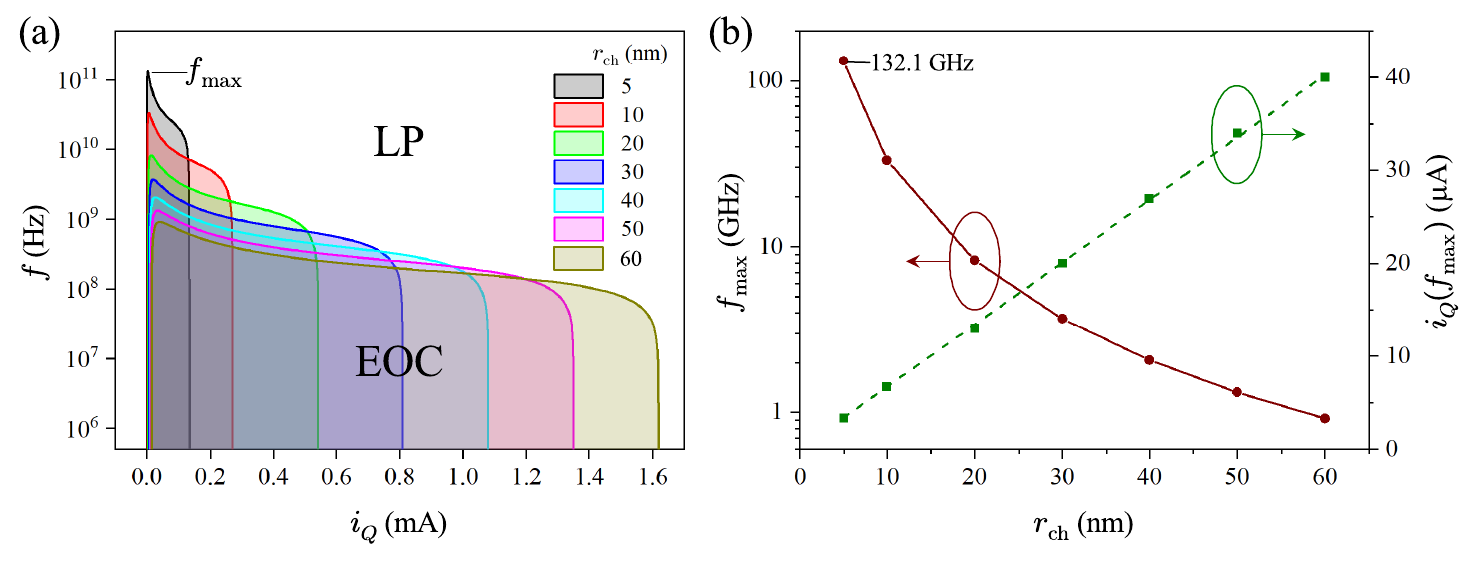}
	\caption{(a) A zoomed view of the $\text{Re}Z=0$ contours for VO$_2$ Mott memristors with channel radius $r_\text{ch}$ in the range of 5~nm to 60~nm. Shaded areas under the contours are the EOC regions wherein $\text{Re}Z(i_Q,f)<0$. The apex of each contour at $f=f_\text{max}$ shows the maximum frequency of the EOC region. (b) Scaling of $f_\text{max}$ (brown) and $i_Q(f_\text{max})$ (green) vs.~$r_\text{ch}$. $f_\text{max}$ increases super-exponentially as $r_\text{ch}$ decreases. $i_Q(f_\text{max})$ scales linearly with $r_\text{ch}$. A linear regression (dashed line) returns a slope of $671\pm3$~(A/m) and coefficient of determination $R^{2}=0.99988$.}
	\label{Fig19_ReZ-SizeScaling}
\end{figure}

To understand the scaling trend of the local activity region versus the device size, we plotted the 2D color scale map of $\text{Re}Z(i_Q,f)$ for VO$_2$ Mott memristors with different combinations of $r_\text{ch}$ and $L_\text{ch}$ sizes. Figure \ref{Fig19_ReZ-SizeScaling} shows the main results of this exercise. We found that Re$Z$ is independent of the VO$_2$ channel length $L_\text{ch}$. This is not unexpected since the VO$_2$ compact model is essentially 2D in nature. Figure \ref{Fig19_ReZ-SizeScaling}(a) is a zoomed view of the $\text{Re}Z=0$ contours for VO$_2$ devices with $r_\text{ch}$ in the range of 5--60~nm. The shaded area under each contour is the EOC region that satisfies $\text{Re}Z(i_Q,f)<0$. The apex of each contour corresponds to the maximum frequency $f_\text{max}$ for the device to remain locally active. Figure \ref{Fig19_ReZ-SizeScaling}(b) shows that $f_\text{max}$ increases super-exponentially as the VO$_2$ channel radius $r_\text{ch}$ shrinks down. For a VO$_2$ device with $r_\text{ch}$ as small as 5~nm, $f_\text{max}$ reaches as high as 132.1 GHz. This is a favorable device scaling in the sense of operation bandwidth for using Mott memristors as locally active components. It also reveals that the steady-state current at $f_\text{max}$ is directly proportional to the radius of the conduction channel $r_\text{ch}$.

\section[5. LA Coupled]{Local analysis of reactively-coupled Mott memristors: 2D relaxation oscillator}
The topological constraint for an isolated 1D Mott memristor limits the dynamics it can describe. It is impossible to exhibit damped or persistent oscillations. However, the topological constraint is lifted once it is coupled to one or more reactive elements, which increases the dimension of the system as a whole and the complexity in dynamics. For continuous dynamical systems, the Poincar\'e–Bendixson theorem says that chaos only arises in three or more dimensions.

For simplicity, we will limit our discussions to 2D cases. Experimentally, it is difficult to characterize an isolated memristor without inadvertently coupling it to one or more reactive elements. On the other hand, such couplings introduce interesting phenomena such as self-excited persistent oscillations or stable limit cycles. Limit cycles belong to an important category of attractors besides fixed points. A nonlinear system consisting of a Mott memristor coupled with reactive elements may exhibit a \textit{local} Hopf-like bifurcation. As a bifurcation parameter is varied, its local stability abruptly switches between a fixed point and a limit cycle around it. Persistent oscillations that arise out of Hopf-like bifurcations are well studied in the Hogdkin-Huxley and FitzHugh-Nagumo models of biological nerve cells~\cite{Hastings74,Troy78,Rinzel80,Dogaru98}, and they are relevant for the intriguing neuronal signaling 
phenomena such as firing of action potentials. However, finding the limit cycle solutions for a dynamical system is generally a very difficult mathematical problem. The unsolved second part of Hilbert's $16^{\text{th}}$ problem is a famous example. The local analysis techniques that we have discussed so far are not sufficient, and one needs to resort to global nonlinear techniques such as nullcline analysis and Lyapunov stability theory. In this section, we apply local analysis to a simple example of a reactively-coupled Mott memristor. Then in the next section we take a cursory glance at global analysis using the same example to illustrate its usefulness.

\subsection[5.1]{Voltage-biased relaxation oscillator circuit}
A voltage-biased Pearson–Anson relaxation oscillator circuit is a simple yet very useful example to illustrate the effect of such external couplings. As shown in figure \ref{Fig20_VO2PACircuit}, if a Mott memristor $M$ is connected to a capacitor $C_\text{p}$ in parallel, and both of them are connected to a resistor $R_\text{s}$ placed in series, then together they form a composite circuit which can be represented as $\{(M// C_\text{p})+R_\text{s}\}$. In practice, one may inadvertently form such a circuit when attempting to test an individual memristor device without explicitly connecting $C_\text{p}$ and $R_\text{s}$. $C_\text{p}$ may come from the geometric capacitance between the two electrodes of a thin-film metal-oxide-metal device, or stray capacitance of coaxial cables. $R_\text{s}$ may arise from the output resistance of a voltage source, resistance of metal lead wires, and contact resistance at the metal-oxide interfaces. If a DC voltage bias $V_\text{dc}$ is applied to one terminal of $R_\text{s}$, and the other terminal of $R_\text{s}$ connected to the memristor is taken as the output node, the $\{(M//C_\text{p})+R_\text{s}\}$ circuit forms a Pearson–Anson (PA) or relaxation oscillator. If the passive elements and voltage bias are appropriately valued, it exhibits persistent self-excited oscillations that will be elaborated below.

\subsection[5.2]{Small-signal analysis: the element combination approach}

\begin{figure}[htb]
	\centering
	\includegraphics[width=0.5\linewidth]{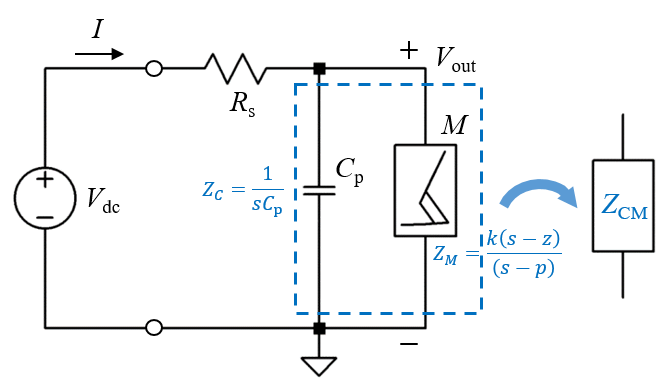}
	\caption{Circuit diagram of a DC voltage biased Pearson-Anson relaxation oscillator formed by a Mott memristor $M$ in parallel with a capacitor $C_\text{p}$, and both connected in series with a resistor $R_\text{s}$. Formulas in blue are the $s$-domain impedances of $M$ and $C_\text{p}$ (with an initial condition $v(0)=0$) used to derive the total impedance $Z_\text{CM}$.}
	\label{Fig20_VO2PACircuit}
\end{figure}

The $\{(M//C_\text{p})+R_\text{s}\}$ Mott memristor PA circuit is a second-order system with two state variables: charge $q_\text{C}$ stored in the capacitor $C_\text{p}$ (or equivalently voltage $v$ across $C_\text{p}$ and $M$), and fraction of metallic phase $x$ in the memristor $M$. This second-order system is described by two coupled differential equations. Its steady states or fixed points can be found by the global nullcline method, which will be covered in the next section. For the sake of continuity in discussion, we assume that fixed points of the PA oscillator are already solved, and focus on what can be told by local analysis for the moment. We can combine $(M//C_\text{p})$ into a composite second-order nonlinear element (dashed box in figure \ref{Fig20_VO2PACircuit}), then apply the small-signal local analyses and Chua’s LA criteria to the system consisting of the composite element in series with $R_\text{s}$.

To perform small-signal analysis, we first need to find out the transfer function of the composite circuit. In the $s$ domain, impedance of a capacitor $C_\text{p}$ is $1/sC_\text{p}$ if assuming an initial condition $v(0)=0$. Impedance $M$ in its pole−zero form is $Z_\text{M}(s;Q)=\frac{k(s-z)}{(s-p)}$ (equation~\ref{eqn31}). One can derive the transfer function $H(s;Q)$ of the PA oscillator about a fixed point Q using the voltage divider formula
\begin{equation}\label{eqn45}
	H(s;Q)=V_\text{out}/V_\text{dc}=Z_\text{CM}/(R_\text{s}+Z_\text{CM})
\end{equation}
where $Z_\text{CM}$ is the total impedance of $C_\text{p}$ in parallel with $M$
\begin{equation}\label{eqn46}
	Z_\text{CM}=Z_\text{C}Z_\text{M}/(Z_\text{C}+Z_\text{M})=\frac{k(s-z)}{kC_\text{p}s^{2}+(1-kC_\text{p}z)s-p}
\end{equation}
Substituting the expression of $Z_\text{CM}$ in the transfer function formula, we get
\begin{equation}\label{eqn47}
	H(s;Q)=\frac{k(s-z)}{kR_\text{s}C_\text{p}s^{2}+(R_\text{s}+k-kR_\text{s}C_\text{p}z)s-(R_\text{s}p+kz)}
\end{equation}
One can see that $H(s;Q)$ of a Mott memristor PA oscillator has the same zero as an uncoupled memristor, but it has a pair of two poles instead of one pole for an uncoupled memristor. 

To simplify the expression of $H(s;Q)$, we define a time constant $\tau_0 \triangleq R_\text{s}C_\text{p}$ and a cutoff frequency $\omega_0 \triangleq (R_\text{s}C_\text{p})^{-1}$. We also substitute $k$ by the positive real memristance function $R_\text{ch}$, and rewrite $H(s;Q)$ in the pole-zero form
\begin{equation}\label{eqn48}
	H(s;Q)=\frac{k'(s-z)}{d_{2}s^2+d_{1}s+d_0}=\frac{k'(s-z)}{(s-p_+)(s-p_-)}
\end{equation}
where
\begin{eqnarray}\label{eqarray9}
	\label{eqn49}
	k' = \omega_0 \\
	\label{eqn50}
	d_2 = 1 \\
	\label{eqn51}
	d_1 = \frac{R_\text{s}+R_\text{ch}-R_\text{ch}\tau_{0}z}{R_\text{ch}\tau_0} = \left(1+\frac{R_\text{s}}{R_\text{ch}}\right)\omega_0-z \\
	\label{eqn52}
	d_0 = \frac{-\left(R_\text{s}p+R_\text{ch}z\right)}{R_\text{ch}\tau_0} = -\frac{R_\text{s}}{R_\text{ch}}\omega_{0}p-\omega_{0}z 
\end{eqnarray}
Here $p$ and $z$ are the pole and zero of the memristor $M$. We then derive the pair of poles $p_\pm$ for the PA oscillator by finding the roots of the quadratic equation $d_{2}s^2+d_{1}s+d_0=0$
\begin{equation}\label{eqn53}
	p_\pm = \frac{-d_1\pm \sqrt{d_1^2-4d_{2}d_0}}{2d_2} = \frac{-d_1\pm \sqrt{d_1^{2}-4d_0}}{2}
\end{equation}
The discriminant $d_1^2-4d_{2}d_0$ of the quadratic equation is expressed as
\begin{equation}\label{eqn54}
	d_1^2-4d_{2}d_0 = z^2 + 2\omega_{0}z\left(1-\frac{R_\text{s}}{R_\text{ch}}\right) + \left(1+\frac{R_\text{s}}{R_\text{ch}}\right)^{2}\omega_0^2 + 4\omega_{0}p\left(\frac{R_\text{s}}{R_\text{ch}}\right)
\end{equation}
If $d_1^2-4d_{2}d_0\geq 0$, $p_\pm$ are positive or negative real numbers. Otherwise if $d_1^2-4d_{2}d_0<0$, $p_\pm$ are a complex conjugate pair. Without loss of generality, here we keep the standard expression for the discriminant of a quadratic equation instead of replacing $d_{2}$ with 1 (equation~(\ref{eqn50})) for the particular case of a Mott memristor PA circuit.

To understand the effects of parameters $R_\text{s}$, $C_\text{p}$ and $V_\text{dc}$ on the dynamical behavior of a Mott memristor PA oscillator, we first calculated the values for the pair of poles $p_\pm$ of its small-signal transfer function $H(s;Q)$, by varying one parameter while fixing the other two parameters, then applied the parametric Nyquist plot technique to gain intuition on the stability of the reduced system.

\begin{figure}[htb]
	\centering
	\includegraphics[width=0.9\linewidth]{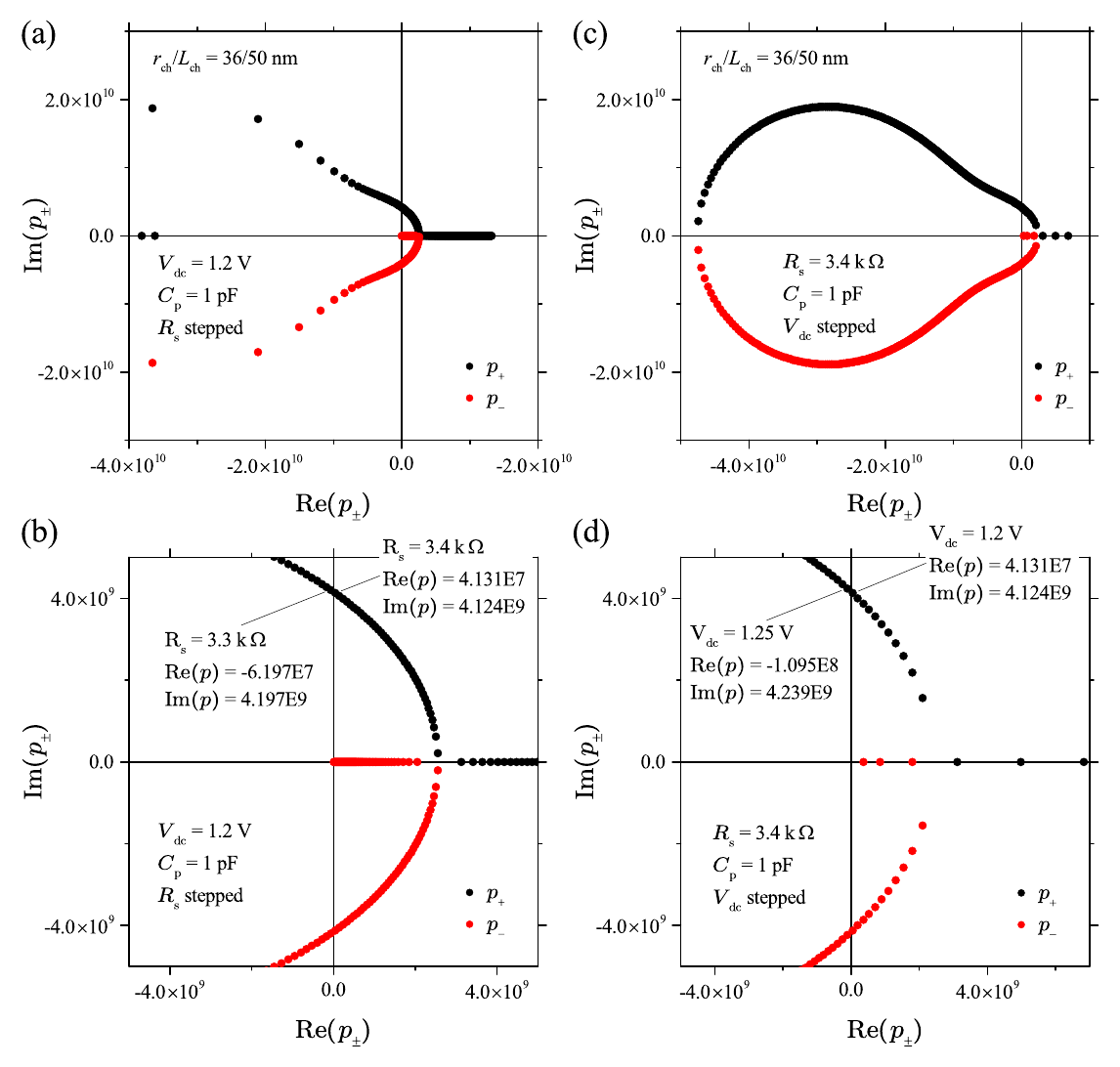}
	\caption{(a) Nyquist plot of Im($p_\pm$) vs.~Re($p_\pm$) for the pair of poles $p_\pm$ of the small-signal transfer function $H(s;Q)$ of the midsize VO$_2$ Mott memristor PA oscillator with $C_\text{p}=1$~pF, $V_\text{dc}=1.2$~V, and $R_\text{s}$ stepped from 100~$\Omega$ to 27~k$\Omega$ at 100~$\Omega$ interval. (b) is a zoomed view of (a), showing that increasing $R_\text{s}$ from 3.3~k$\Omega$ to 3.4~k$\Omega$ turns Re($p_\pm$) from negative to positive. (c) Nyquist plot of the same PA oscillator with $R_\text{s}=3.4~\text{k}\Omega$, $C_\text{p}=1$~pF, and $V_\text{dc}$ stepped from 0.55~V to 13.5~V at 50~mV interval. (d) is a zoomed view of (c), showing that increasing $V_\text{dc}$ from 1.2~V to 1.25~V turns Re($p_\pm$) from positive to negative.}
	\label{Fig21_NyquistVO2PApoles}
\end{figure}

Let us first examine the effect of varying $R_\text{s}$ while fixing $C_\text{p}$ and $V_\text{dc}$. Figure \ref{Fig21_NyquistVO2PApoles}(a) shows Nyquist plot of Im($p_\pm$) vs.~Re($p_\pm$) for the midsize VO$_2$ Mott memristor PA oscillator with $C_\text{p}=1$~pF, $V_\text{dc}=1.2$~V, and $R_\text{s}$ stepped from 100~$\Omega$ to 27~k$\Omega$ at 100~$\Omega$ interval. It reveals three distinctive regions as $R_\text{s}$ increases: (1) $R_\text{s}=100~\Omega-200~\Omega$, $p_+$ and $p_-$ are negative real numbers. (2) $R_\text{s}=300~\Omega−7.5~\text{k}\Omega$, $p_+$ and $p_-$ are a complex conjugate pair. (3) $R_\text{s}=7.6~\text{k}\Omega−27~\text{k}\Omega$, $p_+$ and $p_-$ are positive real numbers. Figure \ref{Fig21_NyquistVO2PApoles}(b) is a zoomed view of figure 19(a), which reveals that increasing $R_\text{s}$ from 3.3~k$\Omega$ to 3.4~k$\Omega$ flips the sign of Re($p_\pm$) from negative to positive, i.e., the pair of poles crosses over from LHP to RHP. If we treat the reduced system as a 1D uncoupled element, and apply the first criterion of Chua’s LA theorem --- the element is locally active if $H(s;Q)$ has a pole in the open RHP $\text{Re}(s)>0$, then the reduced system has a crossover from LP to LA as $R_\text{s}$ increases from 3.3~k$\Omega$ to 3.4~k$\Omega$.

A similar crossover is observed by varying $V_\text{dc}$ while fixing $R_\text{s}$ and $C_\text{p}$. Figure \ref{Fig21_NyquistVO2PApoles}(c) shows Nyquist plot of Im($p_\pm$) vs.~Re($p_\pm$) for the midsize VO$_2$ Mott memristor PA oscillator with $R_\text{s}=3.4~\text{k}\Omega$, $C_\text{p}=1$~pF, and $V_\text{dc}$ stepped from 0.55~V to 13.5~V at 50~mV interval. It reveals two distinctive regions as $V_\text{dc}$ increases: (1) $V_\text{dc}=0.55~\text{V}-0.65~\text{V}$, $p_+$ and $p_-$ are positive real numbers. (2) $V_\text{dc}=0.7~\text{V}-13.5~\text{V}$, $p_+$ and $p_-$ are a complex conjugate pair. Figure \ref{Fig21_NyquistVO2PApoles}(d) is a zoomed view of figure \ref{Fig21_NyquistVO2PApoles}(c), which reveals that increasing $V_\text{dc}$ from 1.2~V to 1.25~V flips the sign of Re($p_\pm$) from positive to negative, i.e., the pair of poles crosses over from RHP to LHP. Applying the first criterion of Chua’s LA theorem in a similar manner, then the reduced system has a crossover from LA to LP as $V_\text{dc}$ increases from 1.2~$\text{V}$ to $1.25~\text{V}$.

Instead of using the sign of Re($p_\pm$) in the Cartesian coordinate as a test for the first LA criterion, one can also use the argument (phase) of a pole in the polar coordinate. The $p_+$ pole is located in the first and second quadrants, including the Re and Im+ axes. Its complex conjugate $p_-$ is located in the third and fourth quadrants, including the Re and Im$-$ axes. We only need to look at arg($p_+$), argument of the $p_+$ pole of $H(s;Q)$, as a test for the first LA criterion. A crossover from LA to LP occurs if arg($p_+$) increases from below $90^\circ$ to above $90^\circ$, i.e., $p_+$ moves from the first quadrant to the second quadrant by crossing the Im+ axis.

\begin{figure}[htb]
	\centering
	\includegraphics[width=0.9\linewidth]{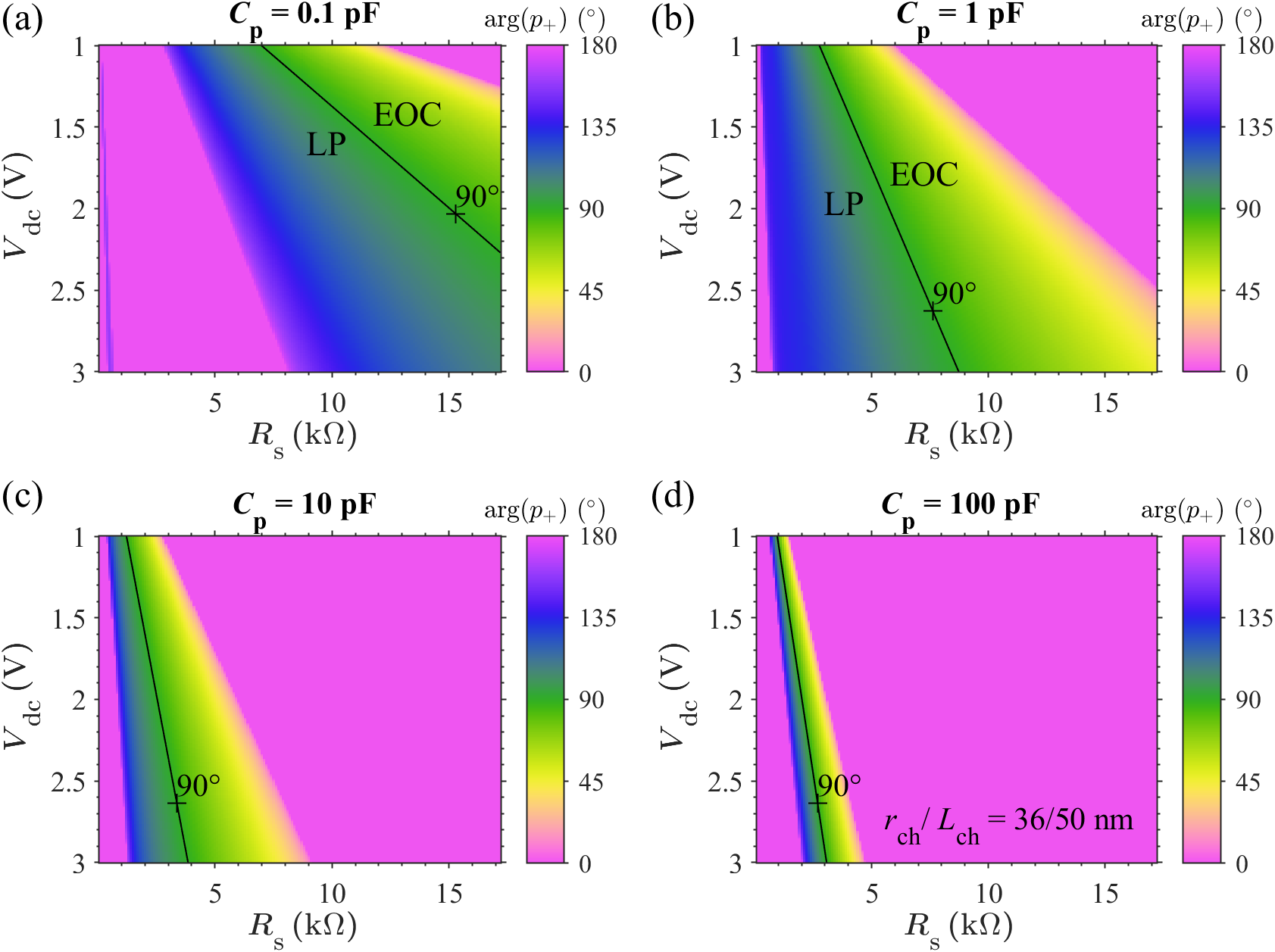}
	\caption{2D color scale maps of $\text{arg}(p_+)(R_\text{s},V_\text{dc})$, argument of the $p_+$ pole of $H(s;Q)$ with $R_\text{s}$ and $V_\text{dc}$ as the $x$ and $y$ coordinates, for the midsize VO$_2$ Mott memristor PA oscillator at (a) $C_\text{p}=0.1$~pF, (b) $C_\text{p}=1$~pF, (c) $C_\text{p}=10$~pF and (d) $C_\text{p}=100$~pF, respectively. In each plot, the LA region ($\text{Re}(p_+)>0$) and the LP region ($\text{Re}(p_+)<0$) are separated by the $90^{\circ}$ contour line (solid black line marked by ``+''). We use a cyclic color map with four distinct colors to allow four orientations or phase angles to be visualized. Both $\text{arg}(p_+)=0^{\circ}$ and $\text{arg}(p_+)=180^{\circ}$ are shown with the same color (pink).}
	\label{Fig22_VO2PA_pole1-Colormap}
\end{figure}

At a fixed $C_\text{p}$ parameter, one can thus visualize the LA and LP operating regions of a Mott memristor PA oscillator by plotting a 2D color scale map of arg($p_+$) with $R_\text{s}$ and $V_\text{dc}$ parameters as the $x$ and $y$ coordinates. The procedure is then repeated at different $C_\text{p}$ values to see how the LA and LP regions evolve as $C_\text{p}$ is adjusted. Figure \ref{Fig22_VO2PA_pole1-Colormap} shows four 2D color scale maps of arg($p_+$) for the midsize VO$_2$ Mott memristor PA oscillator at $C_\text{p}=0.1$~pF (a), 1~pF (b), 10~pF (c) and 100~pF (d), respectively. 

We take figure \ref{Fig22_VO2PA_pole1-Colormap}(a) as an example to discuss their common characteristics. From the complex domain aspect, arg($p_+$) has three distinctive regions if one navigates along the top-right to bottom-left diagonal. At small $V_\text{dc}$ and large $R_\text{s}$ (the top-right pink region), $\text{arg}(p_+)=0^{\circ}$, $p_+$ is a positive real number on the Re+ axis. At larger $V_\text{dc}$ and smaller $R_\text{s}$, $p_+$ is a complex number (the middle colored region) in either the first quadrant (LA) or the second quadrant (LP), divided by the $90^{\circ}$ contour line. At even larger $V_\text{dc}$ and smaller $R_\text{s}$, $\text{arg}(p_+)=180^{\circ}$, $p_+$ is a negative real number on the Re$-$ axis. A conspicuous feature is that all the borderlines between adjacent regions are straight lines that extend from top-left to bottom-right. Despite their appearance, they do not intersect at a common point if extrapolated toward the top-left direction. From the dynamics aspect, there are only two regions, the LA region ($\text{Re}(p_+)>0$) and the LP region ($\text{Re}(p_+)<0$) separated by the $90^{\circ}$ contour line.

The effect of $C_\text{p}$ can be seen by comparing the four color scale maps. As $C_\text{p}$ increases from 0.1~pF to 10~pF, the $90^{\circ}$ LA-LP borderline rotates clockwise, the rotation stalls as $C_\text{p}$ further increases to 100~pF. The $\text{arg}(p_+)=0^{\circ}$ (positive real) region at the top right corner continuously grows with increase in $C_\text{p}$,  while the complex region shrinks with increase in $C_\text{p}$. The $\text{arg}(p_+)=180^{\circ}$ (negative real) region at the bottom left corner initially shrinks significantly as $C_\text{p}$ increases from 0.1~pF to 1~pF, then it recovers a little as $C_\text{p}$ further increases.

Using the element combination approach, one can also examine the case of an inductively coupled Mott memristor, e.g., connecting an external inductor in parallel with a Mott memristor. It is found that the poles of the transfer function of such a composite circuit remain in the LHP, thus the composite circuit does not meet the LA criterion to exhibit instabilities or persistent oscillations~\cite{Brown22a}. This is not surprising since an isolated Mott memristor has an apparent inductive reactance.

\subsection[5.3]{Jacobian matrix method}
For second or higher order nonlinear systems, the Jacobian matrix method is a linearization technique that allows local stability analysis near a hyperbolic fixed point. As an introduction in a nutshell, consider an autonomous system of ODEs $\dot{\textbf{x}}=\textbf{f}(\textbf{x})$, where $\dot{\textbf{x}}$ is the component-wise time derivative for the set of state variables \textbf{x}. \textbf{x} corresponds to a point in an open subset of real $n$-space $E\subset \mathbb{R}^{n}$. $\textbf{f}:E\rightarrow \mathbb{R}^{n}$ is a differentiable function that describes the dynamics of \textbf{x}. \textbf{f} is also called a vector field, since mapping from \textbf{x} to \textbf{f}(\textbf{x}) assigns a vector. For a 2D (planar) system described by $\frac{dx}{dt}=f(x,y)$ and $\frac{dy}{dt}=g(x,y)$, $\textbf{f}(\textbf{x})=(f(x,y),g(x,y))$ can be visualized by a vector based at the point $(x,y)$, whose $x$- and $y$-components are $f(x,y)$ and $g(x,y)$, respectively. The set of solutions $\phi(t,\textbf{x}_0)$ of the initial value problem $\dot{\textbf{x}}=\textbf{f}(\textbf{x})$,  $\textbf{x}(0)=\textbf{x}_0\in E$ is called the flow of the ODE, or the flow of the vector field \textbf{f}. For each initial condition $\textbf{x}_0$, $\phi(t,\textbf{x}_0)$ gives the trajectory of a unique solution of the ODE, which is called the orbit of \textbf{x} under $\phi$.

In autonomous systems, \textbf{f} does not explicitly depend on time. If $\textbf{f}(\textbf{x}_Q)=0$, i.e., its time derivative is zero, then \textbf{x}$_Q$ is a fixed point. The Jacobian matrix \textit{D}\textbf{f}, or simply Jacobian, is the matrix of all the first-order partial derivatives of \textbf{f}(\textbf{x}). The Hartman–Grobman theorem and the stable manifold theorem guarantee that the local qualitative behavior of a nonlinear system $\dot{\textbf{x}}=\textbf{f}(\textbf{x})$ near a hyperbolic fixed point \textbf{x}$_Q$ is determined by a linear system $\dot{\textbf{x}}=A\textbf{x}$, where $A=D\textbf{f}(\textbf{x}_Q)$ is the Jacobian of \textbf{f} at \textbf{x}$_Q$. In other words, the flow of a nonlinear system is topologically conjugate to that of its linearized system
in some neighborhood of a fixed point, so long as it is hyperbolic. If the Jacobian is a square matrix, and none of the eigenvalues of $D\textbf{f}(\textbf{x}_Q)$ is a pure imaginary number, then the fixed point is hyperbolic, and its stability can be told by the signs of the real parts of the eigenvalues, as will be elaborated later.

Next, we take the Jacobian matrix approach to analyze the local stability of a Mott memristor PA oscillator. This second-order system is described by the following equations
\begin{eqnarray}\label{eqarray10}
	\label{eqn55}
	v = R_\text{ch}(x)i_\text{M} \\
	\label{eqn56}
	\frac{dx}{dt} = f_x(x,i_\text{M}) \\
	\label{eqn57}
	\frac{dv}{dt} = \frac{1}{C_\text{p}}\left(\frac{V_\text{dc}-v}{R_\text{s}} - i_\text{M}\right) 
\end{eqnarray}
where $f_x(x,i_\text{M})$ and $R_\text{ch}(x)$ are the kinetic function and memristance function of the memristor $M$, respectively. $i_\text{M}$ is the current flowing through $M$. Substituting it with $v/R_\text{ch}(x)$, we obtain the two coupled ODEs that describe the system dynamics
\begin{eqnarray}\label{eqarray11}
	\label{eqn58}
	\frac{dx}{dt}\triangleq f(x,v) = f_x\left(x,\frac{v}{R_\text{ch}(x)}\right) \\
	\label{eqn59}
	\frac{dv}{dt}\triangleq g(x,v) =  \frac{1}{C_\text{p}}\left(\frac{V_\text{dc}-v}{R_\text{s}} - \frac{v}{R_\text{ch}(x)}\right) 
\end{eqnarray}
In a vector form, the system is described as $\dot{\textbf{x}}=\textbf{f}(\textbf{x})$. Here $\dot{\textbf{x}}=[x,v]^\text{T}$ is the state variable vector, and $\textbf{f}(\textbf{x})=[f(x,v),g(x,v)]^\text{T}$ is the differentiable function that describes the dynamics of \textbf{x}. Around a fixed point $Q$ with a coordinate $(x_Q,v_Q)$, the Jacobian matrix \textit{D}\textbf{f} of the system is a 2\texttimes2 matrix of all the first-order partial derivatives of \textbf{f}(\textbf{x}) that takes the form
\begin{equation}\label{eqn60}
	\left.D\textbf{f}\right|_Q = \begin{bmatrix} \xi_{11} & \xi_{12} \\ \xi_{21} & \xi_{22} \end{bmatrix} =
	\begin{bmatrix}[2] \left.\dfrac{\partial f(x,v)}{\partial x}\right|_Q & \left.\dfrac{\partial f(x,v)}{\partial v}\right|_Q \\ \left.\dfrac{\partial g(x,v)}{\partial x}\right|_Q & \left.\dfrac{\partial g(x,v)}{\partial v}\right|_Q \end{bmatrix}
\end{equation}
The four elements of the Jacobian matrix are derived as
\begin{eqnarray}\label{eqarray12}
	\label{eqn61}
	\xi_{11} = \left.\frac{\partial f(x,v)}{\partial x}\right|_Q = b_{11}(Q) - \frac{b_{12}(Q)a_{11}(Q)}{R_\text{ch}(x_Q)} \\
	\label{eqn62}
	\xi_{12} = \left.\frac{\partial f(x,v)}{\partial v}\right|_Q =  \frac{b_{12}(Q)}{R_\text{ch}(x_Q)} \\
	\label{eqn63}
	\xi_{21} = \left.\frac{\partial g(x,v)}{\partial x}\right|_Q =  \frac{a_{11}(Q)}{R_\text{ch}(x_Q)C_\text{p}} \\
	\label{eqn64}
	\xi_{22} = \left.\frac{\partial g(x,v)}{\partial v}\right|_Q = -\left(\frac{1}{R_\text{s}C_\text{p}} + \frac{1}{R_\text{ch}(x_Q)C_\text{p}}\right) 
\end{eqnarray}
The eigenvalues of the Jacobian around a fixed point $Q$ are calculated by solving its characteristic equation that can be expanded to a quadratic polynomial.
\begin{equation}\label{eqn65}
	\left.(D\textbf{f}-\lambda I \right)|_Q = \lambda^2-\tr(D\textbf{f})\lambda+\det(D\textbf{f}) = \lambda^2-(\xi_{11}+\xi_{22})\lambda+(\xi_{11}\xi_{22}-\xi_{12}\xi_{21})
\end{equation}
where $I$ is the identity matrix. $\tr(D\textbf{f})$ is the trace of the Jacobian and $\det(D\textbf{f})$ is the determinant of it. For simplicity, we use $\tr$ and $\det$ to represent them hereafter. Their expressions are derived as
\begin{eqnarray}\label{eqarray13}
	\label{eqn66}
	\tr = -\left[\omega_1\left(1+\frac{R_1}{R_\text{ch}}\right)+\omega_0\left(1+\frac{R_\text{s}}{R_\text{ch}}\right)\right] \\
	\label{eqn67}
	\det = \omega_1\omega_0\left(1+\frac{R_1}{R_\text{ch}}+\frac{R_\text{s}}{R_\text{ch}}\right) 
\end{eqnarray}
where we define another cutoff frequency $\omega_1\triangleq (R_{1}C_1)^{-1}$ besides the previously defined cutoff frequency $\omega_0 = (R_\text{s}C_\text{p})^{-1}$ to simplify the expressions.

The two roots $\lambda_+$ and $\lambda_-$ of the characteristic equation are
\begin{equation}\label{eqn68}
	\lambda_\pm = \frac{\tr\pm \sqrt{\tr^2-4\det}}{2}
\end{equation}
Note that $\lambda_{+}+\lambda_-=\tr$ and $\lambda_+\lambda_-=\det$, i.e., trace is the sum of eigenvalues, and determinant is the product of them.
The discriminant $\tr^2-4\det$ of the characteristic equation in expanded form is
\begin{equation}\label{eqn69}
	\tr^2-4\det = \omega_1^2(1+\gamma_1)^2 + \omega_0^2(1+\gamma_\text{s})^2 - 2\omega_1\omega_0[(1-\gamma_1)\gamma_\text{s}+(1+\gamma_1)]
\end{equation}
where we define two dimensionless resistance ratios $\gamma_1\triangleq \frac{R_1}{R_\text{ch}}$ and $\gamma_\text{s}\triangleq \frac{R_\text{s}}{R_\text{ch}}$ to simplify the expression.

\subsection[5.4]{Trace-determinant plane classification}
For a 2D homogeneous linear system $\dot{\textbf{x}}=A\textbf{x}$, the parameter space of a trace-determinant (tr-det) plane allows qualitative classification of its fixed points. Here a homogeneous linear system is defined in opposition to a nonhomogeneous linear system $\dot{\textbf{x}}=A\textbf{x}+\textbf{h}(t)$ that includes a vector of functions $\textbf{h}(t)$ independent of solutions and their derivatives. For a 2D nonlinear system after linearization, one needs to be cautious when attempting the tr-det plane method, since linearization may change the type of its fixed points, especially at the borderlines on the tr-det plane. Here we examine the tr-det plane method on a linearized Mott memristor PA oscillator to see if one can gain some insight into the original nonlinear system.

In a tr-det plane, a coordinate (tr,~det) corresponds to a Jacobian matrix with trace tr and determinant det. The location of this point relative to the parabola curve $\tr^2-4\det=0$ determines the geometry of the phase portrait. The sign of the discriminant $\tr^2-4\det$ divides the eigenvalues of $D\textbf{f}$ into the following regions.
\renewcommand{\labelenumi}{\alph{enumi})}
\begin{enumerate}
	\item If $\tr^2-4\det>0$, $\lambda_+$ and $\lambda_-$ are real and distinct
	\item If $\tr^2-4\det<0$, $\lambda_+$ and $\lambda_-$ are complex conjugates with nonzero imaginary part
	\item If $\tr^2-4\det=0$, $\lambda_+$ and $\lambda_-$ are real and repeated (identical)
\end{enumerate}

Within each of these three regions, the tr-det plane further classifies the dynamics and stability of isolated or non-isolated fixed points as enumerated below. The numbers within () that appear out of order are class identifiers (IDs) used in Table~\ref{table3}, which is a tabulated summary of the tr-det plane classification. It is notable that the only stable region in the tr-det plane is the closed second quadrant, i.e., $\tr\leq0$ and $\det\geq0$. If a fixed point is stable, its eigenvalues $\lambda_+$ and $\lambda_-$ must both be negative real: $(\lambda_+, \lambda_-)\leq0$, or they are complex conjugates with negative real part: $\textrm{Re}(\lambda_+,\lambda_-)\leq0$.

Region (a) with two real and distinct eigenvalues $\lambda_+\neq\lambda_-$
\begin{enumerate}
	\item[(1)] $\lambda_+>0>\lambda_-$: unstable saddle point
	\item[(2)] $\lambda_+=0, \lambda_-<0$: stable line of non-isolated fixed points
	\item[(4)] $\lambda_+>0, \lambda_-=0$: unstable line of non-isolated fixed points
	\item[(5)] $0>\lambda_+>\lambda_-$: stable sink
	\item[(13)] $\lambda_+>\lambda_->0$: unstable source
\end{enumerate}

Region (b) with a pair of complex conjugate eigenvalues $\lambda_\pm=\alpha\pm\beta i$, $\beta\neq0$
\begin{enumerate}
	\item[(8)] $\textrm{Re}(\lambda_\pm)<0$: stable spiral sink
	\item[(9)] $\textrm{Re}(\lambda_\pm)=0$: stable center (not asymptotically stable)
	\item[(10)] $\textrm{Re}(\lambda_\pm)>0$: unstable spiral source
\end{enumerate}
Center is a non-hyperbolic fixed point since it has a pair of pure imaginary eigenvalues, therefore the Hartman–Grobman theorem is not applicable. A center is stable (or Lyapunov stable), but not asymptotically stable. A solution that starts close to a center stays close to it, but never converges to it over time, since non-zero pure imaginary eigenvalues correspond to periodic solutions that oscillate without damping.

Region (c) with real and repeated eigenvalues: $\lambda_\pm=\lambda$

\begin{enumerate}
	\item[(3)] $\lambda=0$: parallel lines of non-isolated fixed points, or entire plane
	\item[(6)] $\lambda<0$ and is incomplete: stable degenerate sink
	\item[(7)] $\lambda<0$ and is complete: stable star sink
	\item[(11)] $\lambda>0$ and is incomplete: unstable degenerate source
	\item[(12)] $\lambda>0$ and is complete: unstable star source
\end{enumerate}
Here a nonzero real and repeated eigenvalue is complete if it has two linearly independent eigenvectors, and the fixed point (star) is a proper node. Otherwise if the eigenvalue is incomplete (has only one eigenvector), the fixed point is a degenerate or improper node.

For a linearized 2D nonlinear system, the view is that the tr-det plane predictions for classes 1, 5, 8, 10, and 13 are always correct. Predictions for the other eight classes may not be accurate, but at least it correctly tells about the stability of classes 6, 7, 11, and 12. Prediction of class 9 is accurate if the system is conservative.

% The normal style is for tables to be indented. This is accomplished by using \verb"\begin{indented}" \dots\ \verb"\end{indented}" and putting \verb"\item[]" before the start of the tabular environment. Omit these commands for any tables which will not fit on the page when indented.
\begin{table}
	\caption{\label{table3}Trace-determinant (tr-det) plane classification of fixed points for 2D linear homogeneous systems. In the tr-det plane, the only stable region is the closed second quadrant, i.e., $\tr\leq0$ and $\det\geq0$.} 
	%\begin{indented}
		\lineup
		\renewcommand{\arraystretch}{1.3}
		%\item[]\begin{tabular}{@{}*{5}{c}}
		\begin{tabular}{@{}*{5}{c}}
			\br
			Determinant&Trace&Eigenvalues&Stability and Class&ID \\ 
			\mr
			$\det<0$&any&$\lambda_+>0>\lambda_-$&unstable saddle point&1 \\\hline
			\multirow{3}{*}{$\det=0$} &$\tr<0$&$\lambda_+=0, \lambda_-<0$&stable line of fixed points&2 \\\cline{2-5}
			 &$\tr=0$&$\lambda_+=\lambda_-=0$&parallel lines, or entire plane&3 \\\cline{2-5}
			 &$\tr>0$&$\lambda_+>0, \lambda_-=0$&unstable line of fixed points&4 \\\hline
			\multirow{12}{*}{$\det>0$} &$\tr<-\sqrt{4\det}$&$0>\lambda_+>\lambda_-$&stable node (sink)&5 \\\cline{2-5}
			 &$\tr=-\sqrt{4\det}$&repeated&stable degenerate node&6,7 \\
			  & &$\lambda_\pm=\tr/2<0$&(6), or stable star (7)& \\\cline{2-5}
			 &$-\sqrt{4\det}<\tr<0$&complex conjugate&stable spiral sink&8 \\
			  & &$\textrm{Re}(\lambda_\pm)<0$& & \\\cline{2-5}
			 &$\tr=0$&complex conjugate&stable center&9 \\
			  & &$\textrm{Re}(\lambda_\pm)=0$&(not asymptotically stable)& \\\cline{2-5}
			 &$0<\tr<\sqrt{4\det}$&complex conjugate&unstable spiral source&10 \\
			  & &$\textrm{Re}(\lambda_\pm)>0$& & \\\cline{2-5}
			 &$\tr=\sqrt{4\det}$&repeated&unstable degenerate node&11, 12 \\
			  & &$\lambda_\pm=\tr/2>0$&(11), or unstable star (12)& \\\cline{2-5}
			 &$\tr>\sqrt{4\det}$&$\lambda_+>\lambda_->0$&unstable node (source)&13 \\
			\br
		\end{tabular}
		\renewcommand{\arraystretch}{1}
	%\end{indented}
\end{table}

Now we apply the tr-det plane analysis for the 2D nonlinear system of the VO$_2$ Mott memristor PA oscillator. This system has been analyzed previously using an element combination approach, by treating a Mott memristor $M$ and a capacitor $C_\text{p}$ in parallel as a composite second-order nonlinear element $Z_\text{CM}$, which is connected to a resistor $R_\text{s}$ in series. The small-signal transfer function of this 1D nonlinear system has two poles $p_\pm$ that are a pair of complex conjugate in part of the circuit parameter space. Nyquist plots of figure \ref{Fig21_NyquistVO2PApoles}(a) and \ref{Fig21_NyquistVO2PApoles}(b) show the evolution in the positions of $p_\pm$ as one steps $R_\text{s}$ while keeping $C_\text{p}=1$~pF and $V_\text{dc}=1.2$~V unchanged. Increasing $R_\text{s}$ from 3.3~k$\Omega$ to 3.4~k$\Omega$ flips the sign of Re($p_\pm$) from negative to positive and produces a crossover from LP to LA as per Chua's LA theorem for a 1D uncoupled element.

\begin{figure}[htb]
	\centering
	\includegraphics[width=0.9\linewidth]{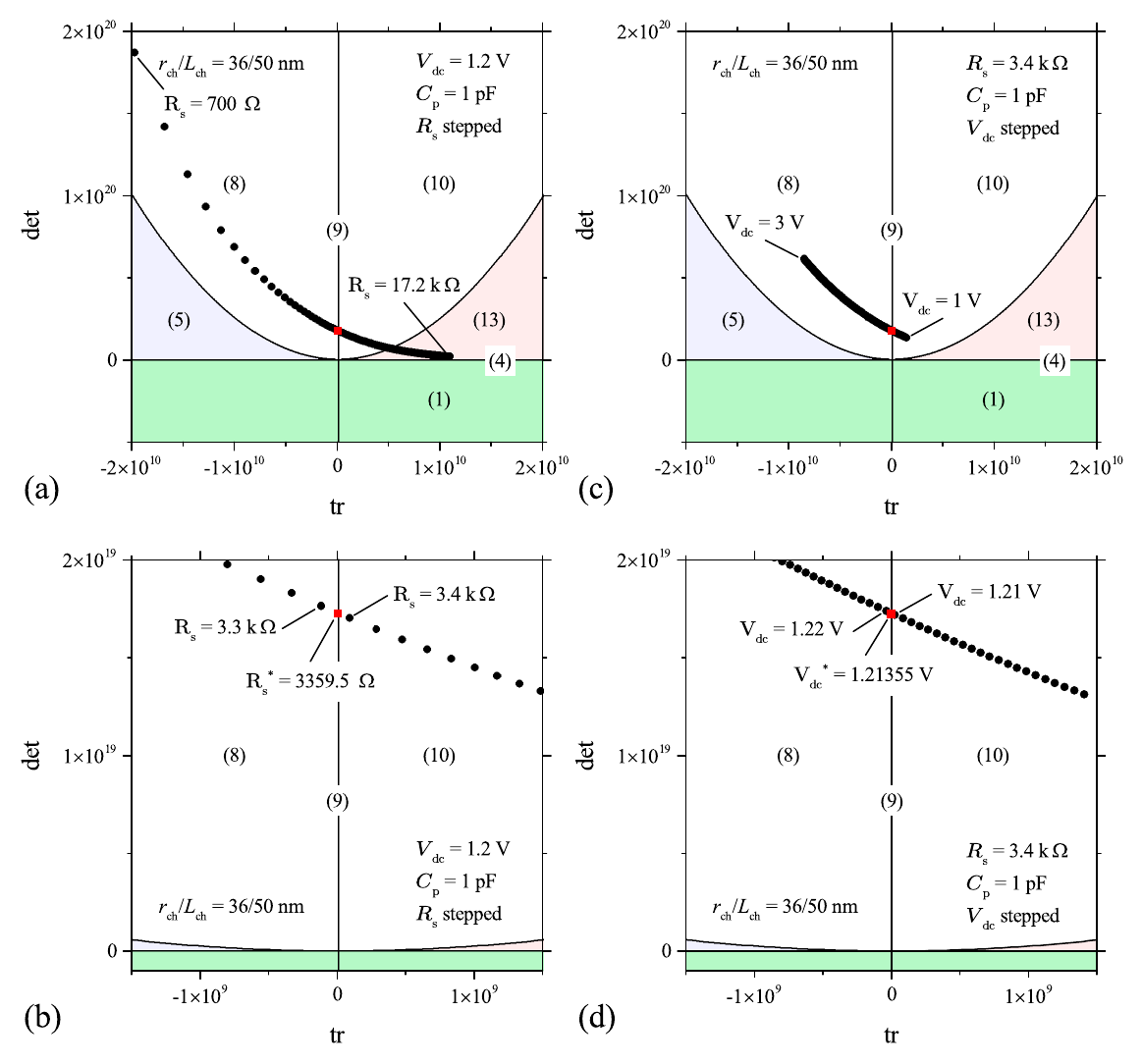}
	\caption{(a) Trace-determinant plane showing the (tr,~det) locus for the Jacobian of the midsize VO$_2$ Mott memristor PA oscillator with $C_\text{p}=1$~pF, $V_\text{dc}=1.2$~V, and bifurcation parameter $R_\text{s}$ stepped from 100~$\Omega$ to 17.2~k$\Omega$ at 100~$\Omega$ interval. (b) is a zoomed view of (a), showing that increasing $R_\text{s}$ from 3.3~k$\Omega$ to 3.4~k$\Omega$ results in a bifurcation from a stable spiral sink (class 8) to an unstable spiral source (class 10). At $R_\text{s}^*=3359.5$~$\Omega$, the fixed point is a stable center (class 9) located on the $\tr = 0$ axis. (c) The (tr,~det) locus for the Jacobian of the same PA oscillator with $R_\text{s}=3.4~\text{k}\Omega$, $C_\text{p}=1$~pF, and bifurcation parameter $V_\text{dc}$ stepped from 1~V to 3~V at 10~mV interval. (d) is a zoomed view of (c), showing that increasing $V_\text{dc}$ from 1.21~V to 1.22~V produces a bifurcation from an unstable spiral source (class 10) to a stable spiral sink (class 8). At $V_\text{dc}^*=1.21355$~V, the fixed point is a stable center (class 9) located on the positive $\det$ axis.}
	\label{Fig23_VO2PA_tr-det}
\end{figure}

Instead of element combination, now we apply the tr-det plane analysis of the Jacobian linearized 2D PA oscillator system. We show that the 2D tr-det plane method elucidates the nature of this crossover in local activity to be a bifurcation which changes the stability of an isolated fixed point. Figure \ref{Fig23_VO2PA_tr-det}(a) plots the (tr,~det) plane that is geometrically divided into different regions by the tr- and det-axes and the $\det=\tr^2/4$ parabola. Each of them are labeled by the class IDs as listed in Table~\ref{table3}. It shows a locus of (tr,~det) calculated from the Jacobian matrix of the midsize VO$_2$ Mott memristor PA oscillator around its fixed points, by fixing $C_\text{p}=1$~pF, $V_\text{dc}=1.2$~V, and stepping $R_\text{s}$ from 100~$\Omega$ to 17.2~k$\Omega$ at 100~$\Omega$ interval. The (tr,~det) points for $100~\Omega\leq R_\text{s}\leq 600~\Omega$ outside the plotted area are all located above the $\det=\tr^2/4$ parabola within the same stable spiral sink (class 8) region. The trajectory of (tr,~det) formed by varying $R_\text{s}$ is nonlinear and convex shaped. Increasing $R_\text{s}$ moves (tr,~det) toward the region of an unstable spiral source (class 10) in the first quadrant and produces a bifurcation as it crosses the positive $\det$ axis. At $R_\text{s}>7.5$~k$\Omega$, (tr,~det) of the fixed point crosses the $\det=\tr^2/4$ parabola into the unstable source (class 13) region, but its stability remains unchanged. Therefore, $R_\text{s}$ is clearly a bifurcation parameter for the 2D PA oscillator.

Figure \ref{Fig23_VO2PA_tr-det}(b) is a zoomed view of (a). It shows that increasing $R_\text{s}$ from 3.3~k$\Omega$ to 3.4~k$\Omega$ produces a stability-change bifurcation from a stable spiral sink (class 8) to an unstable spiral source (class 10), both belonging to Region (b) of $\tr^2-4\det<0$ that has complex conjugate Jacobian eigenvalues. At a critical value of $R_\text{s}^*=3359.5$~$\Omega$, (tr,~det) is located exactly on the $\tr = 0$ axis as the borderline between the unstable first quadrant and stable second quadrant. The tr-det plane predicts that the fixed point is a stable center (class 9). We have learned that predictions about a spiral sink (class 8) and a spiral source (class 10) are always correct for 2D nonlinear systems, but the prediction about a center (class 9) is unproven, since a 2D PA oscillator is not a conservative system.

In figure \ref{Fig23_VO2PA_tr-det}(c), we plot the (tr,~det) locus for the Jacobian of the same PA oscillator around its fixed points, this time by fixing $R_\text{s}=3.4~\text{k}\Omega$, $C_\text{p}=1$~pF, and stepping $V_\text{dc}$ from 1~V to 3~V at 10~mV interval. It shows a similarly shaped convex trajectory of (tr,~det) as the case for stepping $R_\text{s}$. However, the effect of $V_\text{dc}$ is opposite to that of $R_\text{s}$ --- a larger $V_\text{dc}$ moves (tr,~det) toward the stable spiral sink (class 8) region in the second quadrant. Figure \ref{Fig23_VO2PA_tr-det}(d) is a zoomed view of (c), showing that increasing $V_\text{dc}$ from 1.21~V to 1.22~V produces a stability-change bifurcation from an unstable spiral source (class 10) to a stable spiral sink (class 8). At a critical value of $V_\text{dc}^*=1.21355$~V, (tr,~det) is located exactly on the $\tr = 0$ axis. Therefore, $V_\text{dc}$ is also a bifurcation parameter for the 2D PA oscillator. The fact that the critical values of bifurcation parameters $R_\text{s}$ and $V_\text{dc}$ found by the tr-det plane analysis match well with those found by the small-signal Nyquist plot analysis of the element combination approach (see figure \ref{Fig21_NyquistVO2PApoles}) corroborates the validity of both methods.

\begin{figure}[htb]
	\centering
	\includegraphics[width=0.9\linewidth]{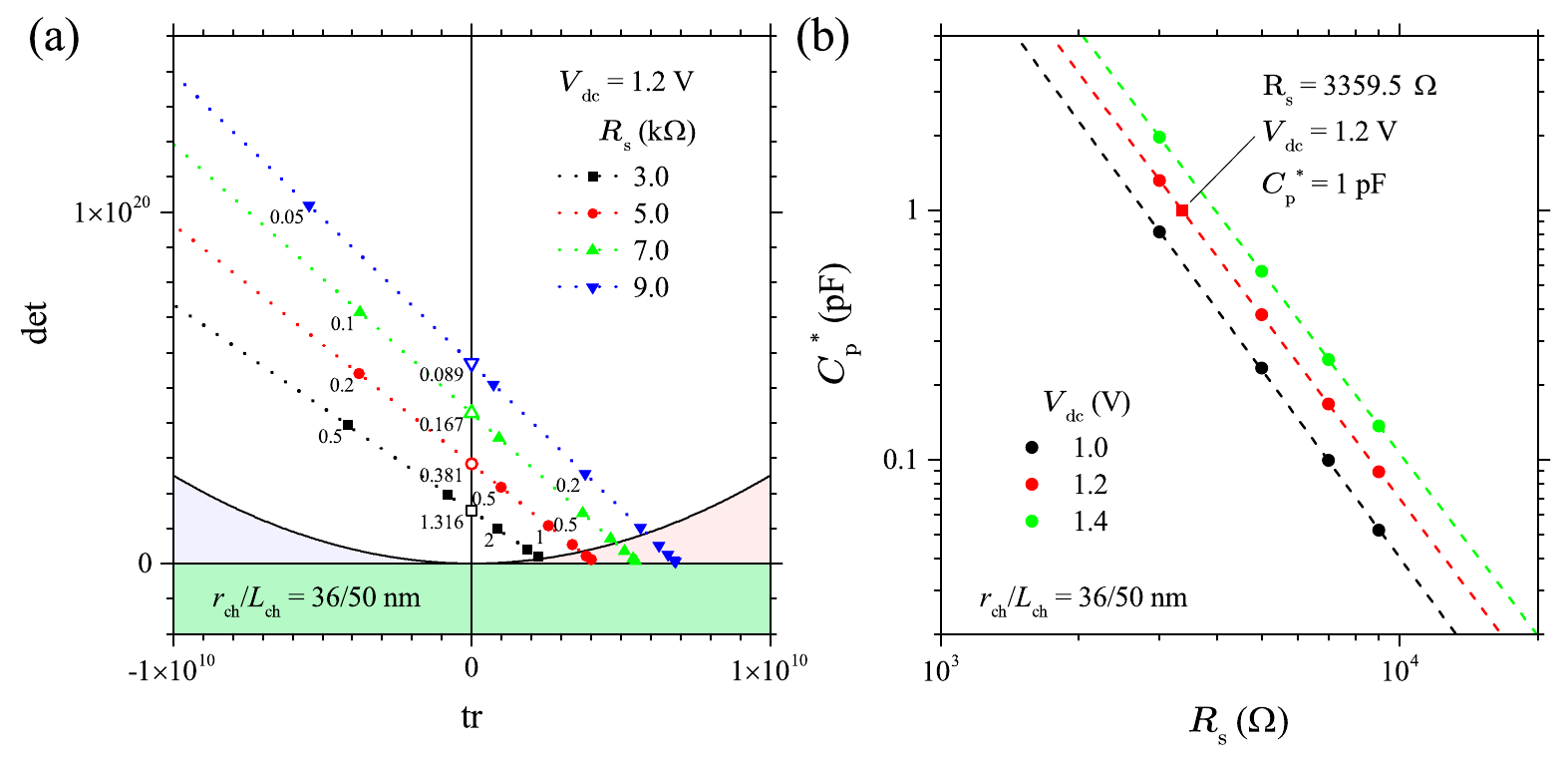}
	\caption{(a) Trace-determinant plane showing four (tr,~det) loci for the Jacobian of the midsize VO$_2$ Mott memristor PA oscillator with $R_\text{s}=3$~k$\Omega$, 5~k$\Omega$, 7~k$\Omega$ and 9~k$\Omega$, respectively, all at $V_\text{dc}=1.2$~V. For each locus, increasing $C_\text{p}$ (from 10~fF to 10~pF in the present case) moves the Jacobian (tr,~det) along a linear trajectory from the stable second quadrant into the unstable first quadrant, which then approaches the positive $\tr$ axis asymptotically. The open symbol that intercepts the positive $\det$ axis shows the critical $C_\text{p}^*$ for the source-sink bifurcation. (b) Log-log plot of the critical $C_\text{p}^*$ vs.~$R_\text{s}$ for three cases of $V_\text{dc}=1.0$~V, 1.2~V and 1.4~V. Dashed lines are power-law fits  $C_\text{p}^*=aR_\text{s}^b$ with an exponent $b\approx-2.5$. Square (red) shows that $C_\text{p}^*=1$~pF if $R_\text{s}=3359.5$~$\Omega$, which is consistent with figure \ref{Fig23_VO2PA_tr-det}(b).}
	\label{Fig24_VO2PA_tr-det_Cp}
\end{figure}

The parallel capacitor $C_\text{p}$ also works as a bifurcation parameter if one fix $R_\text{s}$ and $V_\text{dc}$ and adjust $C_\text{p}$. Figure \ref{Fig24_VO2PA_tr-det_Cp}(a) plots four loci of the Jacobian (tr,~det) for the midsize VO$_2$ Mott memristor PA oscillator, now with $V_\text{dc}=1.2$~V and $R_\text{s}$ fixed at 3~k$\Omega$, 5~k$\Omega$, 7~k$\Omega$ and 9~k$\Omega$, respectively. For each locus, we step up $C_\text{p}$ in the sequence of: 10~fF, 20~fF, 50~fF, 0.1~pF, 0.2~pF, 0.5~pF, 1~pF, 2~pF, 5~pF and 10~pF. Similar to the case of varying $R_\text{s}$, increasing $C_\text{p}$ also moves (tr,~det) from the stable second quadrant into the unstable first quadrant. However, the (tr,~det) locus is linear instead of convex shaped. Equations (\ref{eqn66}) and (\ref{eqn67}) together predict a slope of $-\omega_1\left[1+R_1/\left(R_\text{ch}+R_\text{s}\right)\right]$ for the (tr,~det) locus if $C_\text{p}$ is varied, which matches exactly with the linear regression slopes of the four loci. Equation \ref{eqn67} also predicts that (tr,~det) asymptotically approaches the positive $\tr$ axis as one continuously increases $C_\text{p}$, but never reaches it. Open symbols highlight the critical $C_\text{p}^*$ values for the stability-change bifurcation as the (tr,~det) loci intercept the positive $\det$ axis. It can be seen that a larger fixed $R_\text{s}$ would shift the (tr,~det) locus upward and decrease its critical $C_\text{p}^*$ value. Figure \ref{Fig24_VO2PA_tr-det_Cp}(b) plots the dependence of critical $C_\text{p}^*$ on $R_\text{s}$ in a log-log fashion for three different $V_\text{dc}$ settings at 1.0~V, 1.2~V and 1.4~V. One can tell that $C_\text{p}^*(R_\text{s})$ follows a power law with an exponent close to $\num{-2.5}$. For the trace at $V_\text{dc}=1.2$~V, we added the point of $R_\text{s}=3359.5$~$\Omega$. The power law predicts a $C_\text{p}^*=1$~pF. which is consistent with the critical $(R_\text{s}^*,C_\text{p})$ value for the same bifurcation in the case of  varying $R_\text{s}$ at a fixed $C_\text{p}=1$~pF (see figure \ref{Fig23_VO2PA_tr-det}(b)).

\section[6. GA Coupled]{Global analysis of reactively-coupled Mott memristors: 2D relaxation oscillator}

\subsection[6.1]{Nullclines and direction field}
The local analysis techniques we have discussed so far require a foreknowledge of the fixed points for a 2D nonlinear system. Global analyses, such as nullclines in the phase space of state variables, can be used to analyze a nonlinear system of ODEs and locate its fixed points. For a 2D or planar system, the $x$- (or $y$-) nullcline is defined as the set of points in the phase plane of $(x,y)$ where the time derivative of $x$ (or $y$) vanishes. Therefore the vector field is vertical on the $x$-nullcline and horizontal on the $y$-nullcline. Together, they partition $\mathbb{R}^2$ into different open regions differing on the sign of their time derivatives. One can then determine the direction of vector field in each region. The intersections of the $x$- and $y$-nullclines yield the fixed points. A direction field (also called a slope field) is the scaled version of a vector field, with all the vector lengths normalized to unity. In the 2D phase plane, plotting the $x$- and $y$-nullclines together with the direction field reveals fixed points and offer clues about their dynamical classification as well as orbits of solutions.

For the case of a 2D Mott memristor PA oscillator, the $x$-nullcline $(x_0,v_0)$ is the locus of points where the time derivative of the state variable $x$ for the memristor $M$ vanishes
\begin{equation}\label{eqn70}
	f(x_0,v_0) = f_x\left(x_0,\frac{v_0}{R_\text{ch}(x_0)}\right)=0
\end{equation}
which can be rewritten as
\begin{equation}\label{eqn71}
	v_0 = \left[-A\left(1+Bx_0^2\right)\frac{\ln x_0}{C}\right]^{-0.5}
\end{equation}
Since $(x_0,v_0)$ are steady states of $M$, the $x$-nullcline only depends on the internal characteristics of $M$, and is independent of the external circuit parameters including $R_\text{s}$, $C_\text{p}$ and $V_\text{dc}$. It remains the same as that of an isolated $M$ (see figure \ref{Fig7_StatIV}(c)).

The $v$-nullcline $(x_1,v_1)$ is the locus of points where the time derivative of the state variable $v$ vanishes 
\begin{equation}\label{eqn72}
	g(x_1,v_1) =  \frac{1}{C_\text{p}}\left(\frac{V_\text{dc}-v_1}{R_\text{s}} - \frac{v_1}{R_\text{ch}(x_1)}\right)=0
\end{equation}
which can be rewritten as
\begin{equation}\label{eqn73}
	v_1 = \frac{V_\text{dc}}{1+R_\text{s}A\left(1+Bx_1^2\right)}
\end{equation}
Since $v$ is the voltage across the capacitor $C_\text{p}$ and $M$ in parallel, this means that the charge stored on the capacitor does not change over time, thus there is no current flowing through it. Therefore, $R_\text{s}$ in series with $M$ forms a voltage divider. The $v$-nullcline depends on $V_\text{dc}$ and $R_\text{s}$, but is independent of $C_\text{p}$.

The intersections of the $x$- and $y$-nullclines are the fixed points $Q$ of the 2D system  where  both $x$- and $v$-derivatives vanish
\begin{eqnarray}\label{eqarray14}
	\label{eqn74}
	f(x_Q,v_Q) = f_x\left(x_Q,\frac{v_Q}{R_\text{ch}(x_Q)}\right)=0 \\
	\label{eqn75}
	g(x_Q,v_Q) =  \frac{1}{C_\text{p}}\left(\frac{V_\text{dc}-v_Q}{R_\text{s}} - \frac{v_Q}{R_\text{ch}(x_Q)}\right)=0 
\end{eqnarray}

\begin{figure}[htb]
	\centering
	\includegraphics[width=0.7\linewidth]{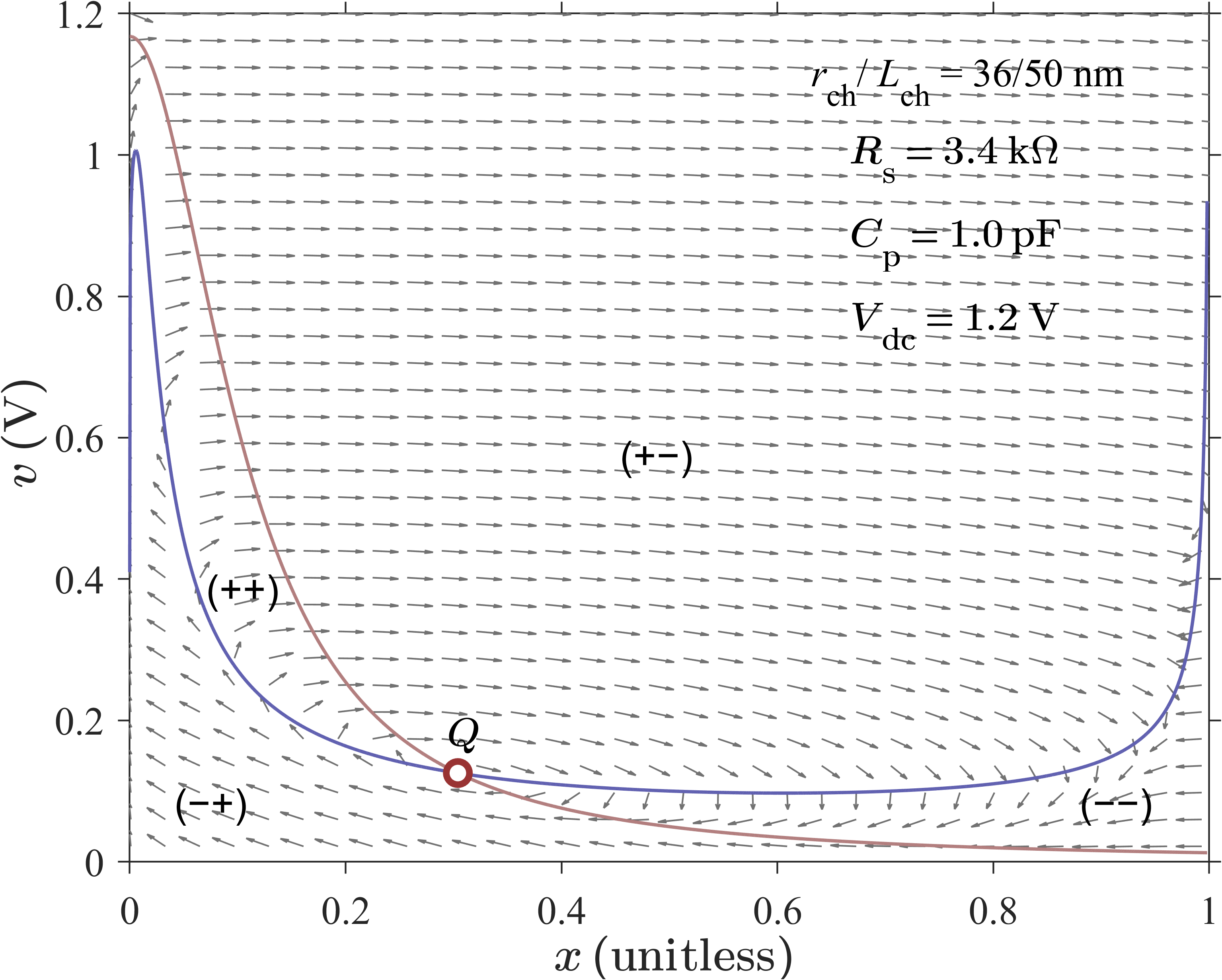}
	\caption{Nullclines and direction field (arrowheads) in the phase plane of the midsize VO$_2$ Mott memristor PA oscillator with $R_\text{s}=3.4$~k$\Omega$, $C_\text{p}=1.0$~pF and $V_\text{dc}=1.2$~V. Under these conditions, the 2D nonlinear system has one fixed point $(x_Q,v_Q)=(0.30396,0.12564)$ at the single intersection of the $x$- (blue violet line) and $v$- (brown line) nullclines. Based on the signs of $dx/dt$ and $dv/dt$, the $x$- and $v$-nullclines partition the $\mathbb{R}^2$ plane into four open regions labeled as $(++)$, $(+-)$, $(-+)$ and $(--)$, respectively.}
	\label{Fig25_VO2PA_NC_1FP}
\end{figure}

Figure \ref*{Fig25_VO2PA_NC_1FP} plots the $x$- and $v$-nullclines and direction field in the phase plane of the midsize VO$_2$ Mott memristor PA oscillator with $R_\text{s}=3.4$~k$\Omega$ and $V_\text{dc}=1.2$~V. Note that the direction field depends on $C_\text{p}$ and is plotted for the case of $C_\text{p}=1.0$~pF. At these settings, the 2D nonlinear system has just one fixed point $(x_Q,v_Q)=(0.30396,0.12564)$ at the single intersection of the $x$-nullcline (blue violet line) and $v$-nullcline (brown line). The $x$- and $v$-nullclines partition the phase plane into four open regions depending on the signs of time derivatives for $x$ and $v$, labeled as $(++)$, $(+-)$, $(-+)$ and $(--)$, respectively. The direction field (arrowheads) shows a conspicuous clockwise rotational pattern around $Q$, suggesting that the orbit of a solution $(x(t),v(t))$ with an initial condition close to $Q$ would rotate around it clockwise. Intuitively, if $Q$ were a stable spiral sink, $(x(t),v(t))$ would approach it along a spiral. If $Q$ were an unstable spiral source, $(x(t),v(t))$ would move away from it along a spiral. However, we will show that it is also possible that $(x(t),v(t))$ forms an isolated periodic orbit that keeps rotating around $Q$. Additional analyses are required besides the nullclines and direction field to know if such a case exists.

\begin{figure}[htb]
	\centering
	\includegraphics[width=0.9\linewidth]{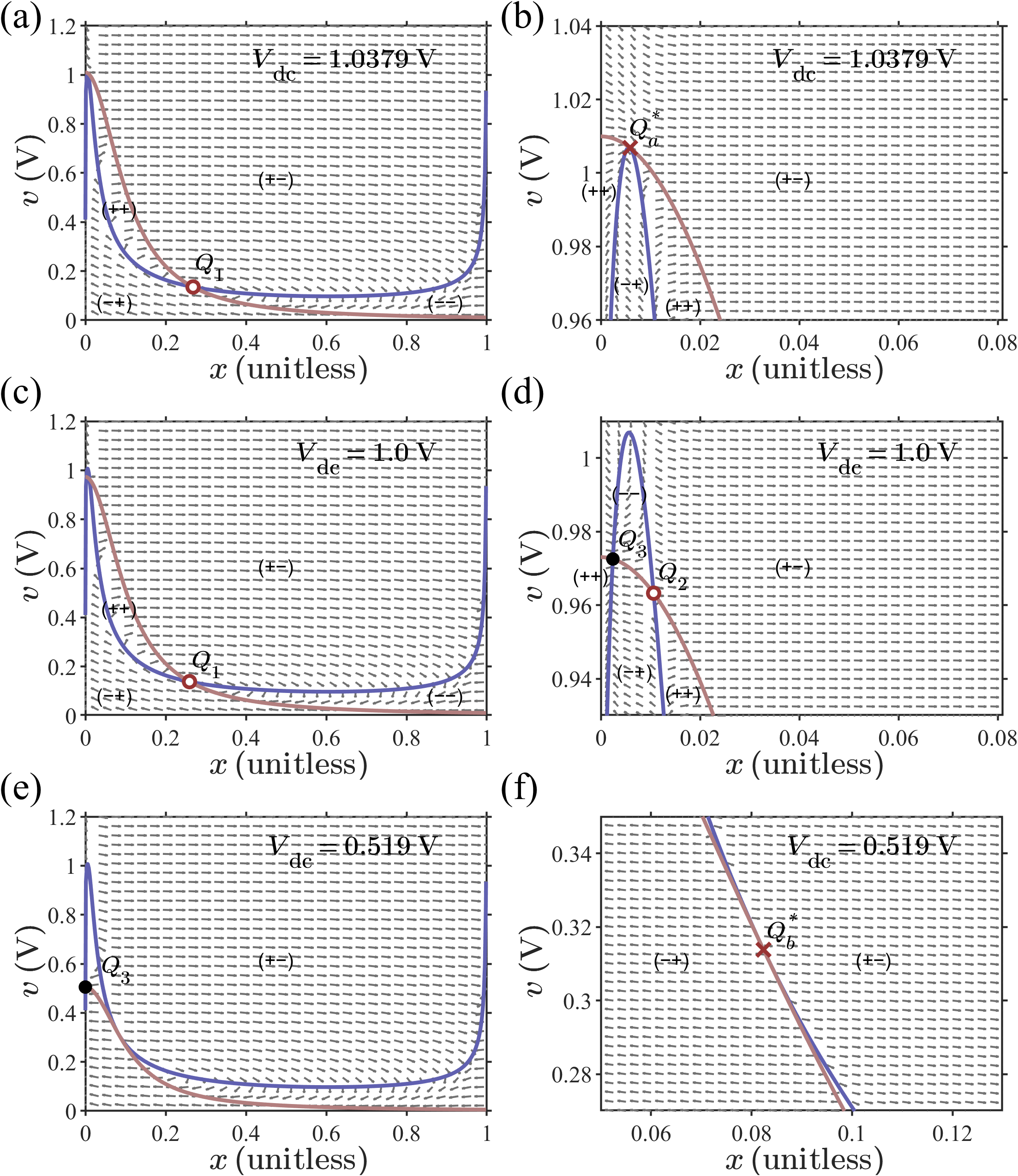}
	\caption{Nullclines and direction fields (arrowheads) in the phase plane of the midsize VO$_2$ Mott memristor PA oscillator with $R_\text{s}=3.4$~k$\Omega$, $C_\text{p}=1.0$~pF and $V_\text{dc}$ at (a) 1.0379~V, (c) 1.0~V, and (e) 0.519~V. (b), (d) and (f) are zoomed views to reveal semi-stable ($\times$), stable ($\bullet$), and unstable ($\circ$) fixed points at the intersections of the $x$- (blue violet line) and $v$- (brown line) nullclines.}
	\label{Fig26_VO2PA_NC_Tiled}
\end{figure}

Since the location of the $v$-nullcline varies with $V_\text{dc}$ and $R_\text{s}$, decreasing $V_\text{dc}$ will shift it downward with respect to the $x$-nullcline, which may change the number of intersections between them. To find out about it, in figure \ref{Fig26_VO2PA_NC_Tiled} we plot three sets of $x$-, $v$-nullclines and direction field for the same model PA oscillator at $V_\text{dc}$ values of 1.0379~V, 1.0~V and 0.519~V, respectively. Figure \ref{Fig26_VO2PA_NC_Tiled}(a) and its zoomed view (b) show the case of $V_\text{dc}=1.0379$~V. At this critical value of $V_\text{dc}$, the $v$-nullcline becomes tangent with the $x$-nullcline near its peak at the PDR-to-NDR crossover $Q_a^*=(0.00589,1.00681)$, increasing the number of fixed points from one to two.

Further reducing $V_\text{dc}$ will split $Q_a^*$ into a pair of unstable ($Q_2$) and stable ($Q_3$) fixed points that move apart from each other as $V_\text{dc}$ further drops. This is characteristic of a 2D saddle-node bifurcation. We are already familiar with the 1D case for an isolated Mott memristor (see figure \ref{Fig6_SaddleNodeBifur}).
To illustrate, Figure \ref{Fig26_VO2PA_NC_Tiled}(c) and its zoomed view (d) show the case of $V_\text{dc}=1.0$~V. Now there are three intersections between the $v$- and $x$-nullclines. Besides the original fixed point $Q_1=(0.25937,0.13823)$ located in the NDR region of $M$, two new fixed points $Q_2=(0.01064,0.96326)$ and $Q_3=(0.00243,0.97254)$ emerge at very small $x$ values. As a result, the $\mathbb{R}^2$ plane is now partitioned into six open regions instead of four. The two additional regions are $(++)$ within the PDR region of $M$, and $(--)$ near the PDR-to-NDR transition of $M$. The direction field gives away that $Q_2$ is an unstable node, while $Q_3$ is a stable node at an intersection in the PDR region (insulating state) of $M$.

$Q_1$ and $Q_2$ approach each other as $V_\text{dc}$ further drops. Figure \ref{Fig26_VO2PA_NC_Tiled}(e) and its zoomed view (f) show that at another critical value of $V_\text{dc}=0.519$~V, the $v$-nullcline becomes tangent with the $x$-nullcline in its NDR region as $Q_1$ and $Q_2$ merge into one fixed point $Q_b^*$. It then disappears if $V_\text{dc}$ continues to drop. At $V_\text{dc}<0.519$~V, only one stable fixed point $Q_3$ survives in the insulating state of $M$.

\subsection[6.2]{2D saddle-node bifurcations by varying $V_\text{dc}$\label{FPsByVdc}}
\begin{figure}[htb]
	\centering
	\includegraphics[width=0.9\linewidth]{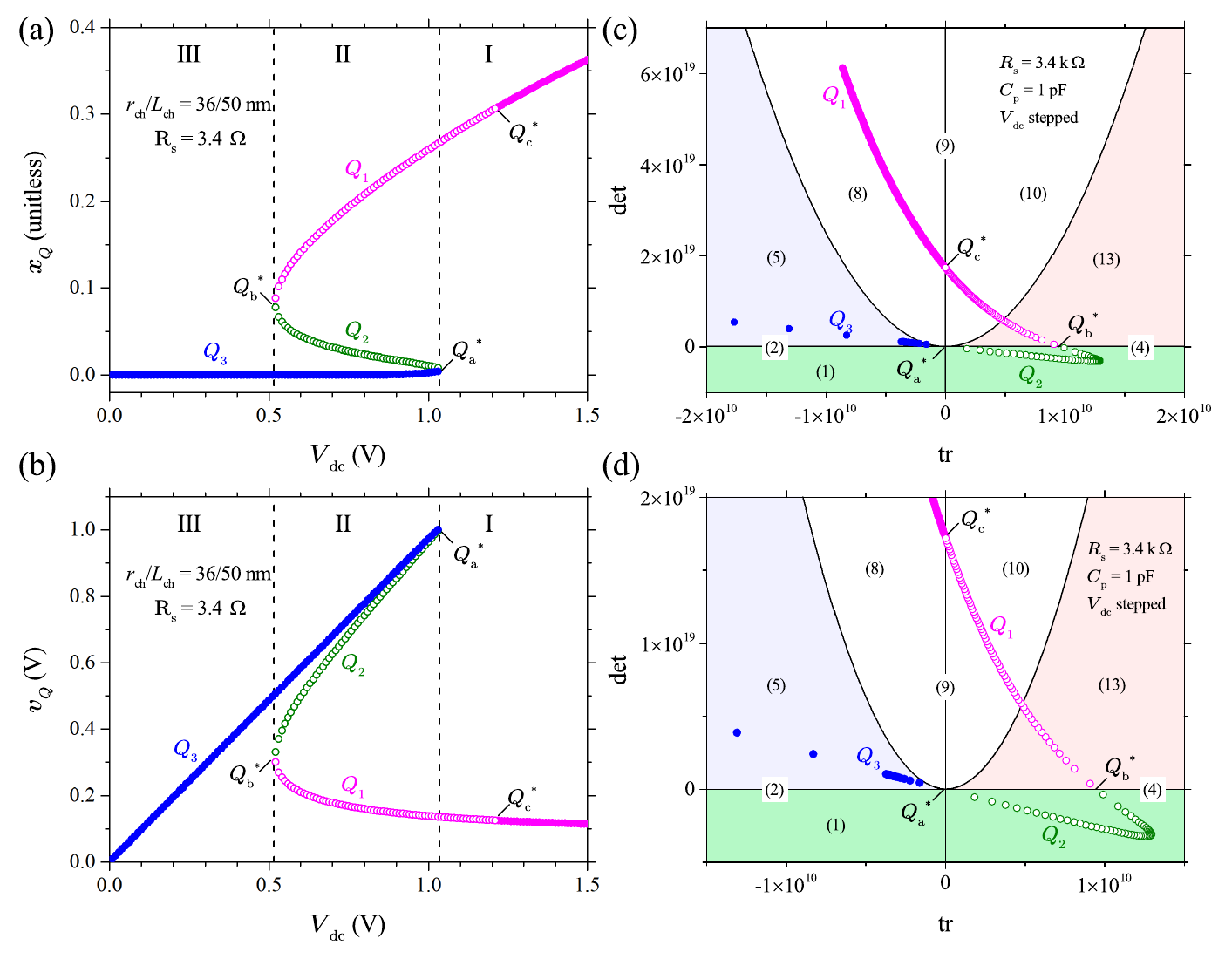}
	\caption{Bifurcation diagrams showing $V_\text{dc}$ dependences of (a) $x_Q$ and (b) $v_Q$ of the fixed point $(x_Q,v_Q)$ for the midsize VO$_2$ Mott memristor PA oscillator, found by the nullclines method. $R_\text{s}=3.4$~k$\Omega$ and $C_\text{p}$ is arbitrary. Stable (unstable) fixed points are represented by solid (open) circles. As $V_\text{dc}$ is stepped from 3~V to 0~V at 0.01~V interval (plotted up to 1.5~V for clarity), initially (region I) the 2D system has only one fixed point $Q_1$, which undergoes a stability-change bifurcation at $V_\text{dc}=1.21355$~V ($Q_c^*$). A saddle-node bifurcation at $V_\text{dc}=1.0379$~V ($Q_a^*$) creates a pair of fixed points $Q_2$ and $Q_3$ in region II. At $V_\text{dc}=0.519$~V ($Q_b^*$), another bifurcation occurs as $Q_1$ and $Q_2$ coalesce and annihilate each other. For even lower $V_\text{dc}$ (region III), only one fixed point $Q_3$ exists. (c) The (tr,~det) loci for the Jacobian of the same fixed points as shown in (a) and (b), calculated with $C_\text{p}=1.0$~pF. (d) is a zoomed view of (c), showing that $Q_a^*$ is located on the negative $\tr$ axis very close to the origin, $Q_b^*$ is on the positive $\tr$ axis, and $Q_c^*$ is on the positive $\det$ axis.}
	\label{Fig27_VO2PA_FPs-Vdc}
\end{figure}

Plotting nullclines and direction fields at different values of $V_\text{dc}$ allowed us to identify two bifurcations, both appear to be 2D saddle-node bifurcation. Next, we apply the bifurcation diagram and tr-det plane methods to clarify their nature. We step the bifurcation parameter $V_\text{dc}$ from 3.0~V to 0~V at an interval of 0.01~V while keep the other parameters unchanged, and solve the fixed points $(x_Q,v_Q)$ by finding all the intersection of $x$- and $v$-nullclines. Figure \ref{Fig27_VO2PA_FPs-Vdc}(a) and (b) are the bifurcation diagrams for $x_Q$ and $v_Q$, respectively. Solid (open) circles are used for stable (unstable) fixed points. There exist three distinctive regions (I, II, III) according to the number of fixed points at a specific $V_\text{dc}$. In region I ($V_\text{dc}>1.0379$~V), there is only one fixed point $Q_1$. The tr-det plane analysis (Figure \ref{Fig23_VO2PA_tr-det}(d)) told us that $Q_1$ has a stability-change bifurcation at $V_\text{dc}=1.21355$~V (labeled as $Q_c^*$) and switches from a stable spiral sink to an unstable spiral source as $V_\text{dc}$ drops below 1.21355~V.

At $V_\text{dc}=1.0379$~V, a new fixed point $Q_a^*$ emerges at $(0.00589,1.00681)$. It then splits into two fixed points $Q_2$ (unstable) and $Q_3$ (stable) which move away from each other as $V_\text{dc}$ further drops. These characteristics resemble a 2D saddle-node bifurcation. In region II $(0.519~\text{V}<V_\text{ch}<1.0379~\text{V})$, the 2D system has three fixed points (stable $Q_3$, unstable $Q_1$ and $Q_2$). As $V_\text{dc}$ continues to decrease, another bifurcation occurs at $V_\text{dc}=0.519$~V, where $Q_1$ and $Q_2$ coalesce into $Q_b^*$ at $(0.08240,0.31379)$ and annihilate each other. This corresponds to the $v$-nullcline becoming tangent with the $x$-nullcline in its NDR region before departing from it. The system only has one stable fixed point $Q_3$ in region III $(0~\text{V}<V_\text{ch}<0.519~\text{V})$, as the $v$-nullcline only intersects with the $x$-nullcline in the insulating region of the Mott memristor.

To further understand the bifurcations at $Q_a^*$ and $Q_b^*$, in figure \ref{Fig27_VO2PA_FPs-Vdc}(c), we plot the (tr,~det) loci for the Jacobian of all the fixed points shown in (a) and (b). Since two of the Jacobian elements are functions of $C_\text{p}$, the calculations are done at $C_\text{p}=1.0$~pF to match with figure \ref{Fig26_VO2PA_NC_Tiled}. The (tr,~det) locus of $Q_1$ was already shown in figure \ref{Fig23_VO2PA_tr-det}(c) and (d) and is re-plotted here with a much wider range of $V_\text{dc}$. At $V_\text{dc}>1.21355$~V, it is in the second quadrant above the $\det=\tr^2/4$ parabola as a stable spiral. At $V_\text{dc}=1.21355$~V, it crosses the positive $\det$ axis at $Q_c^*$ as a center into the first quadrant and switches the stability. For $0.519~\text{V}<V_\text{dc}<1.21355$~V, $Q_1$ remains unstable, first as an unstable spiral, then as an unstable node after crossing the $\det=\tr^2/4$ parabola, before vanishing at $V_\text{dc}=0.519$~V. 

At $V_\text{dc}=1.0379$~V, a saddle-node bifurcation creates a new fixed point $Q_a^*$. Figure \ref{Fig27_VO2PA_FPs-Vdc}(d) as a zoomed view of \ref{Fig27_VO2PA_FPs-Vdc}(c) shows that $Q_a^*$ is located on the negative $\tr$ axis very close to the origin (at $\tr=\num{-2.23772E8}$). It then splits into a pair of fixed points $Q_2$ and $Q_3$. The (tr,~det) locus of $Q_2$ follows a V-shaped trajectory and resides entirely within the fourth quadrant, indicating that $Q_2$ is an unstable saddle point (class 1). The (tr,~det) locus of $Q_3$ is entirely within the second quadrant below the $\det=\tr^2/4$ parabola, indicating that $Q_3$ is a stable sink (class 5). As $V_\text{dc}$ drops from 1.0379~V, $Q_1$ and $Q_2$ approach each other until they coalesce into $Q_b^*$ as $V_\text{dc}$ reaches 0.519~V, indicating that another saddle-node bifurcation occurs. $Q_b^*$ is located on the positive $\tr$ axis. It vanishes at even lower $V_\text{dc}$ values, and the only fixed point left is $Q_3$ in the second quadrant. Interestingly, the saddle-node bifurcation at $Q_b^*$ involves two unstable fixed points $Q_1$ and $Q_2$, rather than a pair of stable and unstable fixed points as in typical cases.

It is worth mentioning that the linearized tr-det plane predictions on the borderline classes (class 2 and 4 on the $\tr$ axis) are incorrect, since $Q_a^*$ and $Q_b^*$ are non-hyperbolic semi-stable fixed points rather than a stable or unstable line of fixed points. It also cannot tell about a possible Hopf bifurcation associated with the non-hyperbolic fixed point $Q_c^*$ (class 9 on the $\det$ axis). Therefore we will revisit this topic in subsection~\ref{HopfByVdc}.

\subsection[6.3]{2D supercritical Hopf-like bifurcation by varying $R_\text{s}$}
The tr-det plane analysis of a linearized VO$_2$ Mott memristor PA oscillator showed that its fixed point can be a non-hyperbolic center as the $R_\text{s}$ or $C_\text{p}$ parameter passes through a critical value. Stability and qualitative behavior of nonlinear systems near a non-hyperbolic fixed point is tricky and requires further theoretical treatment. Here using $R_\text{s}$ as the bifurcation parameter, we illustrate an example of 2D local Hopf-like bifurcation. For nonlinear systems of dimension two or higher, a local Hopf bifurcation, also called Poincar\'e-Andronov-Hopf bifurcation, is the local creation or annihilation of a periodic solution around a fixed point as it switches stability. The bifurcating periodic solution is called a limit cycle, which is an isolated periodic orbit (closed trajectory) with no nearby periodic orbits, such that at least a nearby trajectory spirals into it either as time approaches infinity or as time approaches negative infinity. The orbital stability of a limit cycle is opposite to that of the fixed point it encircles. If a stable limit cycle appears around an unstable fixed point, it is a supercritical Hopf bifurcation. Otherwise if an unstable limit cycle appears around a stable fixed point, then it is a subcritical Hopf bifurcation. The example we will discuss has a supercritical orbital stability.

\subsection[6.4]{Hopf bifurcation theorem}
For a nonlinear system near a non-hyperbolic fixed point in its linear part, the center manifold theorem states that its qualitative behavior can be told by the behavior on the center manifold. The Jacobian matrix of the linearized system defines three main subspaces according to the real part of its eigenvalues. The center subspace is spanned by eigenvectors corresponding to eigenvalues with zero real part. A center manifold is an invariant manifold of the same dimension as and tangent to the center subspace. The stability problem is therefore reduced to lower dimensions. A direct application of the center manifold theorem is the Hopf bifurcation theorem, which allows \textit{analytical} prediction on the existence of limit cycles. A version of the Hopf bifurcation theorem that is generalized to $\mathbb{R}^n$ is briefly introduced here\cite{Marsden76,Guckenheimer83}. Consider a nonlinear system $\dot{\textbf{x}}=\textbf{f}(\textbf{x};\mu)$, $\textbf{x}\in \mathbb{R}^n$, $\mu\in \mathbb{R}$, where $\mu$ is a bifurcation parameter. Assume it has a fixed point $(\textbf{x}_0;\mu)$ so that $\textbf{f}(\textbf{x}_0;\mu)=0$. The eigenvalues of the linearized system $\dot{\textbf{x}}=\textit{D}\textbf{f}(\textbf{x};\mu)$ about this fixed point are $\lambda_\pm(\mu)=\alpha(\mu)\pm\beta(\mu) i$. If both of the following conditions are satisfied at $\mu=\mu_0$:
\begin{enumerate}
	\item $\alpha(\mu_0)=0$, $\beta(\mu_0)\neq0$ (non-hyperbolicity condition), i.e., there is a pair of simple, conjugate pure imaginary eigenvalues and no other pure imaginary eigenvalues,
	\item $\left.\frac{d\alpha(\mu)}{d\mu}\right|_{\mu=\mu_0}=d\neq0$ (transversality condition), i.e., the eigenvalues cross the imaginary axis with finite speed,
\end{enumerate}
then there is a unique center manifold passing through $(\textbf{x}_0;\mu_0)$ in $\mathbb{R}^n\times\mathbb{R}$. 

The third condition (genericity condition) is about the first Poincar\'e-Lyapunov constant $L_1(\mu_0)$, which is the coefficient of cubic terms if the system is transferred to the normal form. If $L_1(\mu_0)\neq0$, then a surface of periodic solutions exists in the center manifold. Approximated to the second order, this surface is a paraboloid tangent to the eigenspace associated with $\lambda_\pm(\mu_0)$. The region for periodic solutions to appear (either as $\mu$ moves into $\mu<\mu_0$ or into $\mu>\mu_0$) as well as the stability of periodic solutions are determined by the signs of $L_1(\mu_0)$ and $d$~\cite{Marsden76}. For the case of $d>0$ that is relevant to our example, if $L_1(\mu_0)<0$, then Hopf bifurcation is supercritical, i.e., a stable limit cycle bifurcates from an unstable fixed point into the region $\mu>\mu_0$. If $L_1(\mu_0)>0$, then Hopf bifurcation is subcritical, i.e., an unstable limit cycle bifurcates from a stable fixed point into the region $\mu<\mu_0$.

\begin{figure}[htb]
	\centering
	\includegraphics[width=0.9\linewidth]{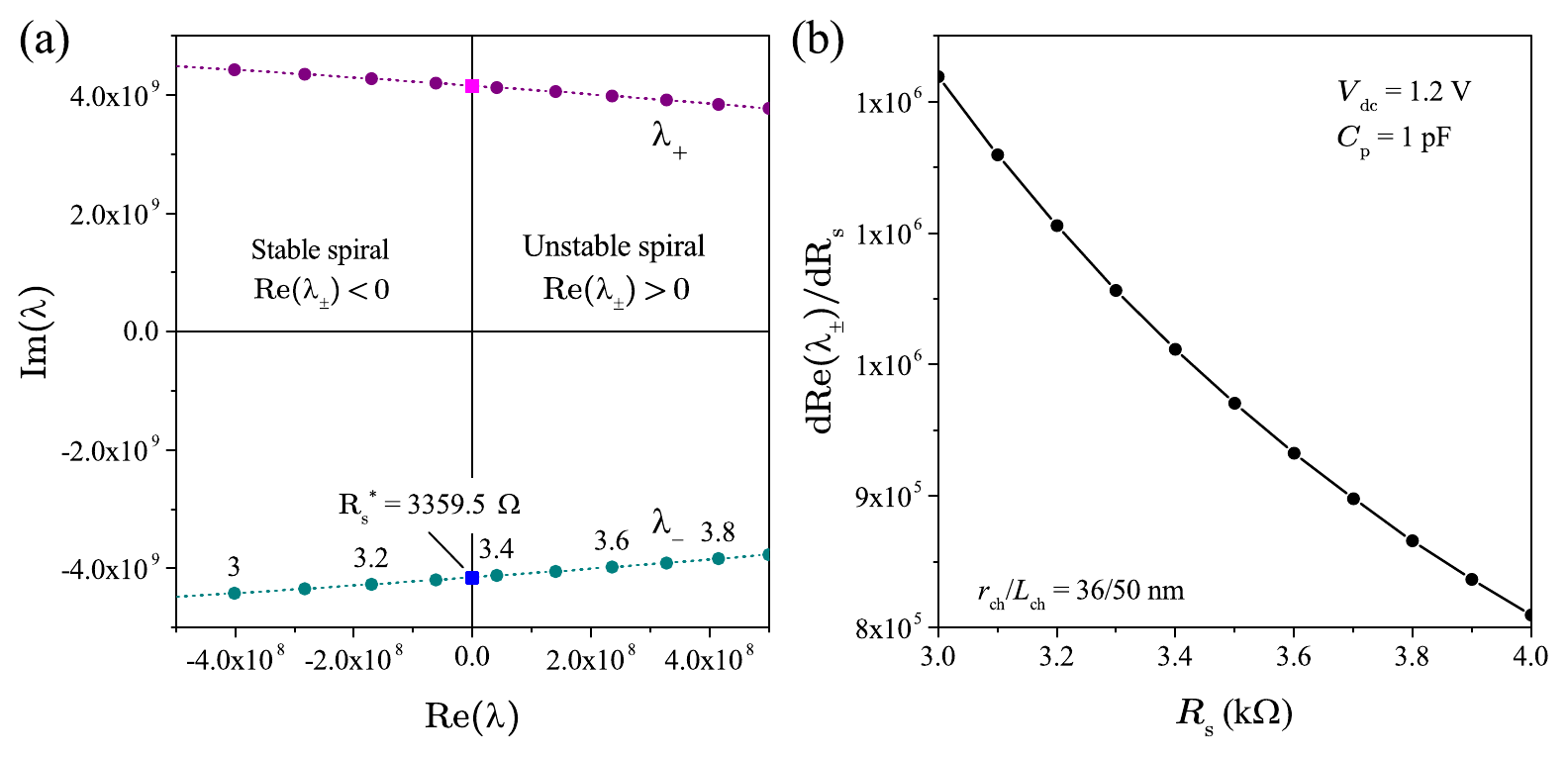}
	\caption{Examination of the non-hyperbolicity and transversality conditions for the Hopf bifurcation theorem. (a) Loci of the conjugate pair of eigenvalues $\lambda_\pm(R_\text{s})$ in the complex plane for the Jacobian of the midsize VO$_2$ Mott memristor PA oscillator with $C_\text{p}=1$~pF, $V_\text{dc}=1.2$~V, and bifurcation parameter $R_\text{s}$ stepped at 100~$\Omega$ interval (labeled as k$\Omega$ for several $\lambda_-$ points). The Jacobian has a stable spiral for Re$(\lambda_\pm)<0$ and an unstable spiral for Re$(\lambda_\pm)>0$. At $R_\text{s}^*=3359.5$~$\Omega$, $\lambda_\pm$ are pure imaginary and the fixed point is a non-hyperbolic center located on the $\tr = 0$ axis. (b) Calculated $d\textrm{Re}(\lambda_\pm)/dR_\text{s}$ vs.~$R_\text{s}$, showing that its value is finite and positive for $R_\text{s}$ that ranges from 3.0~k$\Omega$ to 4.0~k$\Omega$.}
	\label{Fig28_HopfBifTheorem}
\end{figure}

Calculation of $L_1(\mu_0)$ can be a substantial effort, as it involves the second and third order derivatives of the system about the bifurcation point. For the sake of brevity, we leave aside the derivation of the first Poincar\'e-Lyapunov constant $L_1(\mu_0)$ for a Mott memristor PA oscillator, and only examine if it satisfies the non-hyperbolicity and transversality conditions. Figure~\ref{Fig28_HopfBifTheorem}(a) shows the complex-plane loci of the pair of simple, conjugate eigenvalues $\lambda_\pm(R_\text{s})$, calculated by equation~(\ref{eqn68}) for the Jacobian of the midsize VO$_2$ Mott memristor PA oscillator that has been analyzed by the tr-det plane method (see figure~(\ref{Fig23_VO2PA_tr-det})). Fixing $C_\text{p}$, $V_\text{dc}$ and increasing the bifurcation parameter $R_\text{s}$, the fixed point of the linearized system evolves from a stable spiral (Re$(\lambda_\pm)<0$), to a non-hyperboic center (Re$(\lambda_\pm)=0$) at $R_\text{s}^*=3359.5$~$\Omega$, and then to an unstable spiral (Re$(\lambda_\pm)>0$). At $R_\text{s}=R_\text{s}^*$, the system satisfies the non-hyperbolicity condition for a Hopf bifurcation. Figure \ref{Fig28_HopfBifTheorem}(b) plots $d\textrm{Re}(\lambda_\pm)/dR_\text{s}$ vs.~$R_\text{s}$ calculated from $\lambda_\pm(R_\text{s})$, which shows that the derivative of the real part of eigenvalues with respect to the bifurcation parameter $R_\text{s}$ is finite and positive at the non-hyperbolic center ($R_\text{s} = R_\text{s}^*$) and its nearby region. Thus the transversality condition for a Hopf bifurcation is also satisfied.

\subsection[6.5]{Phase portrait analysis of limit cycles}
Now we check the above analytical prediction against numerical calculations. A local Hopf bifurcation can be revealed by numerically solving the coupled ODEs with an arbitrary initial condition $(x_0, v_0)$, then inspecting the orbit of the solution $(x(t),v(t))$ in the phase plane pre-loaded with nullclines and direction field. Such a plot is called a phase portrait. Plotting the time series $x(t)$ and $v(t)$ of the numerical solution helps tell if there are damped oscillations toward a stable fixed point, or persistent self-excited oscillations characteristic of a limit cycle.

\begin{figure}[htb]
	\centering
	\includegraphics[width=0.9\linewidth]{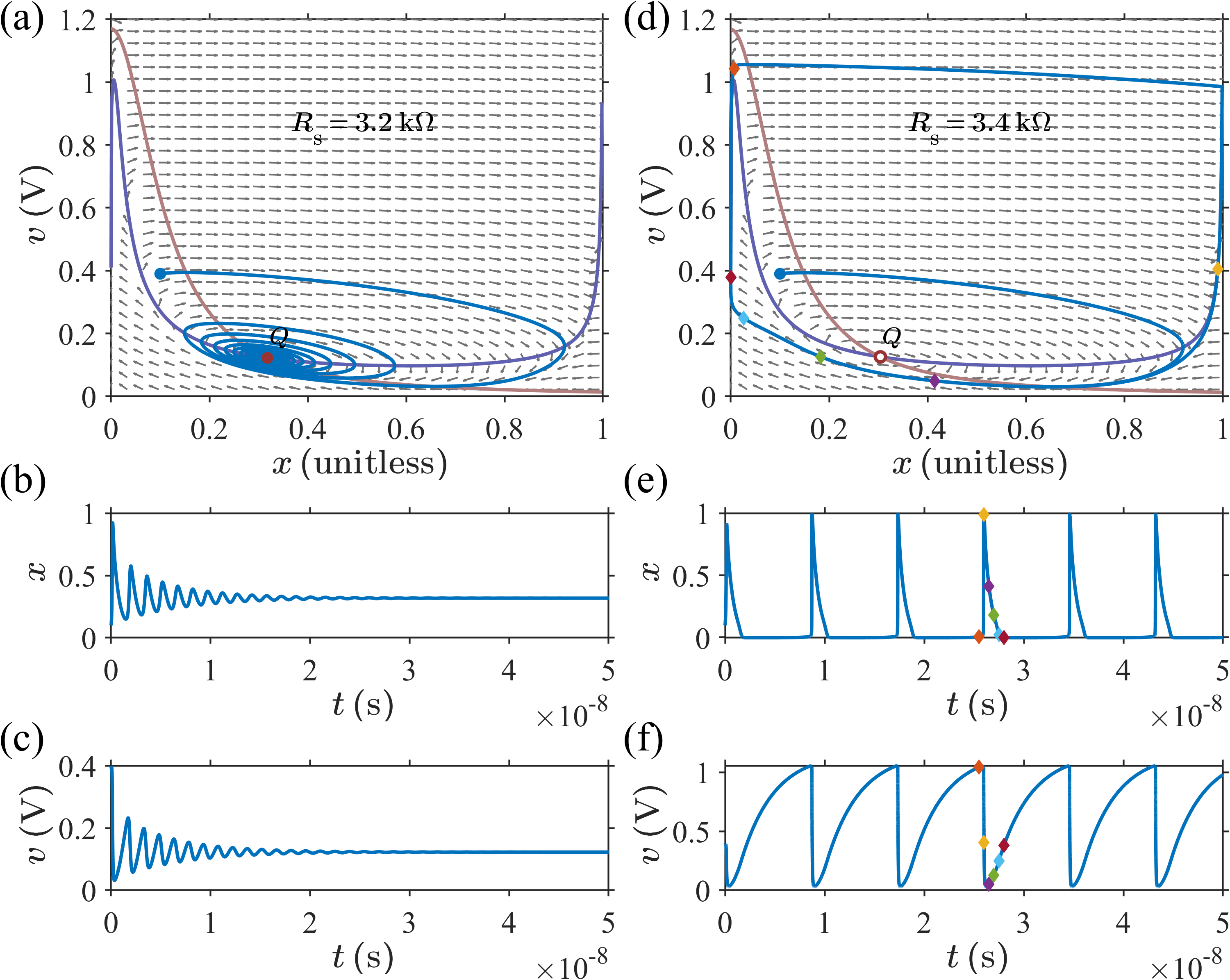}
	\caption{Phase portraits $(x(t),v(t))$ (top row) and the corresponding time series $x(t)$, $v(t)$ (middle and bottom rows) numerically solved by a MATLAB ode45 solver for the midsize VO$_2$ Mott memristor PA oscillator with $C_\text{p}=1.0$~pF, $V_\text{dc}=1.2$~V and the initial condition $(x_0, v_0)=(0.1,0.39)$. Left column ((a)--(c)) is the case for $R_\text{s}=3.2$~k$\Omega$ with a stable fixed point $(x_Q,v_Q) = (0.3178,0.1225)$ at the intersection of the nullclines. The phase portrait and time series corroborate the linear analysis prediction that it is a spiral sink. Right column ((d)--(f)) is the case for $R_\text{s}=3.4$~k$\Omega$ with an unstable fixed point $(x_Q,v_Q)=(0.3040,0.1256)$ (open circle). The phase portrait and time series reveal the birth of a limit cycle encircling $Q$ as it switches stability, characteristic of a Hopf bifurcation. Colored diamonds are solutions equally spaced in time from 25.5~ns to 28~ns at 0.5~ns interval, showing alternative slow-fast motion along the closed limit cycle trajectory.}
	\label{Fig29_VO2PA_PhasePortrait}
\end{figure}

Figure \ref{Fig29_VO2PA_PhasePortrait} compares two sets of phase portraits and the corresponding time series for the midsize VO$_2$ Mott memristor PA oscillator with $C_\text{p}=1.0$~pF, $V_\text{dc}=1.2$~V and initial condition $(x_0, v_0)=(0.1,0.39)$, numerically solved by a MATLAB ode45 solver~\cite{Hong22}. Figure \ref{Fig29_VO2PA_PhasePortrait} (a)--(c) (left column)  show the case for $R_\text{s}=3.2$~k$\Omega$. At this value of $R_\text{s}$, there is a single fixed point $(x_Q,v_Q) = (0.3178,0.1225)$ at the intersection of the nullclines. The linear tr-det plane analysis predicts that it is a stable spiral sink (see figure \ref{Fig23_VO2PA_tr-det}(b) and text). The phase portrait corroborates this prediction, showing that the orbit of the solution (purple trace) converges to $Q$ along a clockwise spiral trajectory. The clockwise rotation of the system state over time is determined by the direction field. The time series $x(t)$ and $v(t)$ shown in the bottom rows exhibit fast damped oscillations that come to rest at $(x_Q,v_Q)$ within about 20~ns.

Figure \ref{Fig29_VO2PA_PhasePortrait} (d)--(f) (right column)  show the case for $R_\text{s}=3.4$~k$\Omega$. Such a small increase in $R_\text{s}$ (by 200~$\Omega$) results in a tiny shift in the location of $Q$ to $(x_Q,v_Q)=(0.3040,0.1256)$. The linear tr-det analysis predicts that $Q$ switches its stability and becomes an unstable spiral source. The orbit initially resembles the case of figure \ref{Fig29_VO2PA_PhasePortrait} (a), but it does not finish even one loop around $Q$ before morphing into a periodic orbit that rotates clockwise about $Q$ with a distorted rectangular shape. The corresponding time series $x(t)$ and $v(t)$ shown in the bottom rows exhibit periodic oscillations --- a pulse train $\tilde{x}(t+T_\text{lc})=\tilde{x}(t)$ and a sawtooth wave $\tilde{v}(t+T_\text{lc})=\tilde{v}(t)$, both launched after a very short transient period. Here $T_\text{lc}$ is the period of the limit cycle. Appearance of a stable limit cycle around a fixed point as it switches from a stable sink to unstable source is the hallmark of a supercritical Hopf bifurcation. We added several colored diamonds to represent solutions $x(t)$ and $v(t)$ equally spaced in time from 25.5~ns to 28~ns at 0.5~ns interval. Their locations on the closed trajectory of limit cycle are clearly unevenly spaced, revealing the alternative slow-fast motion along it as a hallmark for relaxation oscillations.

\begin{figure}[htb]
	\centering
	\includegraphics[width=0.9\linewidth]{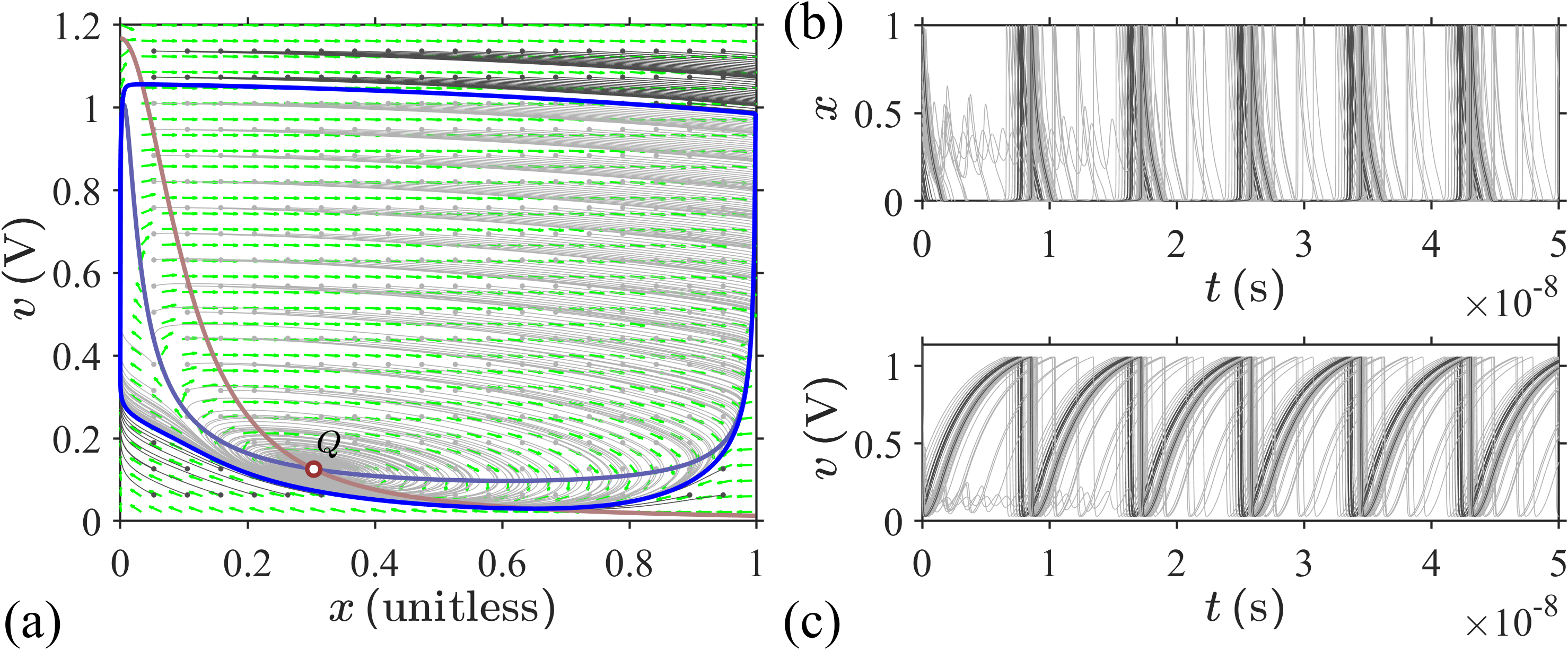}
	\caption{(a) Phase portrait and (b) the corresponding time series $x(t)$, $v(t)$ numerically solved by a MATLAB ode45 solver for the midsize VO$_2$ Mott memristor PA oscillator with $C_\text{p}=1.0$~pF, $V_\text{dc}=1.2$~V and $R_\text{s}=3.4$~k$\Omega$. The system has an unstable fixed point $(x_Q,v_Q)=(0.3040,0.1256)$ (open circle) at the intersection of the nullclines. A total of 324 orbits are solved with their initial conditions (solid dots) located on a regular grid of $(x_0, v_0)$, with 18 $x_0$ levels evenly spaced from 0.05 to 0.95, and 18 $v_0$ levels from 0.06~V to 1.14~V. Orbits that start from within and outside the limit cycle are light and dark gray colored, respectively. Every $(x(t),v(t))$ trajectory converges onto the same limit cycle (blue orbit) encircling $Q$, regardless of the initial condition location. The $x(t)$ and $v(t)$ time series show different oscillation phases depending on the initial condition, but all sharing the same period. The mean (standard deviation) of the oscillation period is 8.621~ns (5~ps).}
	\label{Fig30_VO2PA_LimitCycle}
\end{figure}

To convince ourselves that the periodic orbit revealed by figure \ref{Fig29_VO2PA_PhasePortrait} (d)--(f) is both isolated and stable, i.e., a stable limit cycle, we numerically calculated 324 solutions of the same system with the initial condition $(x_0,v_0)$ distributed on a regularly spaced (18\texttimes18) grid that spans across almost the entire allowable $(x,v)$ phase space. $x_0$ is evenly spaced from 0.05 to 0.95, and $v_0$ from 0.06~V to 1.14~V. Orbits that start from within and outside the limit cycle are light and dark gray colored, respectively. The results are shown in figure \ref{Fig30_VO2PA_LimitCycle}. The phase portrait in figure \ref{Fig30_VO2PA_LimitCycle}(a) offers a sampled view for the flow of this 2D nonlinear system. It shows that, regardless of its initial condition location, the $(x(t),v(t))$ orbit always settles on the same limit cycle $(\tilde{x}(t),\tilde{v}(t))$ (the blue orbit) after a transient movement. If $(x_0,v_0)$ is very close to $Q$, the transient part of the orbit can form many turns of clockwise spiral following the direction field, but the orbit always manage to ``escape'' from $Q$ and becomes a limit cycle encircling it. The time series in figure \ref{Fig30_VO2PA_LimitCycle}(b) reveal that the time elapsed in the transient stage until $x(t)$ (or $v(t)$) become periodic varies across individual solutions, depending on the initial condition $(x_0,v_0)$. The distinctive transient time makes the oscillation waveforms asynchronous across individual orbits, which is equivalent of having different oscillation phases. However, all the time series settle as oscillations sharing the same period. To be precise, the mean (standard deviation) of the oscillation period is 8.621~ns (5~ps), with a coefficient of variation as small as~0.06~\%. The minimum and maximum oscillation period are 8.6~ns and 8.626~ns, respectively. The robustness of a limit cycle against the initial transient may explain why life is full of relaxation oscillators, including the heartbeat~\cite{VanDerPol28}. 

\begin{figure}[htb]
	\centering
	\includegraphics[width=0.9\linewidth]{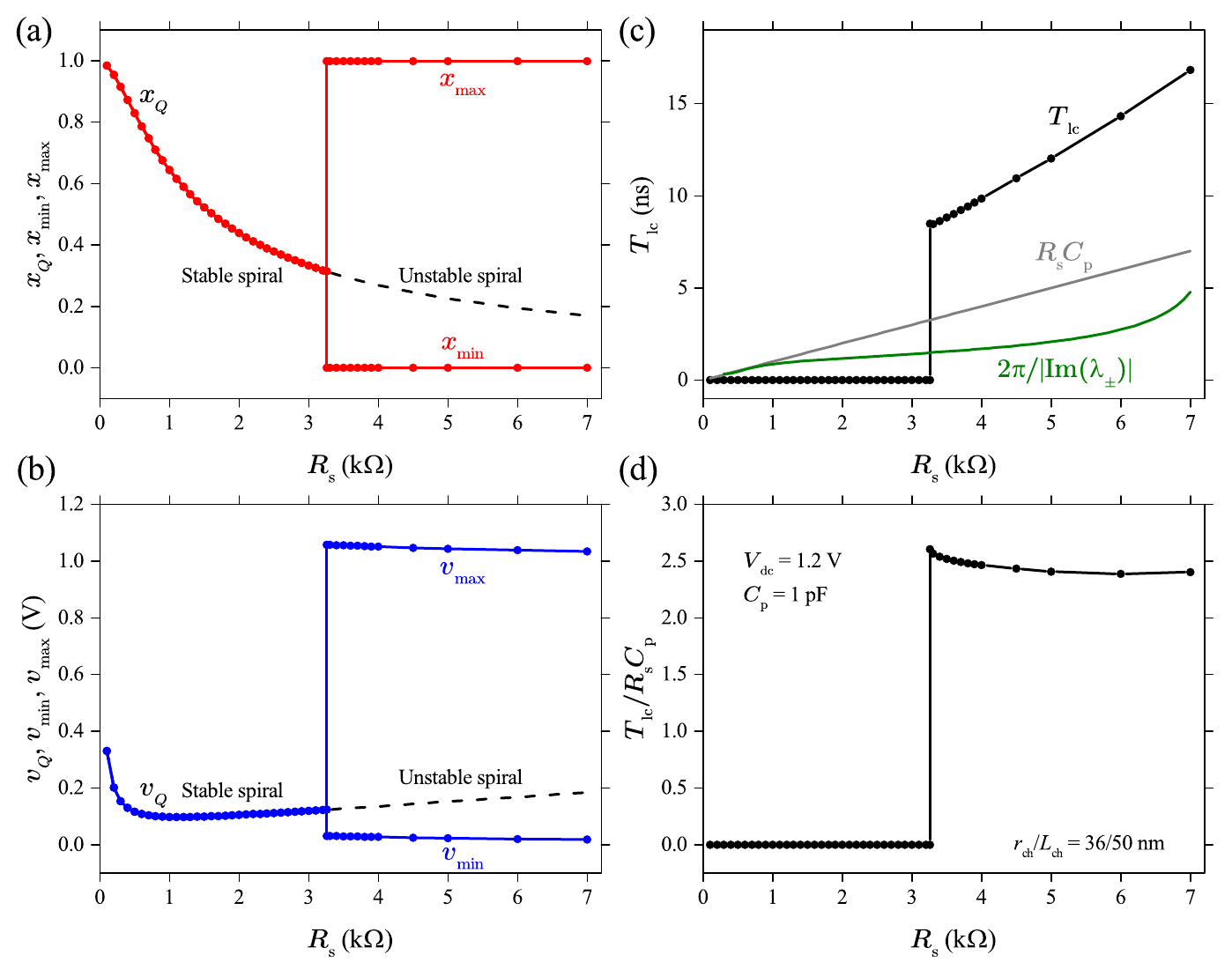}
	\caption{(a) and (b) Numerically-solved bifurcation diagrams of the 2D Hopf-like bifurcation with $R_\text{s}$ as the bifurcation parameter for the midsize VO$_2$ Mott memristor PA oscillator with $C_\text{p}=1.0$~pF and $V_\text{dc}=1.2$~V. The maximum and minimum of $\tilde{x}(t)$ and $\tilde{v}(t)$ limit cycle oscillations are plotted as their bifurcation branches. Dashed lines show the coordinate $(x_Q,v_Q)$ for an unstable spiral at $R_\text{s}\geq R_\text{s}^*$. (c) $R_\text{s}$ dependence of the limit cycle oscillation period $T_\text{lc}$ (black) plotted together with $R_\text{s}C_\text{p}$ (gray) and  $2\pi/\lvert\textrm{Im}(\lambda_\pm)\rvert$ (green). (d) $R_\text{s}$ dependence of the ratio between $T_\text{lc}$ and $R_\text{s}C_\text{p}$.}
	\label{Fig31_HopfBifDiagram_byRs}
\end{figure}

Figure \ref{Fig31_HopfBifDiagram_byRs}(a) and (b) plot the numerically-solved bifurcation diagrams of the 2D Hopf-like bifurcation with $R_\text{s}$ as the bifurcation parameter. We noticed that the critical value $R_\text{s}^*=3258.00799$~$\Omega$ found by numerical calculations is about 3~\% different than the analytical value of 3359.5~$\Omega$ (see figure \ref{Fig28_HopfBifTheorem}), possibly due to rounding or truncation errors. Both $x_Q(R_\text{s})$ and $v_Q(R_\text{s})$ are smooth functions of $R_\text{s}$. For $R_\text{s}<R_\text{s}^*$ (with a difference as small as 10~$\upmu\Omega$), there is a single fixed point $(x_Q,v_Q)$ which is a stable spiral according to the linearization analysis. At $R_\text{s}\geq R_\text{s}^*$, instead of just switching its stability to an unstable spiral (dashed lines), the fixed point bifurcates to a limit cycle. Since a limit cycle is a collection of periodic points $(\tilde{x}(t),\tilde{v}(t))$, we use the maximum and minimum of $\tilde{x}(t)$ and $\tilde{v}(t)$ oscillations to represent their bifurcation branches, and the ranges between maximum and minimum as a measure of the bifurcation amplitude. This definition is not unique. One can borrow the concepts from celestial mechanics and define the unstable spiral as a focus, then the periapsis (minimum) and apoapsis (maximum) distances between a point in the limit cycle orbit and the focus can also be used to represent the bifurcation amplitude. 

A prominent feature about the Mott memristor PA oscillator model is the abrupt appearance or ``hard transition'' of a stable limit cycle that is completely unfolded over an extremely thin bifurcation parameter interval. The amplitude of a classical Hopf bifurcation for smooth systems grows like $\sqrt{\lvert\mu-\mu_0\rvert}$, i.e., the oscillation amplitude is infinitesimal as ${\mu\to\mu_0}$. However, in the present case the oscillations in $\tilde{x}(t)$ and $\tilde{v}(t)$ almost immediately switch to full swing as long as $R_\text{s}$ surpasses $R_\text{s}^*$, then their amplitudes remain essentially unchanged as $R_\text{s}$ further increases. Abrupt appearance of a stable limit cycle was observed in piecewise-linear systems that have a cut-off or saturation region, e.g., a Wien bridge oscillator~\cite{Kriegsmann87,Freire99}. For the present Mott memristor model, the fact that the kinetic function diverges toward negative infinity as $x$ approaches 1.0 (see figure~\ref{Fig3_POP} inset) tells that there is an \textit{implicit saturation} in the model.  A sudden formation of relaxation oscillations, termed ``Canard explosion'', has been observed in chemical and biological systems and analyzed thoroughly in the context of Li\'enard systems, e.g., a van der Pol oscillator~\cite{Krupa01,Rotstein12}. The hard transition in relaxation oscillations forms the basis for understanding the all-or-nothing spike firings in biological neurons that can be considered as reaction-diffusion systems of coupled relaxation oscillators, which has been experimentally demonstrated in Mott memristor based neuromorphic neurons (for examples, see figure~3 in~\cite{Pickett13a} and figure~5 in~\cite{Yi18}).

Figure \ref{Fig31_HopfBifDiagram_byRs}(c) shows the dependence of the limit cycle oscillation period $T_\text{lc}$ on $R_\text{s}$. For $R_\text{s}<R_\text{s}^*$, $T_\text{lc}$ is zero since there is no oscillation. At $R_\text{s}\geq R_\text{s}^*$, $T_\text{lc}$ emerges like a step function and then grows almost linearly with $R_\text{s}$. For comparison, the $R_\text{s}C_\text{p}$ time constant as a function of $R_\text{s}$ is also plotted (gray line). Figure \ref{Fig31_HopfBifDiagram_byRs}(d) shows the ratio between $T_\text{lc}$ and $R_\text{s}C_\text{p}$, which remains almost flat in the bifurcation region with an initial overshoot to 2.6 followed by a gradual descent toward 2.4. Generally, the oscillation period of a Hopf bifurcation approaches $2\pi/\lvert\textrm{Im}(\lambda_\pm)\rvert$ as ${\mu\to\mu_0}$. However, in the present case the calculated $2\pi/\lvert\textrm{Im}(\lambda_\pm)\rvert$ curve (green) is about 1.5~ns at $R_\text{s}\approx R_\text{s}^*$, which is much smaller than $T_\text{lc}\approx8.5$~ns. 

\subsection[6.6]{2D supercritical Hopf-like bifurcation by varying $C_\text{p}$}
From the tr-det plane analysis of a linearized VO$_2$ Mott memristor PA oscillator (see figure \ref{Fig24_VO2PA_tr-det_Cp} and text), we know that $C_\text{p}$ is possibly also a bifurcation parameter as the system's fixed point becomes a non-hyperbolic center as $C_\text{p}$ passes through a critical value. Now we apply the numerical phase portrait method to examine if $C_\text{p}$ is indeed a bifurcation parameter that triggers a 2D local Hopf-like bifurcation.

\begin{figure}[htb]
	\centering
	\includegraphics[width=0.9\linewidth]{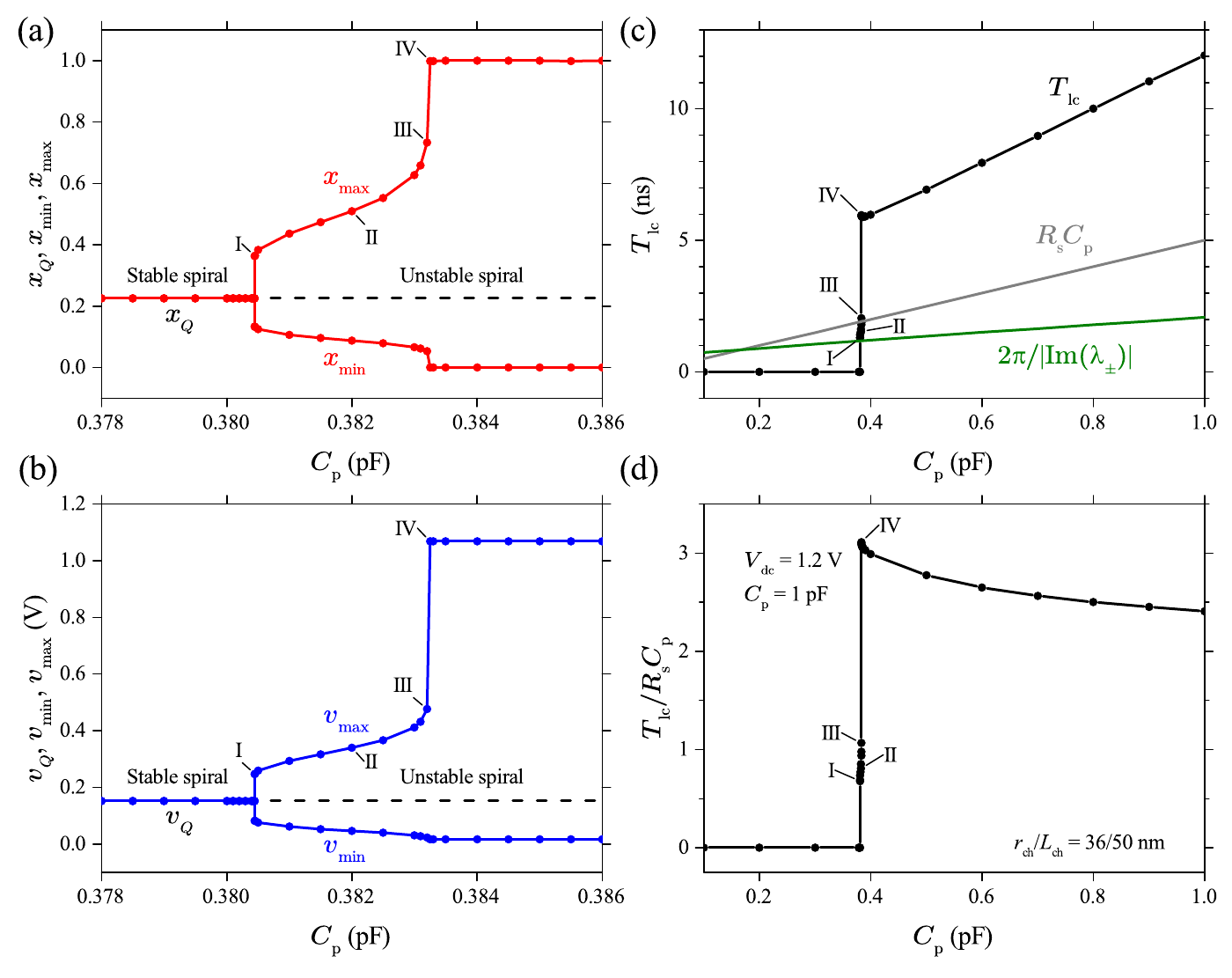}
	\caption{(a) and (b) Numerically-solved bifurcation diagrams of the 2D Hopf-like bifurcation with $C_\text{p}$ as the bifurcation parameter for the midsize VO$_2$ Mott memristor PA oscillator with $R_\text{s}=5.0$~k$\Omega$ and $V_\text{dc}=1.2$~V. The maximum and minimum of $\tilde{x}(t)$ and $\tilde{v}(t)$ limit cycle oscillations are plotted as their bifurcation branches. Dashed lines show the coordinate $(x_Q,v_Q)$ for an unstable spiral for $C_\text{p}$ larger than $C_\text{p}^*=0.380448$~pF. From I to IV, $C_\text{p}$ is 0.380449~pF, 0.382~pF, 0.3832~pF, and 0.38325~pF respectively. (c) $C_\text{p}$ dependence of the limit cycle oscillation period $T_\text{lc}$ (black) plotted together with $R_\text{s}C_\text{p}$ (gray) and  $2\pi/\lvert\textrm{Im}(\lambda_\pm)\rvert$ (green). (d) $C_\text{p}$ dependence of the ratio between $T_\text{lc}$ and $R_\text{s}C_\text{p}$.}
	\label{Fig32_HopfBifDiagram_byCp}
\end{figure}

\begin{figure}[htb]
	\centering
	\includegraphics[width=0.9\linewidth]{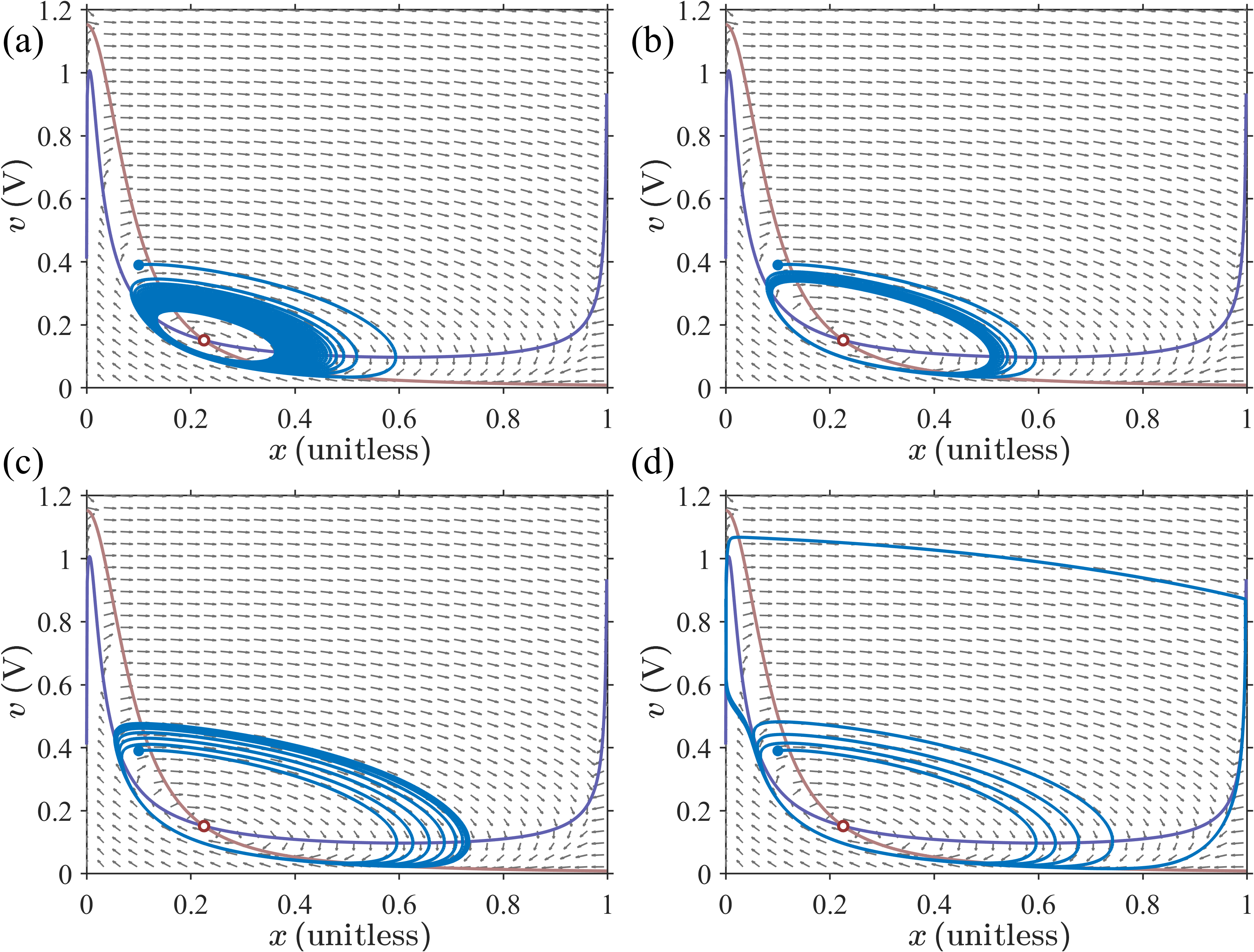}
	\caption{Growth of the 2D Hopf-like bifurcation limit cycle revealed by phase portraits $(x(t),v(t))$ numerically solved for the midsize VO$_2$ Mott memristor PA oscillator with  $R_\text{s}=5.0$~k$\Omega$ and $V_\text{dc}=1.2$~V. From (a) to (d), the bifurcation parameter $C_\text{p}$ is 0.380449~pF, 0.382~pF, 0.3832~pF, and 0.38325~pF respectively, corresponding to points I, II, III, and IV in figure \ref{Fig32_HopfBifDiagram_byCp}. All the solutions start from the same initial condition $(x_0, v_0)=(0.1,0.39)$.}
	\label{Fig33_PP_HopfBifbyCp}
\end{figure}

We numerically solved phase portraits and the corresponding time series for the midsize VO$_2$ Mott memristor PA oscillator with $R_\text{s}=5.0$~k$\Omega$, $V_\text{dc}=1.2$~V, and initial condition $(x_0, v_0)=(0.1,0.39)$. $C_\text{p}$ is varied from 0.1~pF to 1~pF. Figure \ref{Fig32_HopfBifDiagram_byCp}(a) and \ref{Fig32_HopfBifDiagram_byCp}(b) plot the numerically-solved bifurcation diagrams of the 2D Hopf-like bifurcation with $C_\text{p}$ as the bifurcation parameter, which reveal a critical value $C_\text{p}^*=0.380448$~pF to trigger the bifurcation. This value is about 0.3~\% different than the analytical value of 0.381469~pF (see figure \ref{Fig24_VO2PA_tr-det_Cp}), possibly due to rounding or truncation errors. For $C_\text{p}<C_\text{p}^*$ (with a difference as small as 1~attoFarad), there is a single fixed point $(x_Q,v_Q)$ which is a stable spiral according to the linearization analysis. Both $x_Q(R_\text{s})$ and $v_Q(R_\text{s})$ are independent of $C_\text{p}$ as told by the nullcline analysis. At $C_\text{p}\geq C_\text{p}^*$, instead of just switching its stability to an unstable spiral (dashed lines), the fixed point bifurcates to a limit cycle. Compared with the case of $R_\text{s}$-induced Hopf-like bifurcation with a abrupt unfolding, there is a striking difference for $C_\text{p}$-induced Hopf-like bifurcation. Within a narrow range of $C_\text{p}$ (between $C_\text{p}^*$ and ${\sim}0.3832$~pF), the bifurcation amplitude grows more gradually and resembles the general prediction of $\sqrt{\lvert\mu-\mu_0\rvert}$, albeit it still has a abrupt switch on, thus the oscillation amplitude is not infinitesimal as ${\mu\to\mu_0}$. To illustrate the gradual growth of the 2D Hopf-like bifurcation limit cycle, in figure \ref{Fig33_PP_HopfBifbyCp}, we plot the numerically-solved phase portraits $(x(t),v(t))$ at points I, II, III, and IV, corresponding to $C_\text{p}$ at 0.380449~pF, 0.382~pF, 0.3832~pF, and 0.38325~pF respectively. One can tell that the gradual growth of Hopf-like bifurcation gives way to abrupt unfolding upon further increase in $C_\text{p}$ beyond 0.3832~pF.

Figure \ref{Fig32_HopfBifDiagram_byCp}(c) shows the dependence of the limit cycle oscillation period $T_\text{lc}$ on $C_\text{p}$. At $C_\text{p}\approx C_\text{p}^*$, 
the calculated $2\pi/\lvert\textrm{Im}(\lambda_\pm)\rvert$ curve (green) is about 1.18~ns, which is very close to $T_\text{lc}\approx1.28$~ns (point I). This confirms that the oscillation period of $C_\text{p}$-induced Hopf-like bifurcation approaches the general prediction of $2\pi/\lvert\textrm{Im}(\lambda_\pm)\rvert$ as ${\mu\to\mu_0}$. At the upper limit of the gradual growth stage (point III), the oscillation period $T_\text{lc}\approx2.0$~ns is close to the $R_\text{s}C_\text{p}$ time constant. Then it abruptly increases to 6~ns at point IV as the limit cycle expands to full swing. Figure \ref{Fig32_HopfBifDiagram_byCp}(d) shows the ratio between $T_\text{lc}$ and $R_\text{s}C_\text{p}$. In the initial gradual growth stage, this ratio hovers around unity (increases from 0.68 at point I to 1.07 at point III). In the full-swing bifurcation stage, the trend of this ratio versus the bifurcation parameter is similar to the case of $R_\text{s}$, with a larger initial overshoot to 3.1 followed by a gradual descent toward 2.4.

\subsection[6.7]{2D supercritical Hopf-like bifurcation by varying $V_\text{dc}$\label{HopfByVdc}}
Now we revisit the case of varying $V_\text{dc}$ as the bifurcation parameter using the numerical phase portrait method. In subsection \ref{FPsByVdc}, we identified two saddle-node bifurcations using the analytical nullclines and linearized tr-det plane analyses. However, these techniques cannot tell if there exists a Hopf bifurcation or limit cycle around a non-hyperbolic fixed point, such as $Q_c^*$ in figure~\ref{Fig27_VO2PA_FPs-Vdc}. To clarify, numerical phase portrait calculations are needed.

\begin{figure}[htb]
	\centering
	\includegraphics[width=0.9\linewidth]{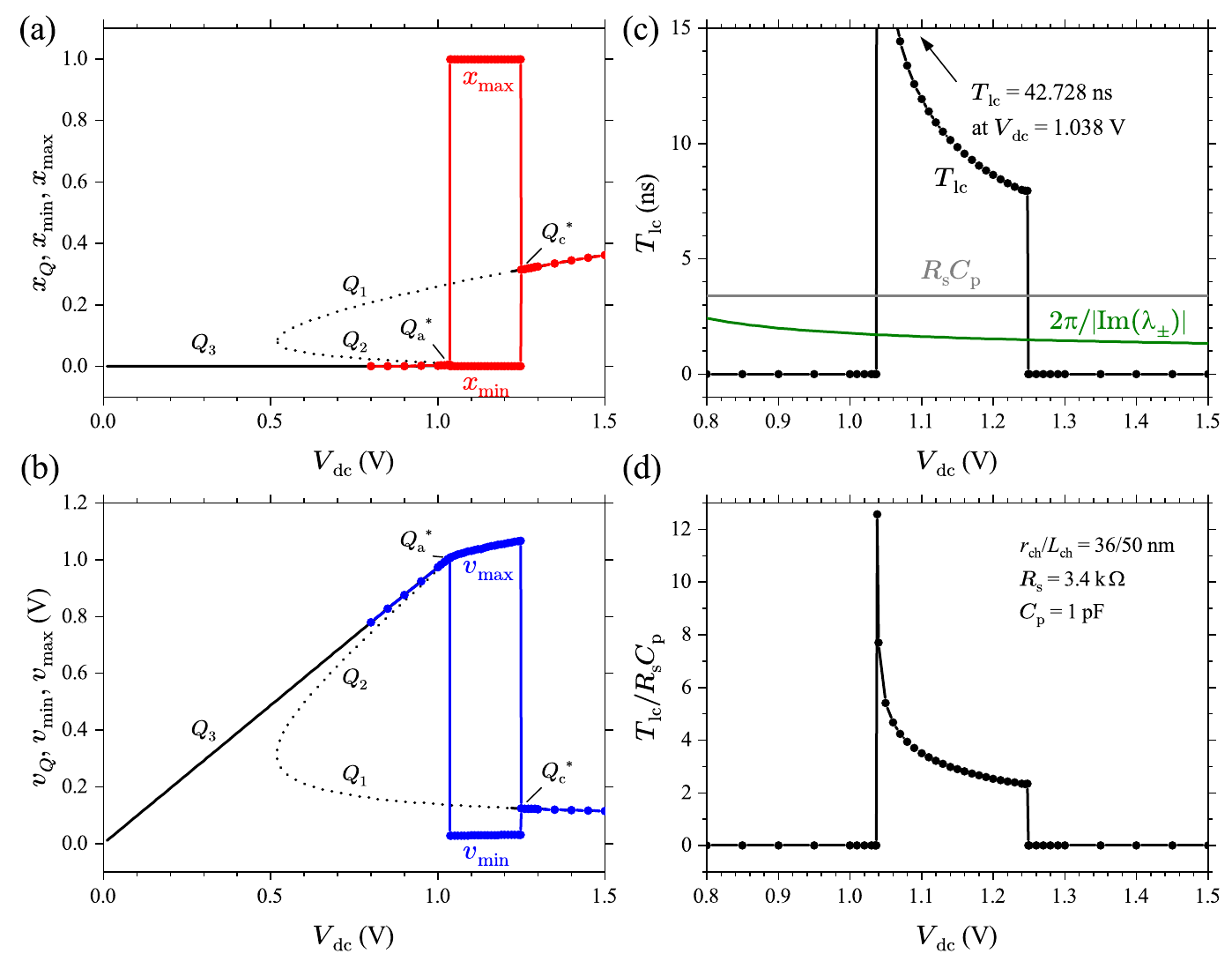}
	\caption{(a) and (b) Numerically-solved bifurcation diagrams of the 2D Hopf-like bifurcations with $V_\text{dc}$ as the bifurcation parameter for the midsize VO$_2$ Mott memristor PA oscillator with $R_\text{s}=3.4$~k$\Omega$ and $C_\text{p}=1.0$~pF. The maximum and minimum of $\tilde{x}(t)$ and $\tilde{v}(t)$ limit cycle oscillations are plotted as their bifurcation branches. Numerical solutions show that a stable limit cycle exists between the non-hyperbolic fixed points $Q_a^*$ and $Q_c^*$ identified in the analytical nullclines and tr-det plane analyses (see figure~\ref{Fig27_VO2PA_FPs-Vdc}). At $V_\text{dc}\leq1.037$~V, the system settles on the stable fixed point $Q_3$ instead of the unstable $Q_1$ or $Q_2$ found by nullclines (short dashed lines). (c) $V_\text{dc}$ dependence of the limit cycle oscillation period $T_\text{lc}$ (black) plotted together with $R_\text{s}C_\text{p}$ (gray) and  $2\pi/\lvert\textrm{Im}(\lambda_\pm)\rvert$ (green). Maximum $T_\text{lc}$ is 42.728~ns at $V_\text{dc}=1.038$~V (above the plot area). (d) $V_\text{dc}$ dependence of the ratio between $T_\text{lc}$ and $R_\text{s}C_\text{p}$.}
	\label{Fig34_HopfBifDiagram_byVdc}
\end{figure}

\begin{figure}[htb]
	\centering
	\includegraphics[width=0.9\linewidth]{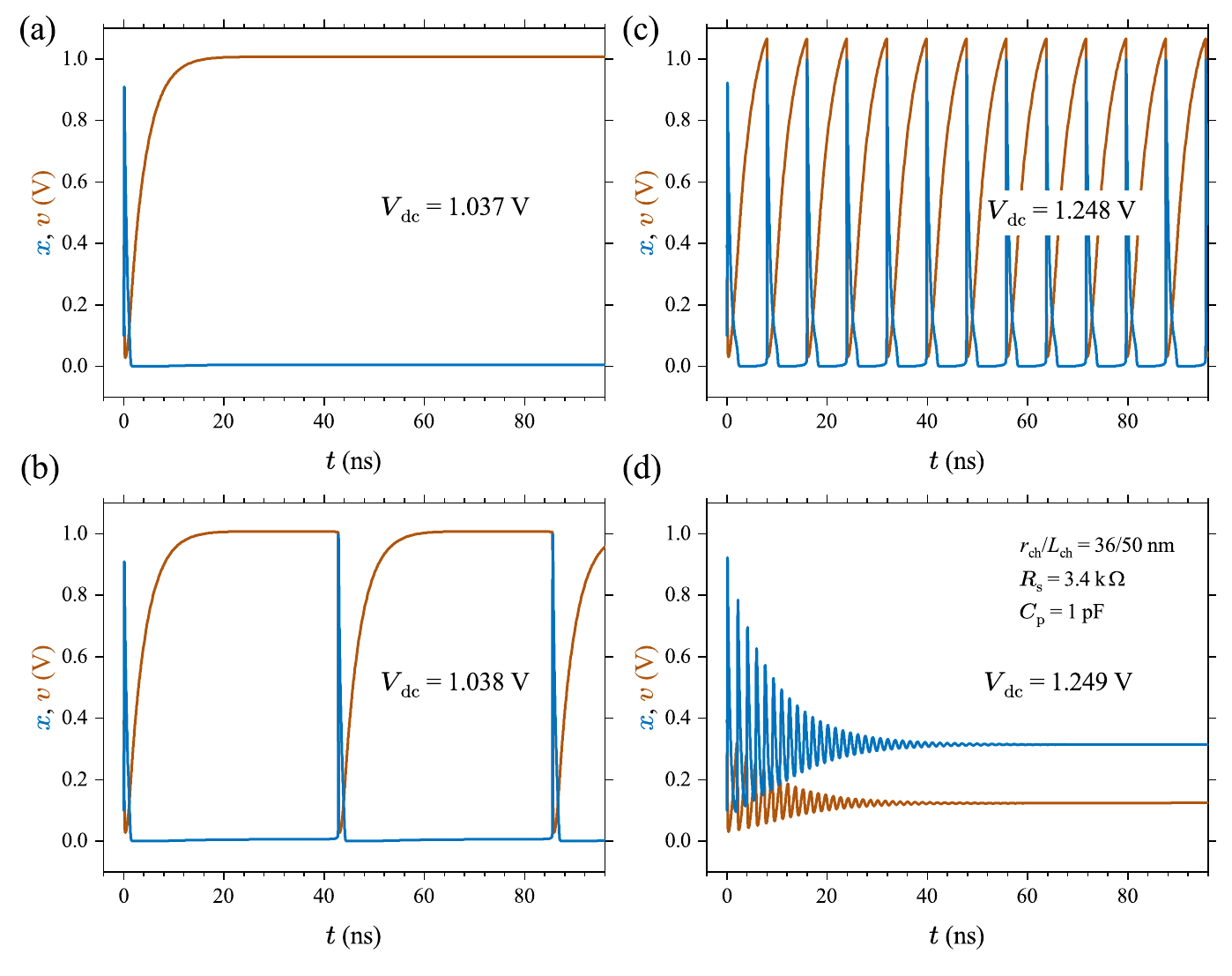}
	\caption{Time series $x(t)$ (blue) and $v(t)$ (brown) numerically solved at $V_\text{dc}$ close to critical levels for the 2D Hopf-like bifurcations of the midsize VO$_2$ Mott memristor PA oscillator with $R_\text{s}=3.4$~k$\Omega$ and $C_\text{p}=1.0$~pF, showing (a) critical damping at $V_\text{dc}=1.037$~V, (b) slow limit cycle oscillations at $V_\text{dc}=1.038$~V, (c) fast limit cycle oscillations at $V_\text{dc}=1.248$~V and (d) underdamping at $V_\text{dc}=1.249$~V. All the solutions start from the same initial condition $(x_0, v_0)=(0.1,0.39)$.}
	\label{Fig35_TS_HopfBifbyVdc}
\end{figure}

We numerically solved the phase portraits and the corresponding time series for the midsize VO$_2$ Mott memristor PA oscillator with $R_\text{s}=3.4$~k$\Omega$, $C_\text{p}=1.0$~pF and initial condition $(x_0, v_0)=(0.1,0.39)$. Figures~\ref{Fig34_HopfBifDiagram_byVdc}(a) and \ref{Fig34_HopfBifDiagram_byVdc}(b) plot the numerically-solved bifurcation diagrams (solid dots). The calculations reveal a stable limit cycle associated with a supercritical Hopf-like bifurcation if $V_\text{dc}$ is within a range bounded by the two non-hyperbolic fixed points $Q_a^*$ and $Q_c^*$, both identified by the analytical methods (see figure~\ref{Fig27_VO2PA_FPs-Vdc}). The numerically-determined critical $V_\text{dc}$ at $Q_a^*$ falls between 1.037~V and 1.038~V, which matches with the analytical result of 1.0379~V. For $Q_c^*$, it is between 1.248~V and 1.249~V, which is about 2.9~\% higher than the analytical result of 1.21355~V. At $V_\text{dc}\leq1.037$~V, the system is critically damped. After a fast transient response, $x(t)$ and $v(t)$ return to the stable steady state $Q_3$  without oscillation. See figure~\ref{Fig35_TS_HopfBifbyVdc}(a) for the case of $V_\text{dc}=1.037$~V. The system never settles on either one of the unstable $Q_1$ or $Q_2$ fixed points (short dashed lines) which are identified by the nullclines. At $V_\text{dc}\geq1.249$~V, the system is underdamped, with $x(t)$ and $v(t)$ oscillating with decaying amplitude to the stable steady state $Q_1$. See figure~\ref{Fig35_TS_HopfBifbyVdc}(d) for the case of $V_\text{dc}=1.249$~V. The system has persistent limit cycle oscillations for 1.038~V$\leq V_\text{dc}\leq1.248$~V (see figure~\ref{Fig35_TS_HopfBifbyVdc}(b) and (c) for the cases of 1.038~V and 1.248~V, respectively). Similar to the case of varying $R_\text{s}$, there is a abrupt unfolding of the bifurcation amplitude at both $Q_a^*$ and $Q_c^*$.

Figure \ref{Fig34_HopfBifDiagram_byVdc}(c) shows the $V_\text{dc}$ dependence of the limit cycle oscillation period $T_\text{lc}$. As $V_\text{dc}$ increases through the lower bound $Q_a^*$ of the limit cycle region, the persistent oscillation is initially extremely slow, i.e., $T_\text{lc}$ significantly overshoots. $T_\text{lc}$ is 12.6 times as much as the $R_\text{s}C_\text{p}$ time constant (42.7~ns vs. 3.4~ns), or 25 times as much as $2\pi/\lvert\textrm{Im}(\lambda_\pm)\rvert$ (42.7~ns vs. 1.7~ns).
As $V_\text{dc}$ increases, $T_\text{lc}$ drops super-exponentially. Figure \ref{Fig34_HopfBifDiagram_byVdc}(d) shows that near the upper bound $Q_c^*$ of the limit cycle region, the ratio between $T_\text{lc}$ and $R_\text{s}C_\text{p}$ descends to 2.34 (7.95~ns vs. 3.4~ns), which is at the same level as the cases of varying $R_\text{s}$ (figure \ref{Fig31_HopfBifDiagram_byRs}(d)) or $C_\text{p}$ (figure \ref{Fig32_HopfBifDiagram_byCp}(d)).

We end this section with a comparison between experimental characteristics of a VO$_2$ PA oscillator circuit and SPICE model simulations built upon the model equations (\ref{eqn1})--(\ref{eqn2}), using parameters listed in table~\ref{table1} and table~\ref{table2}. Details on the implementation of the Mott memristor model in SPICE can be found in the supplementary materials of \cite{Pickett12}. Figure \ref{Fig36_VO2PASimvsExp}(d) shows the circuit schematic labeled with the experimental values. The VO$_2$ nano-crossbar memristor ($X_\text{1}$) has a square junction area of 100$\times$100~nm$^2$ and oxide film thickness of 100~nm, equivalent to a circular channel radius $r_\text{ch}=56$~nm and length $L_\text{ch}=100$~nm in the model. $R_\text{e}=370~\Omega$ is the measured series resistance of metal electrodes. A parallel shunt resistance of 20~k$\Omega$ (not shown) is included in simulations to account for the parasitic insulating-phase conductance present in the VO$_2$ device. The oscillator output voltage $V_\text{out}$ is probed by an input channel of an oscilloscope with high input impedance. The current flowing through $X_\text{1}$ is monitored by a second input channel with 50~$\Omega$ input impedance. Figure \ref{Fig36_VO2PASimvsExp}(a) compares the measured (red) and the simulated (blue) $V_\text{out}$ waveforms, both showing the hallmark sawtooth relaxation oscillations. Figure \ref{Fig36_VO2PASimvsExp}(b) compares the measured and simulated current waveforms. In both (a) and (b), we found excellent agreements between measured and simulated results. Figure \ref{Fig36_VO2PASimvsExp}(c) shows the period $T_\text{n}$ of 24 consecutive oscillation peaks. Measured oscillation period irregularly fluctuates within a range from 72.4~$\upmu$s to 83.9~$\upmu$s, while simulated period is nearly a constant at 76.3~$\upmu$s. Inset of (c) is the recurrence plot (Poincar\'e plot or return map) of adjacent oscillation periods ($T_\text{n}$, $T_\text{n+1}$), showing the irregularities of experimental relaxation oscillations. The randomness in measured oscillation periods manifests that these nanoscaled Mott memristors are intrinsically stochastic, which has been demonstrated in stochastic phase-locked firing (skipping) of neuromorphic neurons built with higher-dimensional VO$_2$ Mott memristor circuits~\cite{Yi18}.

\begin{figure}[htb]
	\centering
	\includegraphics[width=0.9\linewidth]{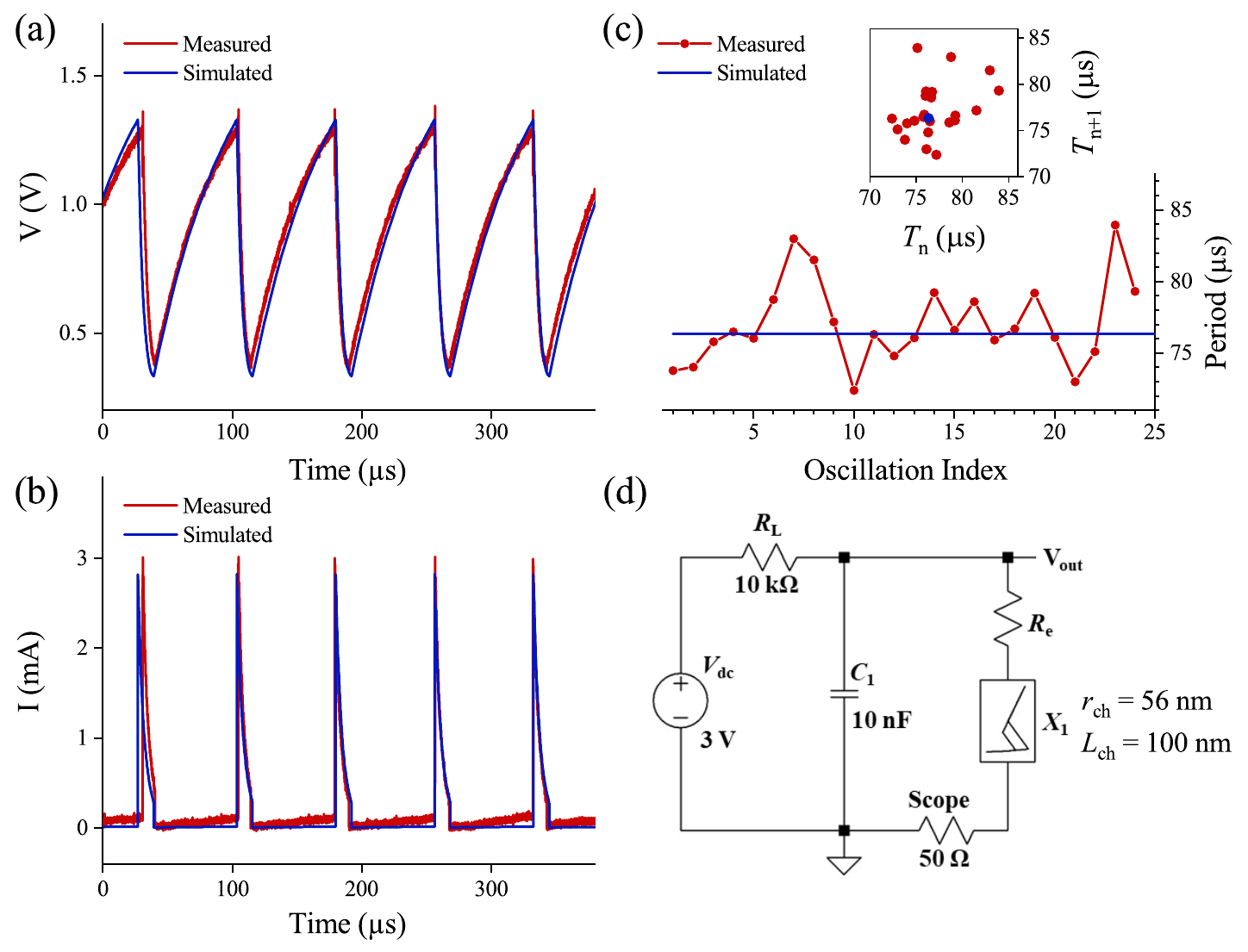}
	\caption{Experimental characteristics of a VO$_2$ PA oscillator compared with SPICE simulations. (a) Measured (red) and simulated (blue) waveforms of the output voltage ($V_\text{out}$). (b) Measured (red) and simulated (blue) waveforms of the current through the memristor ($X_\text{1}$). (c) Measured (red) and simulated oscillation period $T_\text{n}$ ($n=$~1--24). Inset:  recurrence plot ($T_\text{n}$, $T_\text{n+1}$) of the same data. (d) Circuit schematic labeled with the experimental values.}
	\label{Fig36_VO2PASimvsExp}
\end{figure}

\section[7. Conclusion]{Concluding Remarks}
In our view, the implications of locally-active memristors far exceed signal amplification or biological nerve impulse emulation. These scalable nonlinear dynamical elements enable high degree of complexity at the network building block level. From the neuronal dynamics perspective, one can borrow the concept of logical depth and measure the degree of complexity for a neuron model by the approximate number of floating point operations needed to simulate its dynamics for one millisecond duration using a digital computer~\cite{Bennett88}. The biologically plausible HH model takes 1200 FLOP/ms and has the highest degree of complexity among 11 neuron models~\cite{Izhikevich04}. The degree of complexity for a Mott memristor neuron is not less than the HH model given that both of them possess similar amount of neurocomputational properties. Architecturally simple yet dynamically rich neuron nodes may allow computationally efficient small adaptive neural networks that are suited for edge computing scenarios which requires real-time causal reasoning based on time series of unlabeled samples. Such use cases turned out to be rather challenging for today's artificial intelligence systems based on machine learning and computationally expensive offline training on the cloud. As the network scales up, more interesting complexity phenomena may emerge at mesoscopic level of neuron populations out of collective interactions of constituent nodes, such as chaotic attractor itinerancy, self-organization, and synchronization. Understanding these phenomena is crucial for replicating the perception and cognition capabilities of the brain.

\section*{Acknowledgments}
This work was funded by HRL Laboratories, LLC. The author is very grateful to the leaders and colleagues of Sensors and Electronics Laboratory for their unwavering support of this research. Kenneth K. Tsang, James M. Chappell, Stephen K. Lam, Xiwei Bai, Jack A. Crowell and Elias A. Flores are acknowledged for their contributions to the experimental data presented in this manuscript.

\end{document}